\documentclass[aps,rmp,twocolumn,eqsecnum,amssymb]{revtex4}
\usepackage{graphicx}
\begin{document}
\title{The holographic principle}
\author{Raphael Bousso}
\affiliation{Institute for Theoretical Physics, 
University of California, Santa Barbara, California 93106, U.S.A.}
\email{bousso@itp.ucsb.edu}
\begin{abstract}

  There is strong evidence that the area of any surface limits the
information content of adjacent spacetime regions, at $1.4 \times
10^{69}$ bits per square meter.  We review the developments that have
led to the recognition of this entropy bound, placing special emphasis
on the quantum properties of black holes.  The construction of
light-sheets, which associate relevant spacetime regions to any given
surface, is discussed in detail.  We explain how the bound is tested
and demonstrate its validity in a wide range of examples.  

  A universal relation between geometry and information is thus
uncovered.  It has yet to be explained.  The holographic principle
asserts that its origin must lie in the number of fundamental degrees
of freedom involved in a unified description of spacetime and matter.
It must be manifest in an underlying quantum theory of gravity.  We
survey some successes and challenges in implementing the holographic
principle.

\end{abstract}
\maketitle
\tableofcontents

\section{Introduction}
\label{sec-intro}

\subsection{A principle for quantum gravity}
\label{sec-ia}

Progress in fundamental physics has often been driven by the
recognition of a new principle, a key insight to guide the search for
a successful theory.  Examples include the principles of relativity,
the equivalence principle, and the gauge principle.  Such principles
lay down general properties that must be incorporated into the laws of
physics.

A principle can be sparked by contradictions between existing
theories.  By judiciously declaring which theory contains the elements
of a unified framework, a principle may force other theories to be
adapted or superceded.  The special theory of relativity, for example,
reconciles electrodynamics with Galilean kinematics at the expense of
the latter.

A principle can also arise from some newly recognized pattern, an
apparent law of physics that stands by itself, both uncontradicted and
unexplained by existing theories.  A principle may declare this
pattern to be at the core of a new theory altogether.  

In Newtonian gravity, for example, the proportionality of
gravitational and inertial mass in all bodies seems a curious
coincidence that is far from inevitable.  The equivalence principle
demands that this pattern must be made manifest in a new theory.  This
led Einstein to the general theory of relativity, in which the
equality of gravitational and inertial mass is built in from the
start.  Because all bodies follow geodesics in a curved spacetime,
things simply couldn't be otherwise.

The {\em holographic principle\/} belongs in the latter class.  The
unexplained ``pattern'', in this case, is the existence of a precise,
general, and surprisingly strong limit on the information content of
spacetime regions.  This pattern has come to be recognized in stages;
its present, most general form is called the {\em covariant entropy
bound}.  The holographic principle asserts that this bound is not a
coincidence, but that its origin must be found in a new theory.

The covariant entropy bound relates aspects of spacetime geometry to
the number of quantum states of matter.  This suggests that any theory
that incorporates the holographic principle must unify matter,
gravity, and quantum mechanics.  It will be a quantum theory of
gravity, a framework that transcends general relativity and quantum
field theory.  

This expectation is supported by the close ties between
the covariant entropy bound and the semi-classical properties of black
holes.  It has been confirmed---albeit in a limited context---by
recent results in string theory.  

The holographic principle conflicts with received wisdom; in this
sense, it also belongs in the former class.  Conventional theories are
local; quantum field theory, for example, contains degrees of freedom
at every point in space.  Even with a short distance cutoff, the
information content of a spatial region would appear to grow with the
volume.  The holographic principle, on the other hand, implies that
the number of fundamental degrees of freedom is related to the area of
surfaces in spacetime.  Typically, this number is drastically smaller
than the field theory estimate.  

Thus, the holographic principle calls into question not only the
fundamental status of field theory but the very notion of locality.
It gives preference, as we shall see, to the preservation of
quantum-mechanical unitarity.

In physics, information can be encoded in a variety of ways: by the
quantum states, say, of a conformal field theory, or by a lattice of
spins.  Unfortunately, for all its precise predictions about the {\em
number\/} of fundamental degrees of freedom in spacetime, the
holographic principle betrays little about their character.  The
amount of information is strictly determined, but not its form.  It is
interesting to contemplate the notion that pure, abstract information
may underlie all of physics.  But for now, this austerity frustrates
the design of concrete models incorporating the holographic principle.

Indeed, a broader caveat is called for.  The covariant entropy bound
is a compelling pattern, but it may still prove incorrect or merely
accidental, signifying no deeper origin.  If the bound does stem from
a fundamental theory, that relation could be indirect or peripheral,
in which case the holographic principle would be unlikely to guide us
to the core ideas of the theory.  All that aside, the holographic
principle is likely only one of several independent conceptual
advances needed for progress in quantum gravity.

At present, however, quantum gravity poses an immense problem tackled
with little guidance.  Quantum gravity has imprinted few traces on
physics below the Planck energy.  Among them, the information content
of spacetime may well be the most profound.

The direction offered by the holographic principle is impacting
existing frameworks and provoking new approaches.  In particular, it
may prove beneficial to the further development of string theory,
widely (and, in our view, justly) considered the most compelling of
present approaches.

This article will outline the case for the holographic principle whilst
providing a starting point for further study of the literature.  The
material is not, for the most part, technical.  The main mathematical
aspect, the construction of light-sheets, is rather straightforward.
In order to achieve a self-contained presentation, some basic material
on general relativity has been included in an appendix.

In demonstrating the scope and power of the holographic correspondence
between areas and information, our ultimate task is to convey its
character as a law of physics that captures one of the most intriguing
aspects of quantum gravity.  If the reader is led to contemplate the
origin of this particular pattern nature has laid out, our review will
have succeeded.

\subsection{Notation and conventions}
\label{sec-not}

Throughout this paper, Planck units will be used:
\begin{equation}
\hbar = G=c=k=1,
\end{equation}
where $G$ is Newton's constant, $\hbar$ is Planck's constant, $c$
is the speed of light, and $k$ is Boltzmann's constant.  In particular, all
areas are measured in multiples of the square of the Planck length,
\begin{equation}
l_{\rm P}^2 = {G\hbar\over c^3} = 2.59 \times 10^{-66} \mbox{cm}^2.
\label{eq-lpl}
\end{equation}
The Planck units of energy density, mass, temperature, and other
quantities are converted to cgs units, e.g., in Wald (1984), whose
conventions we follow in general.  For a small number of key formulas,
we will provide an alternate expression in which all constants are
given explicitly.

We consider spacetimes of arbitrary dimension $D\geq 4$, unless noted
otherwise.  In explicit examples we often take $D=4$ for definiteness.
The appendix fixes the metric signature and defines ``surface'',
``hypersurface'', ``null'', and many other terms from general
relativity.  The term ``light-sheet'' is defined in Sec.~\ref{sec-bb}.

``GSL'' stands for the generalized second law of thermodynamics
(Sec.~\ref{sec-gsl}).  The number of degrees of freedom of a quantum
system, $N$, is defined as the logarithm of the dimension, ${\cal N}$,
of its Hilbert space in Sec.~\ref{sec-ndof}.  Equivalently, $N$ can be
defined as the number of bits of information times $\ln 2$.

\subsection{Outline}
\label{sec-outline}

In Sec.~\ref{sec-bh}, we review Bekenstein's (1972) notion of black
hole entropy and the related discovery of upper bounds on the entropy
of matter systems.  Assuming weak gravity, spherical symmetry, and
other conditions, one finds that the entropy in a region of space is
limited by the area of its boundary.%
\footnote{The metaphorical name of the principle ('t~Hooft, 1993)
originates here.  In many situations, the covariant entropy bound
dictates that all physics in a region of space is described by data
that fit on its boundary surface, at one bit per Planck area
(Sec.~\ref{sec-spt}).  This is reminiscent of a hologram.  Holography
is an optical technology by which a three-dimensional image is stored
on a two-dimensional surface via a diffraction pattern.  (To avoid any
confusion: this linguistic remark will remain our only usage of the
term in its original sense.)  From the present point of view, the
analogy has proven particularly apt.  In both kinds of ``holography'',
light rays play a key role for the imaging (Sec.~\ref{sec-bb}).
Moreover, the holographic code is not a straightforward projection, as
in ordinary photography; its relation to the three-dimensional image
is rather complicated.  (Most of our intuition in this regard has come
from the AdS/CFT correspondence, Sec.~\ref{sec-ads}.)  Susskind's
(1995) quip that the world is a ``hologram'' is justified by the
existence of preferred surfaces in spacetime, on which all of the
information in the universe can be stored at no more than one bit per
Planck area (Sec.~\ref{sec-screens}).}
Based on this ``spherical entropy bound'', 't~Hooft (1993) and
Susskind (1995b) formulated a holographic principle.  We discuss
motivations for this radical step.

The spherical entropy bound depends on assumptions that are clearly
violated by realistic physical systems.  {\em A priori\/} there is no
reason to expect that the bound has universal validity, nor that it
admits a reformulation that does.  Yet, if the number of degrees of
freedom in nature is as small as 't~Hooft and Susskind asserted, one
would expect wider implications for the maximal entropy of matter.

In Sec.~\ref{sec-seb}, however, we demonstrate that a naive
generalization of the spherical entropy bound is unsuccessful.  The
``spacelike entropy bound'' states that the entropy in a given spatial
volume, irrespective of shape and location, is always less than the
surface area of its boundary.  We consider four examples of realistic,
commonplace physical systems, and find that the spacelike entropy
bound is violated in each one of them.

In light of these difficulties, some authors, forgoing complete
generality, searched instead for reliable conditions under which the
spacelike entropy bound holds.  We review the difficulties faced in
making such conditions precise even in simple cosmological models.

Thus, the idea that the area of surfaces generally bounds the entropy
in enclosed spatial volumes has proven wrong; it can be neither the
basis nor the consequence of a fundamental principle.  This review
would be incomplete if it failed to stress this point.  Moreover, the
ease with which the spacelike entropy bound (and several of its
modifications) can be excluded underscores that a general entropy
bound, if found, is no triviality.  The counterexamples to the
spacelike bound later provide a useful testing ground for the
covariant bound.

Inadequacies of the spacelike entropy bound led Fischler and Susskind
(1998) to a bound involving light cones.  The covariant entropy bound
(Bousso, 1999a), presented in Sec.~\ref{sec-bb}, refines and
generalizes this approach.  Given any surface $B$, the bound states
that the entropy on any light-sheet of $B$ will not exceed the area of
$B$.  Light-sheets are particular hypersurfaces generated by light
rays orthogonal to $B$.  The light rays may only be followed as long
as they are not expanding.  We explain this construction in detail.

After discussing how to define the entropy on a light-sheet, we spell
out known limitations of the covariant entropy bound.  The bound is
presently formulated only for approximately classical geometries, and
one must exclude unphysical matter content, such as large negative
energy.  We conclude that the covariant entropy bound is well-defined
and testable in a vast class of solutions.  This includes all
thermodynamic systems and cosmologies presently known or considered
realistic.

In Sec.~\ref{sec-lsd} we review the geometric properties of
light-sheets, which are central to the operation of the covariant
entropy bound.  Raychaudhuri's equation is used to analyse the effects
of entropy on light-sheet evolution.  By construction, a light-sheet
is generated by light rays that are initially either parallel or
contracting.  Entropic matter systems carry mass, which causes the
bending of light.  

This means that the light rays generating a light-sheet will be
focussed towards each other when they encounter entropy.  Eventually
they self-intersect in a caustic, where they must be terminated
because they would begin to expand.  This mechanism would provide an
``explanation'' of the covariant entropy bound if one could show that
the mass associated with entropy is necessarily so large that
light-sheets focus and terminate before they encounter more entropy
than their initial area.

Unfortunately, present theories do not impose an independent,
fundamental lower bound on the energetic price of entropy.  However,
Flanagan, Marolf, and Wald (2000) were able to identify conditions on
entropy density which are widely satisfied in nature and which are
sufficient to guarantee the validity of the covariant entropy bound.
We review these conditions.

The covariant bound can also be used to obtain sufficient criteria
under which the spacelike entropy bound holds.  Roughly, these
criteria can be summarized by demanding that gravity be weak.
However, the precise condition requires the construction of
light-sheets; it cannot be formulated in terms of intrinsic properties
of spatial volumes.  

The event horizon of a black hole is a light-sheet of its final
surface area.  Thus, the covariant entropy bound includes to the
generalized second law of thermodynamics in black hole formation as a
special case.  More broadly, the generalized second law, as well as
the Bekenstein entropy bound, follow from a strengthened version of
the covariant entropy bound.

In Sec.~\ref{sec-tests}, the covariant entropy bound is applied to a
variety of thermodynamic systems and cosmological spacetimes.  This
includes all of the examples in which the spacelike entropy bound is
violated.  We find that the covariant bound is satisfied in each case.

In particular, the bound is found to hold in strongly gravitating
regions, such as cosmological spacetimes and collapsing objects.
Aside from providing evidence for the general validity of the bound,
this demonstrates that the bound (unlike the spherical entropy bound)
holds in a regime where it cannot be derived from black hole
thermodynamics.

In Sec.~\ref{sec-hp}, we arrive at the holographic principle.  We note
that the covariant entropy bound holds with remarkable generality but
is not logically implied by known laws of physics.  We conclude that
the bound has a fundamental origin.  As a universal limitation on the
information content of Lorentzian geometry, the bound should be
manifest in a quantum theory of gravity.  We formulate the holographic
principle and list some of its implications.  The principle poses a
challenge for local theories.  It suggests a preferred role for null
hypersurfaces in the classical limit of quantum gravity.

In Sec.~\ref{sec-hsht} we analyze an example of a holographic theory.
Quantum gravity in certain asymptotically Anti-de~Sitter spacetimes is
fully defined by a conformal field theory.  The latter theory contains
the correct number of degrees of freedom demanded by the holographic
principle.  It can be thought of as living on a kind of holographic
screen at the boundary of spacetime and containing one bit of
information per Planck area.

Holographic screens with this information density can be constructed
for arbitrary spacetimes---in this sense, the world is a hologram.  In
most other respects, however, global holographic screens do not
generally support the notion that a holographic theory is a
conventional field theory living at the boundary of a spacetime.  

At present, there is much interest in finding more general holographic
theories.  We discuss the extent to which string theory, and a number
of other approaches, have realized this goal.  A particular area of
focus is de~Sitter space, which exhibits an absolute entropy bound.
We review the implications of the holographic principle in such
spacetimes.

\subsection{Related subjects and further reading}
\label{sec-further}

The holographic principle has developed from a large set of ideas and
results, not all of which seemed mutually related at first.  This is
not a historical review; we have aimed mainly at achieving a coherent,
modern perspective on the holographic principle.  We do not give equal
emphasis to all developments, and we respect the historical order only
where it serves the clarity of exposition.  Along with length
constraints, however, this approach has led to some omissions and
shortcomings, for which we apologize.

We have chosen to focus on the covariant entropy bound because it can
be tested using only quantum field theory and general relativity.  Its
universality motivates the holographic principle independently of any
particular ansatz for quantum gravity (say, string theory) and without
additional assumptions (such as unitarity).  It yields a precise and
general formulation.

Historically, the idea of the holographic principle was tied, in part,
to the debate about information loss in black holes\footnote{See, for
example, Hawking (1976b, 1982), Page (1980, 1993), Banks, Susskind,
and Peskin (1984), 't~Hooft (1985, 1988, 1990), Polchinski and
Strominger (1994), Strominger (1994).}  and to the notion of black
hole complementarity.\footnote{See, e.g., 't~Hooft (1991), Susskind,
Thorlacius, and Uglum (1993), Susskind (1993b), Stephens, 't~Hooft,
and Whiting (1994), Susskind and Thorlacius (1994).  For recent
criticism, see Jacobson (1999).}  Although we identify some of the
connections, our treatment of these issues is far from comprehensive.
Reviews include Thorlacius (1995), Verlinde (1995), Susskind and Uglum
(1996), Bigatti and Susskind (2000), and Wald (2001).

Some aspects of what we now recognize as the holographic principle
were encountered, at an early stage, as features of string theory.
(This is as it should be, since string theory is a quantum theory of
gravity.)  In the infinite momentum frame, the theory admits a
lower-dimensional description from which the gravitational dynamics of
the full spacetime arises non-trivially (Giles and Thorn, 1977; Giles,
McLerran, and Thorn, 1978; Thorn, 1979, 1991, 1995, 1996; Klebanov and
Susskind, 1988).  Susskind (1995b) placed this property of string
theory in the context of the holographic principle and related it to
black hole thermodynamics and entropy limitations.

Some authors have traced the emergence of the holographic principle
also to other approaches to quantum gravity; see Smolin (2001) for a
discussion and further references.

By tracing over a region of space one obtains a density matrix.
Bombelli {\em et al.} (1986) showed that the resulting entropy is
proportional to the boundary area of the region.  A more general
argument was given by Srednicki (1993).  Gravity does not enter in
this consideration.  Moreover, the entanglement entropy is generally
unrelated to the size of the Hilbert space describing either side of
the boundary.  Thus, it is not clear to what extent this suggestive
result is related to the holographic principle.

This is not a review of the AdS/CFT correspondence (Maldacena, 1998;
see also Gubser, Klebanov, and Polyakov, 1998; Witten, 1998).  This
rich and beautiful duality can be regarded (among its many interesting
aspects) as an implementation of the holographic principle in a
concrete model.  Unfortunately, it applies only to a narrow class of
spacetimes of limited physical relevance.  By contrast, the
holographic principle claims a far greater level of generality---a
level at which it continues to lack a concrete implementation.

We will broadly discuss the relation between the AdS/CFT
correspondence and the holographic principle, but we will not dwell on
aspects that seem particular to AdS/CFT.  (In particular, this means
that the reader should not expect a discussion of every paper
containing the word ``holographic'' in the title!)  A detailed
treatment of AdS/CFT would go beyond the purpose of the present text.
An extensive review has been given by Aharony {\em et al.} (2000).

The AdS/CFT correspondence is closely related to some recent models of
our 3+1 dimensional world as a defect, or brane, in a 4+1 dimensional
AdS space.  In the models of Randall and Sundrum (1999a,b), the
gravitational degrees of freedom of the extra dimension appear on the
brane as a dual field theory under the AdS/CFT correspondence.  While
the holographic principle can be considered a prerequisite for the
existence of such models, their detailed discussion would not
significantly strengthen our discourse.  Earlier seminal papers in
this area include Ho\v{r}ava and Witten (1996a,b).

A number of authors (e.g., Brustein and Veneziano, 2000; Verlinde,
2000; Brustein, Foffa, and Veneziano, 2001; see Cai, Myung, and Ohta,
2001, for additional references) have discussed interesting bounds
which are not directly based on the area of surfaces.  Not all of
these bounds appear to be universal.  Because their relation to the
holographic principle is not entirely clear, we will not attempt to
discuss them here.  Applications of entropy bounds to string cosmology
(e.g., Veneziano, 1999a; Bak and Rey, 2000b; Brustein, Foffa, and
Sturani, 2000) are reviewed by Veneziano (2000).

The holographic principle has sometimes been said to exclude certain
physically acceptable solutions of Einstein's equations because they
appeared to conflict with an entropy bound.  The covariant bound has
exposed these tensions as artifacts of the limitations of earlier
entropy bounds.  Indeed, this review bases the case for a holographic
principle to a large part on the very generality of the covariant
bound.  However, the holographic principle does limit the
applicability of quantum field theory on cosmologically large scales.
It calls into question the conventional analysis of the cosmological
constant problem (Cohen, Kaplan, and Nelson, 1999; Ho\v{r}ava, 1999;
Banks, 2000a; Ho\v{r}ava and Minic, 2000; Thomas, 2000).  It has also
been applied to the calculation of anisotropies in the cosmic
microwave background (Hogan, 2002a,b).  The study of cosmological
signatures of the holographic principle may be of great value, since
it is not clear whether more conventional imprints of short-distance
physics on the early universe are observable even in principle (see,
e.g., Kaloper {\em et al.}, 2002, and references therein).

Most attempts at implementing the holographic principle in a unified
theory are still in their infancy.  It would be premature to attempt a
detailed review; some references are given in Sec.~\ref{sec-toe}.

Other recent reviews overlapping with some or all of the topics
covered here are Bigatti and Susskind (2000), Bousso (2000a), 't~Hooft
(2000b), Bekenstein (2001) and Wald (2001).  Relevant textbooks include
Hawking and Ellis (1973); Misner, Thorne, and Wheeler (1973); Wald
(1984, 1994); Green, Schwarz, and Witten (1987); and Polchinski
(1998).

\section{Entropy bounds from black holes}
\label{sec-bh}

This section reviews black hole entropy, some of the entropy bounds
that have been inferred from it, and their relation to 't~Hooft's
(1993) and Susskind's (1995b) proposal of a holographic principle.

The entropy bounds discussed in this section are ``universal''
(Bekenstein, 1981) in the sense that they are independent of the
specific characteristics and composition of matter systems.  Their
validity is not truly universal, however, because they apply only when
gravity is weak.

We consider only Einstein gravity.  For black hole thermodynamics in
higher-derivative gravity, see, e.g., Myers and Simon (1988), Jacobson
and Myers (1993), Wald (1993), Iyer and Wald (1994, 1995), Jacobson,
Kang, and Myers (1994), and the review by Myers
(1998).\footnote{Abdalla and Correa-Borbonet (2001) have commented on
entropy bounds in this context.}

\subsection{Black hole thermodynamics}
\label{sec-bhtd}

The notion of black hole entropy is motivated by two results in
general relativity.

\subsubsection{Area theorem}
\label{sec-areathm}

The {\sl area theorem\/} (Hawking, 1971) states that {\em the area of
a black hole event horizon never decreases with time:}
\begin{equation}
dA \geq 0.
\label{eq-areathm}
\end{equation}  
Moreover, if two black holes merge, the area of the new black hole
will exceed the total area of the original black holes.  

For example, an object falling into a Schwarzschild black hole will
increase the mass of the black hole, $M$.%
\footnote{This assumes that the object has positive mass.  Indeed, the
assumptions in the proof of the theorem include the null energy
condition.  This condition is given in the Appendix, where the
Schwarzschild metric is also found.}
Hence the horizon area, $A=16\pi M^2$ in $D=4$, increases.  On the
other hand, one would not expect the area to decrease in any classical
process, because the black hole cannot emit particles.

The theorem suggests an analogy between black hole area and
thermodynamic entropy.

\subsubsection{No-hair theorem}
\label{sec-no-hair} 
 
Work of Israel (1967, 1968), Carter (1970), Hawking (1971, 1972), and
others, implies the curiously named {\sl no-hair theorem}: {\em A
stationary black hole is characterized by only three quantities: mass,
angular momentum, and charge.}\footnote{Proofs and further details can
be found, e.g., in Hawking and Ellis (1973), or Wald (1984).  This
form of the theorem holds only in $D=4$.  Gibbons, Ida, and Shiromizu
(2002) have recently given a uniqueness proof for static black holes
in $D>4$.  Remarkably, Emparan and Reall (2001) have found a
counterexample to the stationary case in $D=5$.  This does not affect
the present argument, in which the no hair theorem plays a heuristic
role.}

Consider a complex matter system, such as a star, that collapses to
form a black hole.  The black hole will eventually settle down into a
final, stationary state.  The no-hair theorem implies that this state
is unique.

From an outside observer's point of view, the formation of a black
hole appears to violate the second law of thermodynamics.  The phase
space appears to be drastically reduced.  The collapsing system may
have arbitrarily large entropy, but the final state has none at all.
Different initial conditions will lead to indistinguishable results.

A similar problem arises when a matter system is dropped into an
existing black hole.  Geroch has proposed a further method for
violating the second law, which exploits a classical black hole to
transform heat into work; see Bekenstein (1972) for details.

\subsubsection{Bekenstein entropy and the generalized second law}
\label{sec-gsl}

Thus, the no-hair theorem poses a paradox, to which the area theorem
suggests a resolution.  When a thermodynamic system disappears behind
a black hole's event horizon, its entropy is lost to an outside
observer.  The area of the event horizon will typically grow when the
black hole swallows the system.  Perhaps one could regard this area
increase as a kind of compensation for the loss of matter entropy?

Based on this reasoning, Bekenstein (1972, 1973, 1974) suggested that
a black hole actually carries an entropy equal to its horizon area,
$S_{\rm BH} = \eta A$, where $\eta$ is a number of order unity.  In
Sec.~\ref{sec-bh-radiation} it will be seen that $\eta={1\over 4}$
(Hawking, 1974):
\begin{equation}
S_{\rm BH} = {A\over 4}.
\label{eq-eta}
\end{equation}
[In full, $S_{\rm BH} = kA c^3 / (4G \hbar)$.]  The entropy of a black
hole is given by a quarter of the area of its horizon in Planck units.
In ordinary units, it is the horizon area divided by about
$10^{-69}$\/m$^2$.

Moreover, Bekenstein (1972, 1973, 1974) proposed that the second law
of thermodynamics holds only for the {\em sum\/} of black hole entropy
and matter entropy:
\begin{equation}
dS_{\rm total} \geq 0.
\label{eq-gsl}
\end{equation}
For ordinary matter systems alone, the second law need not hold.  But
if the entropy of black holes, Eq.~(\ref{eq-eta}), is included in the
balance, the total entropy will never decrease.  This is referred to
as the {\em generalized second law\/} or {\em GSL}.

The content of this statement may be illustrated as follows.  Consider
a thermodynamic system ${\cal T}$, consisting of well-separated,
non-interacting components.  Some components, labeled ${\cal C}_i$,
may be thermodynamic systems made from ordinary matter, with entropy
$S({\cal C}_i)$.  The other components, ${\cal B}_j$, are black holes,
with horizon areas $A_j$.  The total entropy of ${\cal T}$ is given by
\begin{equation}
S_{\rm total}^{\rm initial} = S_{\rm matter} + S_{\rm BH}.
\end{equation}
Here, $S_{\rm matter} = \sum S({\cal C}_i)$ is the total entropy of
all ordinary matter.  $S_{\rm BH} = \sum {A_j\over 4}$ is the total
entropy of all black holes present in ${\cal T}$.

Now suppose the components of ${\cal T}$ are allowed to interact until a
new equilibrium is established.  For example, some of the matter
components may fall into some of the black holes.  Other matter
components might collapse to form new black holes.  Two or more black
holes may merge.  In the end, the system ${\cal T}$ will consist of a
new set of components $\hat{\cal C}_i$ and $\hat{\cal B}_j$, for which
one can again compute a total entropy, $S_{\rm total}^{\rm final}$.
The GSL states that
\begin{equation}
S_{\rm total}^{\rm final} \geq S_{\rm total}^{\rm initial}.
\end{equation}

What is the microscopic, statistical origin of black hole entropy?  We
have learned that a black hole, viewed from the outside, is unique
classically.  The Bekenstein-Hawking formula, however, suggests that
it is compatible with $e^{S_{\rm BH}}$ independent quantum states.
The nature of these quantum states remains largely mysterious.  This
problem has sparked sustained activity through various different
approaches, too vast in scope to sketch in this review.  

However, one result stands out because of its quantitative accuracy.
Recent developments in string theory have led to models of limited
classes of black holes in which the microstates can be identified and
counted (Strominger and Vafa, 1996; for a review, see, e.g., Peet,
2000).  The formula $S=A/4$ was precisely confirmed by this
calculation.

\subsubsection{Hawking radiation}
\label{sec-bh-radiation}

Black holes clearly have a mass, $M$.  If Bekenstein entropy, $S_{\rm
BH}$, is to be taken seriously, then the first law of thermodynamics
dictates that black holes must have a temperature, $T$:
\begin{equation}
dM = T dS_{\rm BH}.
\label{eq-fl}
\end{equation}
Indeed, Einstein's equations imply an analogous ``first law of black
hole mechanics'' (Bardeen, Carter, and Hawking, 1973).  The entropy is
the horizon area, and the surface gravity of the black hole, $\kappa$,
plays the role of the temperature:
\begin{equation}
dM = \frac{\kappa}{8\pi} dA.
\end{equation}
For a definition of $\kappa$, see Wald (1984); e.g., a Schwarzschild
black hole in $D=4$ has $\kappa = (4M)^{-1}$.

It may seem that this has taken the thermodynamic analogy a step too
far.  After all, a blackbody with non-zero temperature must radiate.
But for a black hole this would seem impossible.  Classically, no
matter can escape from it, so its temperature must be exactly zero.

This paradox was resolved by the discovery that black holes do in fact
radiate via a quantum process.  Hawking (1974, 1975) showed by a
semi-classical calculation that a distant observer will detect a
thermal spectrum of particles coming from the black hole, at a
temperature
\begin{equation}
T = \frac{\kappa}{2\pi}.
\end{equation}
For a Schwarzschild black hole in $D=4$, this temperature is $\hbar
c^3 / (8 \pi G k M)$, or about $10^{26}$ Kelvin divided by the mass of
the black hole in grams.  Note that such black holes have negative
specific heat.

The discovery of Hawking radiation clarified the interpretation of the
thermodynamic description of black holes.  What might otherwise have
been viewed as a mere analogy (Bardeen, Carter, and Hawking, 1973) was
understood to be a true physical property.  The entropy and
temperature of a black hole are no less real than its mass.  

In particular, Hawking's result affirmed that the entropy of black
holes should be considered a genuine contribution to the total entropy
content of the universe, as Bekenstein (1972, 1973, 1974) had
anticipated.  Via the first law of thermodynamics, Eq.~(\ref{eq-fl}),
Hawking's calculation fixes the coefficient $\eta$ in the Bekenstein
entropy formula, Eq.~(\ref{eq-eta}), to be $1/4$.

A radiating black hole loses mass, shrinks, and eventually disappears
unless it is stabilized by charge or a steady influx of energy.  Over
a long time of order $M^{D-1\over D-3}$, this process converts the
black hole into a cloud of radiation.  (See Sec.~\ref{sec-bhc} for the
question of unitarity in this process.)

It is natural to study the operation of the GSL in the two types of
processes discussed in Sec.~\ref{sec-no-hair}.  We will first discuss
the case in which a matter system is dropped into an existing black
hole.  Then we will turn to the process in which a black hole is
formed by the collapse of ordinary matter.  In both cases, ordinary
entropy is converted into horizon entropy.  

A third process, which we will not discuss in detail, is the Hawking
evaporation of a black hole.  In this case, the horizon entropy is
converted back into radiation entropy.  This type of process was not
anticipated when Bekenstein (1972) proposed black hole entropy and the
GSL.  It is all the more impressive that the GSL holds also in this
case (Bekenstein, 1975; Hawking, 1976a).  Page (1976) has estimated
that the entropy of Hawking radiation exceeds that of the evaporated
black hole by 62\%.

\subsection{Bekenstein bound}
\label{sec-bekbound} 

When a matter system is dropped into a black hole, its entropy is lost
to an outside observer.  That is, the entropy $S_{\rm matter}$ starts
at some finite value and ends up at zero.  But the entropy of the
black hole increases, because the black hole gains mass, and so its
area $A$ will grow.  Thus it is at least conceivable that the total
entropy, $S_{\rm matter} + \frac{A}{4}$, does not decrease in the
process, and that therefore the generalized second law of
thermodynamics, Eq.~(\ref{eq-gsl}), is obeyed.

Yet it is by no means obvious that the generalized second law will
hold.  The growth of the horizon area depends essentially on the mass
that is added to the black hole; it does not seem to care about the
entropy of the matter system.  If it were possible to have matter
systems with arbitrarily large entropy at a given mass and size, the
generalized second law could still be violated.

The thermodynamic properties of black holes developed in the previous
subsection, including the assignment of entropy to the horizon, are
sufficiently compelling to be considered laws of nature.  Then one may
turn the above considerations around and demand that the generalized
second law hold in all processes.  One would expect that this would
lead to a universal bound on the entropy of matter systems in terms of
their extensive parameters.  

For any weakly gravitating matter system in asymptotically flat space,
Bekenstein (1981) has argued that the GSL implies the following bound:
\begin{equation}
S_{\rm matter} \leq 2 \pi E R.
\label{eq-bekbound}
\end{equation}
[In full, $S \leq 2 \pi k E R / (\hbar c)$; note that Newton's
constant does not enter.]  Here, $E$ is the total mass-energy of the
matter system.  The circumferential radius $R$ is the radius of the
smallest sphere that fits around the matter system (assuming that
gravity is sufficiently weak to allow for a choice of time slicing
such that the matter system is at rest and space is almost Euclidean).

We will begin with an argument for this bound in arbitrary spacetime
dimension $D$ that involves a strictly classical analysis of the {\em
Geroch process}, by which a system is dropped into a black hole from
the vicinity of the horizon.  We will then show, however, that a
purely classical treatment is not tenable.  The extent to which
quantum effects modify, or perhaps invalidate, the derivation of the
Bekenstein bound from the GSL is controversial.  The gist of some of
the pertinent arguments will be given here, but the reader is referred
to the literature for the subtleties.

\subsubsection{Geroch process}
\label{sec-geroch} 

Consider a weakly gravitating stable thermodynamic system of total
energy $E$.  Let $R$ be the radius of the smallest $D-2$ sphere
circumscribing the system.  To obtain an entropy bound, one may move
the system from infinity into a Schwarzschild black hole whose radius,
$b$, is much larger than $R$ but otherwise arbitrary.  One would like
to add as little energy as possible to the black hole, so as to
minimize the increase of the black hole's horizon area and thus to
optimize the tightness of the entropy bound.  Therefore, the strategy
is to extract work from the system by lowering it slowly until it is
just outside the black hole horizon, before one finally drops it in.

The mass added to the black hole is given by the energy $E$ of the
system, redshifted according to the position of the center of mass at
the drop-off point, at which the circumscribing sphere almost touches
the horizon.  Within its circumscribing sphere, one may orient the
system so that its center of mass is ``down'', i.e., on the side of the
black hole.  Thus the center of mass can be brought to within a proper
distance $R$ from the horizon, while all parts of the system remain
outside the horizon.  Hence, one must calculate the redshift factor at
radial proper distance $R$ from the horizon.

The Schwarzschild metric is given by
\begin{equation}
ds^2 = -V(r) dt^2 + V(r)^{-1} dr^2 + r^2 d\Omega_{D-2}^2,
\label{eq-schsch}
\end{equation}
where
\begin{equation}
V(r) = 1 - \left( \frac{b}{r} \right)^{D-3} \equiv \left[ \chi(r)
\right]^2
\label{eq-vofr3}
\end{equation}
defines the redshift factor, $\chi$ (Myers and Perry, 1986).  The
black hole radius is related to the mass at infinity, $M$, by
\begin{equation}
b^{D-3} = \frac{16 \pi M}{(D-2) {\cal A}_{D-2}},
\label{eq-bm}
\end{equation}
where ${\cal A}_{D-2} = 2 \pi^{D-1\over 2}/\Gamma({D-1\over 2})$ is
the area of a unit $D-2$ sphere.  The black hole has horizon area
\begin{equation}
A = {\cal A}_{D-2} b^{D-2}.
\label{eq-ab}
\end{equation}

Let $c$ be the radial coordinate distance from the horizon:
\begin{equation}
c = r-b.
\end{equation}
Near the horizon, the redshift
factor is given by
\begin{equation}
\chi^2(c) = (D-3) \frac{c}{b},
\end{equation}
to leading order in $c/b$.  The proper distance $l$ is related to the
coordinate distance $c$ as follows:
\begin{equation}
l(c) = \int_0^c \frac{dc}{\chi(c)} = 2 \left( \frac{bc}{D-3}
\right)^{1/2}.
\end{equation}
Hence,
\begin{equation}
\chi(l) = \frac{D-3}{2b}\, l.
\end{equation}

The mass added to the black hole is 
\begin{equation}
\delta M \leq E\, \chi(l)\left|_R =  \frac{D-3}{2b}\, ER. \right.
\end{equation}
By Eqs.~(\ref{eq-bm}), (\ref{eq-ab}), and (\ref{eq-eta}), the black
hole entropy increases by
\begin{equation}
\delta S_{\rm BH} = \frac{dS_{\rm BH}}{dM}\, \delta M \leq 2 \pi ER.
\end{equation}
By the generalized second law, this increase must at least compensate
for the lost matter entropy: $\delta S_{\rm BH} - S_{\rm matter} \geq
0$.  Hence,
\begin{equation}
S_{\rm matter} \leq 2 \pi ER.
\label{eq-bekbound2}
\end{equation}

\subsubsection{Unruh radiation}
\label{sec-unruh} 

The above derivation of the Bekenstein bound, by a purely classical
treatment of the Geroch process, suffers from the problem that it can
be strengthened to a point where it yields an obviously false
conclusion.  Consider a system in a rectangular box whose height, $h$,
is much smaller than its other dimensions.  Orient the system so that
the small dimension is aligned with the radial direction, and the long
dimensions are parallel to the horizon. The minimal distance between
the center of mass and the black hole horizon is then set by the
height of the box, and will be much smaller than the circumferential
radius.  In this way, one can ``derive'' a bound of the form
\begin{equation}
S_{\rm matter} \leq \pi E h.
\label{eq-bekbound-wrong}
\end{equation}
The right hand side goes to zero in the limit of vanishing height, at
fixed energy of the box.  But the entropy of the box does not go to
zero unless {\em all\/} of its dimensions vanish.  If only the height
goes to zero, the vertical modes become heavy and have to be excluded.
But entropy will still be carried by light modes living in the other
spatial directions.

Unruh and Wald (1982, 1983) have pointed out that a system held at
fixed radius just outside a black hole horizon undergoes acceleration,
and hence experiences Unruh radiation (Unruh, 1976).  They argued that
this quantum effect will change both the energetics (because the
system will be buoyed by the radiation) and the entropy balance in the
Geroch process (because the volume occupied by the system will be
replaced by entropic quantum radiation after the system is dropped
into the black hole).  Unruh and Wald concluded that the Bekenstein
bound is neither necessary nor sufficient for the operation of the
GSL.  Instead, they suggested that the GSL is automatically protected
by Unruh radiation as long as the entropy of the matter system does
not exceed the entropy of unconstrained thermal radiation of the same
energy and volume.  This is plausible if the system is indeed weakly
gravitating and if its dimensions are not extremely unequal.

Bekenstein (1983, 1994a), on the other hand, has argued that Unruh
radiation merely affects the lowest layer of the system and is
typically negligible.  Only for very flat systems, Bekenstein (1994a)
claims that the Unruh-Wald effect may be important.  This would
invalidate the derivation of Eq.~(\ref{eq-bekbound-wrong}) in the
limit where this bound is clearly incorrect.  At the same time, it
would leave the classical argument for the Bekenstein bound,
Eq.~(\ref{eq-bekbound2}), essentially intact.  As there would be an
intermediate regime where Eq.~(\ref{eq-bekbound-wrong}) applies,
however, one would not expect the Bekenstein bound to be optimally
tight for non-spherical systems.

The question of whether the GSL implies the Bekenstein bound remains
controversial (see, e.g., Bekenstein, 1999, 2001; Pelath and Wald,
1999; Wald, 2001; Marolf and Sorkin, 2002).

The arguments described here can also be applied to other kinds of
horizons.  Davies (1984) and Schiffer (1992) considered a Geroch
process in de~Sitter space, respectively extending the Unruh-Wald and
the Bekenstein analysis to the cosmological horizon.  Bousso (2001)
has shown that the GSL implies a Bekenstein-type bound for dilute
systems in asymptotically de~Sitter space, with the assumption of
spherical symmetry but not necessarily of weak gravity.  In this case
one would not expect quantum buoyancy to play a crucial role.

\subsubsection{Empirical status}

Independently of its logical relation to the GSL, one can ask whether
the Bekenstein bound actually holds in nature.  Bekenstein (1981,
1984) and Schiffer and Bekenstein (1989) have made a strong case that
all physically reasonable, weakly gravitating matter systems satisfy
Eq.~(\ref{eq-bekbound}); some come within an order of magnitude of
saturation.  This empirical argument has been called into question by
claims that certain systems violate the Bekenstein bound; see, e.g.,
Page (2000) and references therein.  Many of these counter-examples,
however, fail to include the whole gravitating mass of the system in
$E$.  Others involve questionable matter content, such as a very large
number of species (Sec.~\ref{sec-species}).  Bekenstein (2000c) gives
a summary of alleged counter-examples and their refutations, along
with a list of references to more detailed discussions.  If the
Bekenstein bound is taken to apply only to complete, weakly
gravitating systems that can actually be constructed in nature, it has
not been ruled out (Flanagan, Marolf, and Wald, 2000; Wald, 2001).

The application of the bound to strongly gravitating systems is
complicated by the difficulty of defining the radius of the system in
a highly curved geometry.  At least for spherically symmetric systems,
however, this is not a problem, as one may define $R$ in terms of the
surface area.  A Schwarzschild black hole in four dimensions has
$R=2E$.  Hence, its Bekenstein entropy, $S=A/4=\pi R^2$, exactly
saturates the Bekenstein bound (Bekenstein, 1981).  In $D>4$, black
holes come to within a factor $\frac{2}{D-2}$ of saturating the bound
(Bousso, 2001).

\subsection{Spherical entropy bound}
\label{sec-spheb}

Instead of dropping a thermodynamic system into an existing black hole
via the Geroch process, one may also consider the {\em Susskind
process}, in which the system is {\em converted\/} to a black hole.
Susskind (1995b) has argued that the GSL, applied to this
transformation, yields the {\sl spherical entropy bound}
\begin{equation}
S_{\rm matter} \leq \frac{A}{4},
\label{eq-spheb}
\end{equation}
where $A$ is a suitably defined area enclosing the matter system.  

The description of the Susskind process below is influenced by the
analysis of Wald (2001).

\subsubsection{Susskind process}
\label{sec-susskind}

Let us consider an isolated matter system of mass $E$ and entropy
$S_{\rm matter}$ residing in a spacetime ${\cal M}$.  We require that
the asymptotic structure of ${\cal M}$ permits the formation of black
holes.  For definiteness, let us assume that ${\cal M}$ is
asymptotically flat.  We define $A$ to be the area of the
circumscribing sphere, i.e., the smallest sphere that fits around the
system.  Note that $A$ is well-defined only if the metric near the
system is at least approximately spherically symmetric.  This will be
the case for all spherically symmetric systems, and for all weakly
gravitating systems, but not for strongly gravitating systems lacking
spherical symmetry.  Let us further assume that the matter system is
stable on a timescale much greater than $A^{1/2}$.  That is, it
persists and does not expand or collapse rapidly, so that the
time-dependence of $A$ will be negligible.

The system's mass must be less than the mass $M$ of a black hole of
the same surface area.  Otherwise, the system could not be
gravitationally stable, and from the outside point of view it would
already be a black hole.  One would expect that the system can be
converted into a black hole of area $A$ by collapsing a shell of mass
$M-E$ onto the system.%
\footnote{This assumes that the shell can actually be brought to
within $A$ without radiating or ejecting shell mass or system mass.
For two large classes of systems, Bekenstein (2000a,b) obtains
Eq.~(\ref{eq-spheb}) under weaker assumptions.}

Let the shell be well-separated from the system initially.  Its
entropy, $S_{\rm shell}$, is non-negative.  The total initial entropy
in this thermodynamic process is given by
\begin{equation} 
S_{\rm total}^{\rm initial} = S_{\rm matter}+S_{\rm shell}.
\end{equation}
The final state is a black hole, with entropy
\begin{equation}
S_{\rm total}^{\rm final} = S_{\rm BH} = \frac{A}{4}.
\end{equation}
By the generalized second law of thermodynamics, Eq.~(\ref{eq-gsl}),
the initial entropy must not exceed the final entropy.  Since $S_{\rm
shell}$ is obviously non-negative, Eq.~(\ref{eq-spheb}) follows.

\subsubsection{Relation to the Bekenstein bound}

Thus, the spherical entropy bound is obtained directly from the GSL
via the Susskind process.  Alternatively, and with similar
limitations, one can obtain the same result from the Bekenstein bound,
if the latter is assumed to hold for strongly gravitating systems.
The requirement that the system be gravitationally stable implies $2M
\leq R$ in four dimensions.  From Eq.~(\ref{eq-bekbound}), one thus
obtains:
\begin{equation}
S \leq 2\pi M R \leq \pi R^2 = \frac{A}{4}.
\end{equation}
This shows that the spherical entropy bound is weaker than the
Bekenstein bound, in situations where both can be applied.  

The spherical entropy bound, however, is more closely related to the
holographic principle.  It can be cast in a covariant and general form
(Sec.~\ref{sec-bb}).  An interesting open question is whether one can
reverse the logical direction and derive the Bekenstein bound from the
covariant entropy bound under suitable assumptions
(Sec.~\ref{sec-gia}).

In $D>4$, gravitational stability and the Bekenstein bound imply only
$S \leq \frac{D-2}{8} A$ (Bousso, 2001).  The discrepancy may stem
from the extrapolation to strong gravity and/or the lack of a reliable
calibration of the prefactor in the Bekenstein bound.

\subsubsection{Examples}
\label{sec-gb}

The spherical entropy bound is best understood by studying a number of
examples in four spacetime dimensions.  We follow 't~Hooft (1993) and
Wald (2001).

It is easy to see that the bound holds for black holes.  By
definition, the entropy of a single Schwarzschild black hole, $S_{\rm
BH} = A/4$, precisely saturates the bound.  In this sense, a black
hole is the most entropic object one can put inside a given spherical
surface ('t~Hooft, 1993).

Consider a system of several black holes of masses $M_i$, in $D=4$.
Their total entropy will be given by
\begin{equation}
S = 4 \pi \sum M_i^2.
\end{equation}
From the point of view of a distant observer, the system must not
already be a larger black hole of mass $\sum M_i$.  Hence, it must be
circumscribed by a spherical area
\begin{equation}
A \geq 16 \pi \left( \sum M_i \right)^2 > 16 \pi \sum M_i^2 = 4S.
\end{equation}
Hence, the spherical entropy bound is satisfied with room to spare.

Using ordinary matter instead of black holes, it turns out to be
difficult even to approach saturation of the bound.  In order to
obtain a stable, highly entropic system, a good strategy is to make it
from massless particles.  Rest mass only enhances gravitational
instability without contributing to the entropy.  Consider, therefore,
a gas of radiation at temperature $T$, with energy $E$, confined in a
spherical box of radius $R$.  We must demand that the system is not a
black hole: $R \geq 2E$.  For an order-of-magnitude estimate of the
entropy, we may neglect the effects of self-gravity and treat the
system as if it lived on a flat background.

The energy of the ball is related to its temperature as
\begin{equation}
E \sim Z R^3 T^4,
\end{equation}
where $Z$ is the number of different species of particles in the gas.
The entropy of the system is given by
\begin{equation}
S \sim Z R^3 T^3.
\end{equation}
Hence, the entropy is related to the size and energy as
\begin{equation}
S \sim Z^{1/4} R^{3/4} E^{3/4}.
\end{equation}
Gravitational stability then implies that
\begin{equation}
S \lesssim Z^{1/4} A^{3/4}.
\label{eq-radball}
\end{equation}
In order to compare this result to the spherical entropy bound, $S
\leq A/4$, recall that we are using Planck units.  For any geometric
description to be valid, the system must be much larger than the
Planck scale:
\begin{equation}
A \gg 1.
\end{equation}
A generous estimate for the number of species in nature is $Z \sim
O(10^3)$.  Hence, $Z^{1/4} A^{3/4}$ is much smaller than $A$ for all
but the smallest, nearly Planck size systems, in which the present
approximations cannot be trusted in any case.  For a gas ball of size
$R\gg 1$, the spherical entropy bound will be satisfied with a large
factor, $R^{1/2}$, to spare.

\subsubsection{The species problem}
\label{sec-species}

An interesting objection to entropy bounds is that one can write down
perfectly well-defined field theory Lagrangians with an arbitrarily
large number of particle species (Sorkin, Wald, and Zhang, 1981; Unruh
and Wald, 1982).  In the example of Eq.~(\ref{eq-radball}), a
violation of the spherical entropy bound for systems up to size $A$
would require
\begin{equation}
Z \gtrsim A.
\end{equation}
For example, to construct a counterexample of the size of a proton,
one would require $Z \gtrsim 10^{40}$.  It is trivial to write down a
Lagrangian with this number of fields.  But this does not mean that
the entropy bound is wrong.

In nature, the effective number of matter fields is whatever it is; it
cannot be tailored to the specifications of one's favorite
counterexample.  The spherical bound is a statement about nature.  If
it requires that the number of species is not exponentially large,
then this implication is certainly in good agreement with observation.
At any rate it is more plausible than the assumption of an
exponentially large number of light fields.  

Indeed, an important lesson learned from black holes and the
holographic principle is that nature, at a fundamental level, will not
be described by a local field theory living on some background
geometry (Susskind, Thorlacius, and Uglum, 1993).

The spherical entropy bound was derived from the generalized second
law of thermodynamics (under a set of assumptions).  Could one not,
therefore, use the GSL to rule out large $Z$?  Consider a radiation
ball with $Z\gtrsim A$ massless species, so that $S>A$.  The system is
transformed to a black hole of area $A$ by a Susskind process.
However, Wald (2001) has shown that the apparent entropy decrease is
irrelevant, because the black hole is catastrophically unstable.  In
Sec.~\ref{sec-bh-radiation}, the time for the Hawking evaporation of a
black hole was estimated to be $A^{3/2}$ in $D=4$.  This implicitly
assumed a small number of radiated species.  But for large $Z$, one
must take into account that the radiation rate is actually
proportional to $Z$.  Hence, the evaporation time is given by
\begin{equation}
t_0 \sim \frac{A^{3/2}}{Z}.
\end{equation}
With $Z \gtrsim A$, one has $t_0 \lesssim A^{1/2}$.  The time needed
to form a black hole of area $A$ is at least of order $A^{1/2}$, so
the black hole in question evaporates faster than it forms.

Wald's analysis eliminates the possibility of using the GSL to exclude
large $Z$ for the process at hand.  But it produces a different,
additional argument against proliferating the number of species.
Exponentially large $Z$ would render black holes much bigger than the
Planck scale completely unstable.  Let us demand, therefore, that
super-Planckian black holes be at least metastable.  Then $Z$ cannot
be made large enough to construct a counterexample from
Eq.~(\ref{eq-radball}).  From a physical point of view, the
metastability of large black holes seems a far more natural assumption
than the existence of an extremely large number of particle species.

Further arguments on the species problem (of which the
possible renormalization of Newton's constant with $Z$ has received
particular attention) are found in Bombelli {\em et al.} (1986),
Bekenstein (1994b, 1999, 2000c), Jacobson (1994), Susskind and Uglum
(1994, 1996), Frolov (1995), Brustein, Eichler, and Foffa (2000),
Veneziano (2001), Wald (2001), and Marolf and Sorkin (2002).

\section{Towards a holographic principle}
\label{sec-towards}

\subsection{Degrees of freedom}
\label{sec-ndof}

How many degrees of freedom are there in nature, at the most
fundamental level?  The holographic principle answers this question in
terms of the area of surfaces in spacetime.  Before reaching this
rather surprising answer, we will discuss a more traditional way one
might have approached the question.  Parts of this analysis follow
't~Hooft (1993) and Susskind (1995b).

For the question to have meaning, let us restrict to a finite region
of volume $V$ and boundary area $A$.  Assume, for now, that gravity is
not strong enough to blur the definition of these quantities, and that
spacetime is asymptotically flat.  Application of the spherical
entropy bound, Eq.~(\ref{eq-spheb}), will force us to consider the
circumscribing sphere of the region.  This surface will coincide with
the boundary of the region only if the boundary is a sphere, which we
shall assume.

In order to satisfy the assumptions of the spherical entropy bound we
also demand that the metric of the enclosed region is not strongly
time-dependent, in the sense described at the beginning of
Sec.~\ref{sec-susskind}.  In particular, this means that $A$ will not
be a trapped surface in the interior of a black hole.

Let us define the {\em number of degrees of freedom\/} of a
quantum-mechanical system, $N$, to be the logarithm of the dimension
${\cal N}$ of its Hilbert space ${\cal H}$:
\begin{equation}
N = \ln {\cal N} = \ln \dim( {\cal H} ).
\end{equation}
Note that a harmonic oscillator has $N=\infty$ with this definition.
The number of degrees of freedom is equal (up to a factor of $\ln 2$)
to the number of bits of information needed to characterize a state.
For example, a system with 100 spins has ${\cal N}=2^{100}$ states,
$N=100 \ln 2$ degrees of freedom, and can store 100 bits of
information.

\subsection{Fundamental system}
\label{sec-funds}

Consider a spherical region of space with no particular restrictions
on matter content.  One can regard this region as a quantum-mechanical
system and ask how many different states it can be in.  In other
words, what is the dimension of the quantum Hilbert space describing
all possible physics confined to the specified region, down to the
deepest level?

Thus, our question is not about the Hilbert space of a specific
system, such as a hydrogen atom or an elephant.  Ultimately, all these
systems should reduce to the constituents of a fundamental theory.
The question refers directly to these constituents, given only the
size\footnote{The precise nature of the geometric boundary conditions
is discussed further in Sec.~\ref{sec-edef}.} of a region.  Let us
call this system the {\em fundamental system}.  

How much complexity, in other words, lies at the deepest level of
nature?  How much information is required to specify {\em any\/}
physical configuration completely, as long as it is contained in a
prescribed region?

\subsection{Complexity according to local field theory}
\label{sec-fromft}

In the absence of a unified theory of gravity and quantum fields, it
is natural to seek an answer from an approximate framework.  Suppose
that the ``fundamental system'' is local quantum field theory on a
classical background spacetime satisfying Einstein's equations
(Birrell and Davies, 1982; Wald, 1994).  A quantum field theory
consists of one or more oscillators at every point in space.  Even a
single harmonic oscillator has an infinite-dimensional Hilbert space.
Moreover, there are infinitely many points in any volume of space, no
matter how small. Thus, the answer to our question appears to be
$N = \infty$.  However, so far we have disregarded the
effects of gravity altogether.

A finite estimate is obtained by including gravity at least in a
crude, minimal way.  One might expect that distances smaller than the
Planck length, $l_{\rm P} = 1.6 \times 10^{-33}$\/cm, cannot be
resolved in quantum gravity.  So let us discretize space into a Planck
grid and assume that there is one oscillator per Planck volume.
Moreover, the oscillator spectrum is discrete and bounded from below
by finite volume effects.  It is bounded from above because it must be
cut off at the Planck energy, $M_{\rm P} = 1.3 \times 10^{19}$\/GeV.
This is the largest amount of energy that can be localized to a Planck
cube without producing a black hole.  Thus, the total number of
oscillators is $V$ (in Planck units), and each has a finite number of
states, $n$.  (A minimal model one might think of is a Planckian
lattice of spins, with $n=2$.)  Hence, the total number of independent
quantum states in the specified region is
\begin{equation}
{\cal N} \sim n^V.
\label{eq-states-volume}
\end{equation}
The number of degrees of freedom is given by
\begin{equation}
N \sim V \ln n \gtrsim V.
\label{eq-ndof-volume}
\end{equation}
This result successfully captures our prejudice that the degrees of
freedom in the world are local in space, and that, therefore,
complexity grows with volume.  It turns out, however, that this view
conflicts with the laws of gravity.

\subsection{Complexity according to the spherical entropy bound}
\label{sec-fromspheb}

Thermodynamic entropy has a statistical interpretation.  Let $S$ be
the thermodynamic entropy of an isolated system at some specified
value of macroscopic parameters such as energy and volume.  Then $e^S$
is the number of independent quantum states compatible with these
macroscopic parameters.  Thus, entropy is a measure of our ignorance
about the detailed microscopic state of a system.  One could relax the
macroscopic parameters, for example by requiring only that the energy
lie in some finite interval.  Then more states will be allowed, and
the entropy will be larger.

The question at the beginning of this section was ``How many
independent states are required to describe all the physics in a
region bounded by an area $A$?''  Recall that all thermodynamic
systems should ultimately be described by the same underlying theory,
and that we are interested in the properties of this ``fundamental
system''.  We are now able to rephrase the question as follows: ``What
is the entropy, $S$, of the `fundamental system', given that only the
boundary area is specified?''  Once this question is answered, the
number of states will simply be ${\cal N} = e^S$, by the argument
given in the previous paragraph.

In Sec.~\ref{sec-spheb} we obtained the spherical entropy bound,
Eq.~(\ref{eq-spheb}), from which the entropy can be determined without
any knowledge of the nature of the ``fundamental system''.  The bound,
\begin{equation}
S \leq \frac{A}{4},
\end{equation}
makes reference only to the boundary area; it does not care about the
microscopic properties of the thermodynamic system.  Hence, it applies
to the ``fundamental system'' in particular.  A black hole that just
fits inside the area $A$ has entropy
\begin{equation}
S_{\rm BH} = \frac{A}{4},
\end{equation}
so the bound can clearly be saturated with the given boundary
conditions.  Therefore, the number of degrees of freedom in a region
bounded by a sphere of area $A$ is given by
\begin{equation}
N = \frac{A}{4};
\label{eq-ndof-area}
\end{equation}
the number of states is
\begin{equation}
{\cal N} = e^{A/4}.
\label{eq-states-area}
\end{equation}

We assume that all physical systems are larger than the Planck scale.
Hence, their volume will exceed their surface area, in Planck units.
(For a proton, the volume is larger than the area by a factor of
$10^{20}$; for the earth, by $10^{41}$.)  The result obtained from the
spherical entropy bound is thus at odds with the much larger number
of degrees of freedom estimated from local field theory.  Which of the
two conclusions should we believe?

\subsection{Why local field theory gives the wrong answer}
\label{sec-qftwrong}

We will now argue that the field theory analysis overcounted available
degrees of freedom, because it failed to include properly the effects
of gravitation.  We assume $D=4$ and neglect factors of order unity.
(In $D>4$ the gist of the discussion is unchanged though some of the
powers are modified.)

The restriction to a finite spatial region provides an infrared
cut-off, precluding the generation of entropy by long wavelength
modes.  Hence, most of the entropy in the field theory estimate comes
from states of very high energy.  But a spherical surface cannot
contain more mass than a black hole of the same area.  According to
the Schwarzschild solution, Eq.~(\ref{eq-schsch}), the mass of a black
hole is given by its radius.  Hence, the mass $M$ contained within a
sphere of radius $R$ obeys
\begin{equation}
M \lesssim R
\label{eq-mlessthanr}.
\end{equation}

The ultra-violet cutoff imposed in Sec.~\ref{sec-fromft} reflected
this, but only on the smallest scale ($R=1$).  It demanded only that
each Planck volume must not contain more than one Planck mass.  For
larger regions this cutoff would permit $M \sim R^3$, in violation of
Eq.~(\ref{eq-mlessthanr}).  Hence our cut-off was too lenient to
prevent black hole formation on larger scales.

For example, consider a sphere of radius $R=1\,$cm, or $10^{33}$ in
Planck units.  Suppose that the field energy in the enclosed region
saturated the naive cut-off in each of the $\sim 10^{99}$ Planck
cells.  Then the mass within the sphere would be $\sim 10^{99}$.  But
the most massive object that can be localized to the sphere is a black
hole, of radius and mass $10^{33}$.

Thus, most of the states included by the field theory estimate are too
massive to be gravitationally stable.  Long before the quantum fields
can be excited to such a level, a black hole would form.%
\footnote{Thus, black holes provide a natural covariant cut-off which
becomes stronger at larger distances.  It differs greatly from the
fixed distance or fixed energy cutoffs usually considered in quantum
field theory.}
If this black hole is still to be contained within a specified sphere
of area $A$, its entropy may saturate but not exceed the spherical
entropy bound.
 
Because of gravity, not all degrees of freedom that field theory
apparently supplies can be used for generating entropy, or storing
information.  This invalidates the field theory estimate,
Eq.~(\ref{eq-ndof-volume}), and thus resolves the apparent
contradiction with the holographic result, Eq.~(\ref{eq-ndof-area}).

Note that the present argument does not provide independent
quantitative confirmation that the maximal entropy is given by the
area.  This would require a detailed understanding of the relation
between entropy, energy, and gravitational back-reaction in a given
system.

\subsection{Unitarity and a holographic interpretation}
\label{sec-hsi} 

Using the spherical entropy bound, we have concluded that $A/4$
degrees of freedom are sufficient to fully describe any stable region
in asymptotically flat space enclosed by a sphere of area $A$.  In a
field theory description, there are far more degrees of freedom.
However, we have argued that any attempt to excite more than $A/4$ of
these degrees of freedom is thwarted by gravitational collapse.  From
the outside point of view, the most entropic object that fits in the
specified region is a black hole of area $A$, with $A/4$ degrees of
freedom.

A conservative interpretation of this result is that the demand for
gravitational stability merely imposes a practical limitation for the
information content of a spatial region.  If we are willing to pay the
price of gravitational collapse, we can excite more than $A/4$ degrees
of freedom---though we will have to jump into a black hole to verify
that we have succeeded.  With this interpretation, all the degrees of
freedom of field theory should be retained.  The region will be
described by a quantum Hilbert space of dimension $e^V$.

The following two considerations motivate a rejection of this
interpretation.  Both arise from the point of view that physics in
asymptotically flat space can be consistently described by a
scattering matrix.  The S-matrix provides amplitudes between initial
and final asymptotic states defined at infinity.  Intermediate black
holes may form and evaporate, but as long as one is not interested in
the description of an observer falling into the black hole, an
S-matrix description should be satisfactory from the point of view of
an observer at infinity.

One consideration concerns economy.  A fundamental theory should not
contain more than the necessary ingredients.  If $A/4$ is the amount
of data needed to describe a region completely, that should be the
amount of data used. This argument is suggestive; however, it could be
rejected as merely aesthetical and gratuitously radical.

A more compelling consideration is based on unitarity.
Quantum-mechanical evolution preserves information; it takes a pure
state to a pure state.  But suppose a region was described by a
Hilbert space of dimension $e^V$, and suppose that region was
converted to a black hole.  According to the Bekenstein entropy of a
black hole, the region is now described by a Hilbert space of
dimension $e^{A/4}$.  The number of states would have decreased, and
it would be impossible to recover the initial state from the final
state.  Thus, unitarity would be violated.  Hence, the Hilbert space
must have had dimension $e^{A/4}$ to start with.

The insistence on unitarity in the presence of black holes led
't~Hooft (1993) and Susskind (1995b) to embrace a more radical,
``holographic'' interpretation of Eq.~(\ref{eq-ndof-area}).  

{\sl Holographic principle (preliminary formulation)}. {\em
A region with boundary of area $A$ is fully described by no more than
$A/4$ degrees of freedom, or about 1 bit of information per Planck
area.  A fundamental theory, unlike local field theory, should
incorporate this counterintuitive result.}

\subsection{Unitarity and black hole complementarity}
\label{sec-bhc}

The unitarity argument would be invalidated if it turned out that
unitarity is not preserved in the presence of black holes.  Indeed,
Hawking (1976b) has claimed that the evaporation of a black hole---its
slow conversion into a cloud of radiation---is not a unitary process.
In semi-classical calculations, Hawking radiation is found to be
exactly thermal, and all information about the ingoing state appears
lost.  Others (see Secs.~\ref{sec-further}, \ref{sec-strings}) argued,
however, that unitarity must be restored in a complete quantum gravity
theory.

The question of unitarity of the S-matrix arises not only when a black
hole forms, but again, and essentially independently, when the black
hole evaporates.  The holographic principle is necessary for unitarity
at the first stage.  But if unitarity were later violated during
evaporation, it would have to be abandoned, and the holographic
principle would lose its basis.

It is not understood in detail how Hawking radiation carries away
information. Indeed, the assumption that it does seems to lead to a
paradox, which was pointed out and resolved by Susskind, Thorlacius,
and Uglum (1993).  When a black hole evaporates unitarily, the same
quantum information would seem to be present both inside the black
hole (as the original matter system that collapsed) and outside, in
the form of Hawking radiation.  The simultaneous presence of two
copies appears to violate the linearity of quantum mechanics, which
forbids the ``xeroxing'' of information.

One can demonstrate, however, that no single observer can see both
copies of the information.  Obviously an infalling observer cannot
escape the black hole to record the outgoing radiation.  But what
prevents an outside observer from first obtaining, say, one bit of
information from the Hawking radiation, only to jump into the black
hole to collect a second copy?

Page (1993) has shown that more than half of a system has to be
observed to extract one bit of information.  This means that an
outside observer has to linger for a time compared to the evaporation
time scale of the black hole ($M^3$ in $D=4$) in order to gather a
piece of the ``outside data'', before jumping into the black hole to
verify the presence of the same data inside.

However, the second copy can only be observed if it has not already
hit the singularity inside the black hole by the time the observer
crosses the horizon.  One can show that the energy required for a
single photon to evade the singularity for so long is exponential in
the square of the black hole mass.  In other words, there is far too
little energy in the black hole to communicate even one bit of
information to an infalling observer in possession of outside data.

The apparent paradox is thus exposed as the artifact of an
operationally meaningless, global point of view.  There are two
complementary descriptions of black hole formation, corresponding to
an infalling and and an outside observer.  Each point of view is
self-consistent, but a simultaneous description of both is neither
logically consistent nor practically testable.  Black hole
complementarity thus assigns a new role to the observer in quantum
gravity, abandoning a global description of spacetimes with horizons.

Further work on black hole complementarity includes 't~Hooft (1991),
Susskind (1993a,b, 1994), Stephens, 't~Hooft, and Whiting (1994),
Susskind and Thorlacius (1994), Susskind and Uglum, (1994).  Aspects
realized in string theory are also discussed by Lowe, Susskind, and
Uglum (1994), Lowe {\em et al.} (1995); see Sec.~\ref{sec-strings}.
For a review, see, e.g., Thorlacius (1995), Verlinde (1995), Susskind
and Uglum (1996), and Bigatti and Susskind (2000).

Together, the holographic principle and black hole complementarity
form the conceptual core of a new framework for black hole formation
and evaporation, in which the unitarity of the S-matrix is retained at
the expense of locality.\footnote{In this sense, the holographic
principle, as it was originally proposed, belongs in the first class
discussed in Sec.~\ref{sec-ia}.  However, one cannot obtain its modern
form (Sec.~\ref{sec-hp}) from unitarity.  Hence we resort to the
covariant entropy bound in this review.  Because the bound can be
tested using conventional theories, this also obviates the need to
assume particular properties of quantum gravity in order to induce the
holographic principle.}

In the intervening years, much positive evidence for unitarity has
accumulated.  String theory has provided a microscopic, unitary
quantum description of some black holes (Strominger and Vafa, 1996;
see also Callan and Maldacena, 1996; Sec.~\ref{sec-strings}).
Moreover, there is overwhelming evidence that certain asymptotically
Anti-de Sitter spacetimes, in which black holes can form and
evaporate, are fully described by a unitary conformal field theory
(Sec.~\ref{sec-ads}).

Thus, a strong case has been made that the formation and evaporation
of a black hole is a unitary process, at least in asymptotically flat
or AdS spacetimes.

\subsection{Discussion}
\label{sec-limitations}

In the absence of a generally valid entropy bound, the arguments for a
holographic principle were incomplete, and its meaning remained
somewhat unclear.  Neither the spherical entropy bound,
nor the unitarity argument which motivates its elevation to a
holographic principle, are applicable in general spacetimes.  

An S-matrix description is justified in a particle accelerator, but
not in gravitational physics.  In particular, realistic universes do
not permit an S-matrix description.  (For recent discussions see,
e.g., Banks, 2000a; Fischler, 2000a,b; Bousso, 2001a; Fischler {\em et
al.}, 2001; Hellerman, Kaloper, and Susskind, 2001.)  Even in
spacetimes that do, observers don't all live at infinity.  Then the
question is not so much whether unitarity holds, but how it can be
defined.

As black hole complementarity itself insists, the laws of physics must
also describe the experience of an observer who falls into a black
hole.  The spherical entropy bound, however, need not apply inside
black holes.  Moreover, it need not hold in many other important
cases, in view of the assumptions involved in its derivation.  For
example, it does not apply in cosmology, and it cannot be used when
spherical symmetry is lacking.  In fact, it will be seen in
Sec.~\ref{sec-seb} that the entropy in spatial volumes can exceed the
boundary area in all of these cases.

Thus, the holographic principle could not, at first, establish a
general correspondence between areas and the number of fundamental
degrees of freedom.  But how can it point the way to quantum gravity,
if it apparently does not apply to many important solutions of the
classical theory?

The AdS/CFT correspondence (Sec.~\ref{sec-ads}), holography's most
explicit manifestation to date, was a thing of the future when the
holographic principle was first proposed.  So was the covariant
entropy bound (Secs.~\ref{sec-bb}--\ref{sec-tests}), which exposes the
apparent limitations noted above as artifacts of the original,
geometrically crude formulation.  The surprising universality of the
covariant bound significantly strengthens the case for a holographic
principle (Sec.~\ref{sec-hp}).

As 't~Hooft and Susskind anticipated, the conceptual revisions
required by the unitarity of the S-matrix have proven too profound to
be confined to the narrow context in which they were first recognized.
We now understand that areas should generally be associated with
degrees of freedom in adjacent spacetime regions.  Geometric
constructs that precisely define this relation---light-sheets---have
been identified (Fischler and Susskind, 1998; Bousso, 1999a).  The
holographic principle may have been an audacious concept to propose.
In light of the intervening developments, it has become a difficult
one to reject.

\section{A spacelike entropy bound?}
\label{sec-seb}

The heuristic derivation of the spherical entropy bound rests on a
large number of fairly strong assumptions.  Aside from suitable
asymptotic conditions, the surface $A$ has to be spherical, and the
enclosed region must be gravitationally stable so that it can be
converted to a black hole.

Let us explore whether the spherical entropy bound, despite these
apparent limitations, is a special case of a more general entropy
bound.  We will present two conjectures for such a bound.  In this
section, we will discuss the spacelike entropy bound, perhaps the most
straightforward and intuitive generalization of
Eq.~(\ref{eq-spheb}).  We will present several counterexamples to
this bound and conclude that it does not have general validity.
Turning to a case of special interest, we will find that it is
difficult to precisely define the range of validity of the spacelike
entropy bound even in simple cosmological spacetimes.

\subsection{Formulation}
\label{sec-seb-form}

One may attempt to extend the scope of Eq.~(\ref{eq-spheb}) simply by
dropping the assumptions under which it was derived (asymptotic
structure, gravitational stability, and spherical symmetry).  Let us
call the resulting conjecture the {\em spacelike entropy bound}: {\em
The entropy contained in any spatial region will not exceed the area
of the region's boundary.} More precisely, the {\sl spacelike entropy
bound} is the following statement (Fig.~\ref{fig-spheb}):
\begin{figure}
\includegraphics[width=7cm]{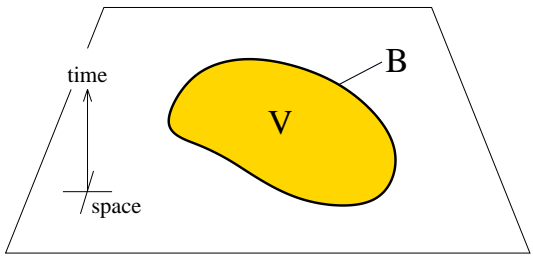}
\caption{A hypersurface of equal time.  The spacelike entropy bound
attempts to relate the entropy in a spatial region, $V$, to the area
of its boundary, $B$.  This is not successful.}
\label{fig-spheb}
\end{figure}

Let $V$ be a compact portion of a hypersurface of equal time in the
spacetime ${\cal M}$.\footnote{Here $V$ is used both to denote a
spatial region, and its volume.  Note that we use more careful
notation to distinguish a surface ($B$) from its area ($A$).}  Let
$S(V)$ be the entropy of all matter systems in $V$.  Let $B$ be the
boundary of $V$ and let $A$ be the area of the boundary of $V$.  Then
\begin{equation}
S(V) \leq {A[B(V)]\over 4}.
\label{eq-seb}
\end{equation}

\subsection{Inadequacies}
\label{sec-fail}

The spacelike entropy bound is not a successful conjecture.
Eq.~(\ref{eq-seb}) is contradicted by a large variety of
counterexamples.  We will begin by discussing two examples from
cosmology.  Then we will turn to the case of a collapsing star.
Finally, we will expose violations of Eq.~(\ref{eq-seb}) even for all
isolated, spherical, weakly gravitating matter systems.

\subsubsection{Closed spaces}
\label{sec-fail1}

It is hardly necessary to describe a closed universe in detail to see
that it will lead to a violation of the spacelike holographic
principle.  It suffices to assume that the spacetime ${\cal M}$
contains a closed spacelike hypersurface, ${\cal V}$.  (For example,
there are realistic cosmological solutions in which ${\cal V}$ has the
topology of a three-sphere.)  We further assume that ${\cal V}$
contains a matter system that does not occupy all of ${\cal V}$, and
that this system has non-zero entropy $S_0$.

Let us define the volume $V$ to be the whole hypersurface, except for
a small compact region $Q$ outside the matter system.  Thus, $S_{\rm
matter}(V) = S_0 >0$.  The boundary $B$ of $V$ coincides with the
boundary of $Q$.  Its area can be made arbitrarily small by
contracting $Q$ to a point.  Thus one obtains $S_{\rm
matter}(V)>A[B(V)])$, and the spacelike entropy bound,
Eq.~(\ref{eq-seb}), is violated.

\subsubsection{The Universe}
\label{sec-fail2}

On large scales, the universe we inhabit is well approximated as a
three-dimensional, flat, homogeneous and isotropic space, expanding in
time.  Let us pick one homogeneous hypersurface of equal time, ${\cal
V}$.  Its entropy content can be characterized by an average ``entropy
density'', $\sigma$, which is a positive constant on ${\cal V}$.
Flatness implies that the geometry of ${\cal V}$ is Euclidean
${\mathbb R}^3$.  Hence, the volume and area of a two-sphere grow in
the usual way with the radius:
\begin{equation}
V={4\pi\over 3} R^3,~~~A[B(V)] = 4\pi R^2.
\end{equation}

The entropy in the volume $V$ is given by
\begin{equation}
S_{\rm matter}(V) = \sigma V = {\sigma\over 6\sqrt{\pi}} A^{3/2}.
\end{equation}
Recall that we are working in Planck units.  By taking the radius of
the sphere to be large enough,
\begin{equation}
R \geq {3\over 4\sigma},
\end{equation}
one finds a volume for which the spacelike entropy bound,
Eq.~(\ref{eq-seb}), is violated (Fischler and Susskind, 1998).

\subsubsection{Collapsing star}
\label{sec-fail3}

Next, consider a spherical star with non-zero entropy
$S_0$.  Suppose the star burns out and undergoes catastrophic
gravitational collapse.  From an outside observer's point of view, the
star will form a black hole whose surface area will be at least
$4S_0$, in accordance with the generalized second law of
thermodynamics.

However, we can follow the star as it falls through its own horizon.
From collapse solutions (see, e.g., Misner, Thorne, and Wheeler,
1973), it is known that the star will shrink to zero radius and end in
a singularity.  In particular, its surface area becomes arbitrarily
small: $A \to 0$.  By the second law of thermodynamics, the entropy in
the enclosed volume, i.e., the entropy of the star, must still be at
least $S_0$.  Once more, the spacelike entropy bound fails (Easther
and Lowe, 1999).

As in the previous two examples, this failure does not concern the
spherical entropy bound, even though spherical symmetry may hold.  We
are considering a regime of dominant gravity, in violation of the
assumptions of the spherical bound.  In the interior of a black hole,
both the curvature and the time-dependence of the metric are large.

\subsubsection{Weakly gravitating system}
\label{sec-fail4}

The final example is the most subtle.  It shows that the spacelike
entropy bound can be violated by the very systems for which the
spherical entropy bound is believed to hold: spherical, weakly
gravitating systems.  This is achieved merely by a non-standard
coordinate choice that breaks spherical symmetry and measures a
smaller surface area.

Consider a weakly gravitating spherical thermodynamic system in
asymptotically flat space.  Note that this class includes most
thermodynamic systems studied experimentally; if they are not
spherical, one redefines their boundary to be the circumscribing
sphere.

\begin{figure}
\includegraphics[width=7.5cm]{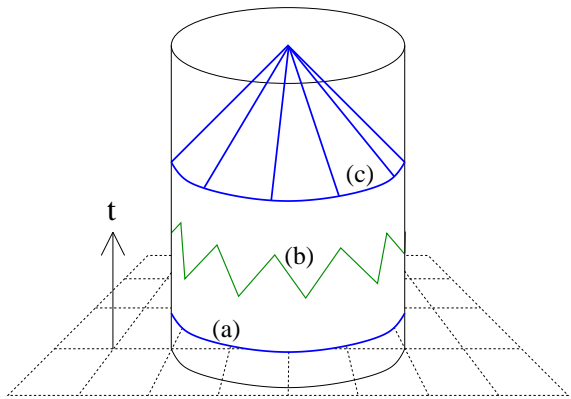}
\caption{The worldvolume of a ball of gas, with one spatial dimension
suppressed.  (a) A time slice in the rest frame of the system is shown
as a flat plane.  It intersects the boundary of system on a spherical
surface, whose area exceeds the system's entropy.  (b) In a different
coordinate system, however, a time slice intersects the boundary on
Lorentz-contracted surfaces whose area can be made arbitrarily small.
Thus the spacelike entropy bound is violated. (c) The light-sheet of a
spherical surface is shown for later reference (Sec.~\ref{sec-els}).
Light-sheets of wiggly surfaces may not penetrate the entire system
(Sec.~\ref{sec-nearnull}).---The solid cylinder depicted here can also
be used to illustrate the conformal shape of Anti-de~Sitter space
(Sec.~\ref{sec-ads}).}
\label{fig-ball}
\end{figure}

A coordinate-independent property of the system is its world volume,
$W$.  For a stable system with the spatial topology of a
three-dimensional ball ($\mathbf{D}^3$), the topology of $W$ is given
by ${\mathbb R}\times \mathbf{D}^3$ (Fig.~\ref{fig-ball}).

The volume of the ball of gas, at an instant of time, is geometrically
the intersection of the world volume $W$ with an equal time
hypersurface $t=0$:
\begin{equation}
V \equiv W \cap \{t=0\}.
\label{eq-vwt}
\end{equation}
The boundary of the volume $V$ is a surface $B$ given by
\begin{equation}
B = \partial W \cap \{t=0\}.
\label{eq-bvwt}
\end{equation}

The time coordinate $t$, however, is not uniquely defined.  One
possible choice for $t$ is the proper time in the rest frame of the
weakly gravitating system (Fig.~\ref{fig-ball}a).  With this choice,
$V$ and $B$ are metrically a ball and a sphere, respectively.  The
area $A(B)$ and the entropy $S_{\rm matter}(V)$ were calculated in
Sec.~\ref{sec-gb} for the example of a ball of gas.  They were found
to satisfy the spacelike entropy bound, Eq.~(\ref{eq-seb}).

From the point of view of general relativity, there is nothing special
about this choice of time coordinate. The laws of physics must be {\em
covariant}, i.e., invariant under general coordinate transformations.
Thus Eq.~(\ref{eq-seb}) must hold also for a volume $V'$ associated
with a different choice of time coordinate, $t'$.  In particular, one
may choose the $t'=$const hypersurface to be rippled like a fan.  Then
its intersection with $\partial W$, $B'$, will be almost null almost
everywhere, like the zigzag line circling the worldvolume in
Fig.~\ref{fig-ball}b.  The boundary area so defined can be made
arbitrarily small (Jacobson, 1999; Flanagan, Marolf, and Wald, 2000;
Smolin, 2001).%
  \footnote{The following construction exemplifies this for a
  spherical system.  Consider the spatial $D-2$ sphere $B$ defined by
  $t=0$ and parametrized by standard spherical coordinates
  $(\theta_1,\ldots,\theta_{D-3},\varphi)$.  Divide $B$ into $2n$
  segments of longitude defined by ${k\over 2n} \leq {\varphi\over
  2\pi} < {k+1\over 2n}$ with $k=0\ldots 2n-1$.  By translation of $t$
  this segmentation carries over to $\partial W$. For each even (odd)
  segment, consider a Lorentz observer boosted with velocity $\beta$
  in the positive (negative) $\varphi$ direction at the midpoint of
  the segment on the equator of $B$.  The time foliations of these
  $2n$ observers, restricted respectively to each segment and joined
  at the segment boundaries, define global equal time hypersurfaces.
  The slices can be smoothed at the segment boundaries and in the
  interior of $W$ without affecting the conclusions.  After picking a
  particular slice, $t'=0$, a volume $V'$ and its boundary $B'$ can be
  defined in analogy with Eqs.~(\ref{eq-vwt}) and (\ref{eq-bvwt}).
  Since $V'$ contains the entire thermodynamic system, the entropy is
  not affected by the new coordinate choice:
  $S_{\rm matter}(V') = S_{\rm matter}(V)$.
  Because of Lorentz contraction, the proper area $A(B')$ is
  smaller than $A(B)$.  Indeed, by taking $\beta \to 1$ and
  $n\to\infty$ one can make $A(B')$ arbitrarily small:
  $A(B')~~\stackrel{n\to\infty}{\longrightarrow}~~
  A(B) \sqrt{1-\beta^2}~~
  \stackrel{\beta\to 1}{\longrightarrow}~~ 0.$
  An analogous construction for a square system takes a simpler form;
  see Sec.~\ref{sec-nearnull}.}
This construction has shown that a spherical system with non-zero
entropy $S_{\rm matter}$ can be enclosed within a surface of area
$A(B')<S_{\rm matter}$, and the spacelike entropy bound,
Eq.~(\ref{eq-seb}), is again violated.  

How is this possible?  After all, the spherical entropy bound should
hold for this system, because it can be converted into a spherical
black hole of the same area.  However, this argument implicitly
assumed that the boundary of a spherically symmetric system is a
sphere (and therefore agrees with the horizon area of the black hole
after the conversion).  With the non-standard time coordinate $t'$,
however, the boundary is not spherically symmetric, and its area is
much smaller than the final black hole area. (The latter is unaffected
by slicing ambiguities because a black hole horizon is a null
hypersurface.)

\subsection{Range of validity}
\label{sec-range}

In view of these problems, it is clear that the spacelike entropy
bound cannot be maintained as a fully general conjecture holding for
all volumes and areas in all spacetimes.  Still, the spherical entropy
bound, Eq.~(\ref{eq-spheb}), clearly holds for many systems that do
not satisfy its assumptions, suggesting that those assumptions may be
unnecessarily restrictive.

For example, the earth is part of a cosmological spacetime that is
not, as far as we know, asymptotically flat.  However, the earth does
not curve space significantly.  It is well separated from other matter
systems.  On time and distance scales comparable to the earth's
diameter, the universe is effectively static and flat.  In short, it
is clear that the earth will obey the spacelike entropy bound.%
\footnote{Pathological slicings such as the one in
Sec.~\ref{sec-fail4} must still be avoided.  Here we define the
earth's surface area by the natural slicing in its approximate Lorentz
frame.}

The same argument can be made for the solar system, and even for the
milky way.  As we consider larger regions, however, the effects of
cosmological expansion become more noticable, and the flat space
approximation is less adequate.  An important question is whether a
definite line can be drawn.  In cosmology, is there a largest region
to which the spacelike entropy bound can be reliably applied?  If so,
how is this region defined?  Or does the bound gradually become less
accurate at larger and larger scales?%
\footnote{The same questions can be asked of the Bekenstein bound,
Eq.~(\ref{eq-bekbound}).  Indeed, Bekenstein (1989), who proposed its
application to the past light cone of an observer, was the first to
raise the issue of the validity of entropy bounds in cosmology.}

Let us consider homogeneous, isotropic universes, known as
Friedmann-Robertson-Walker (FRW) universes (Sec.~\ref{sec-cosmo}).
Fischler and Susskind (1998) abandoned the spacelike formulation
altogether (Sec.~\ref{sec-mot}).  For adiabatic FRW universes,
however, their proposal implied that the spacelike entropy bound
should hold for spherical regions smaller than the particle horizon
(the future light cone of a point at the big bang).

Restriction to the particle horizon turns out to be sufficient for the
validity of the spacelike entropy bound in simple flat and open
models; thus, the problem in Sec.~\ref{sec-fail2} is resolved.
However, it does not prevent violations in closed or collapsing
universes.  The particle horizon area vanishes when the light cone
reaches the far end of a closed universe---this is a special case of
the problem discussed in Sec.~\ref{sec-fail1}.  An analogue of the
problem of Sec.~\ref{sec-fail3} can arise also.  Generally, closed
universes and collapsing regions exhibit the greatest difficulties for
the formulation of entropy bounds, and many authors have given them
special attention.

Davies (1988) and Brustein (2000) proposed a generalized second law
for cosmological spacetimes.  They suggested that contradictions in
collapsing universes may be resolved by augmenting the area law with
additional terms.  Easther and Lowe (1999) argued that the second law
of thermodynamics implies a holographic entropy bound, at least for
flat and open universes, in regions not exceeding the Hubble
horizon.\footnote{The Hubble radius is defined to be ${a\over da/dt}$,
where $a$ is the scale factor of the universe; see Eq.~(\ref{eq-FRW1})
below.}  Similar conclusions were reached by Veneziano (1999b),
Kaloper and Linde (1999), and Brustein (2000).

Bak and Rey (2000a) argued that the relevant surface is the apparent
horizon, defined in Sec.~\ref{sec-ah}.  This is a minor distinction
for typical flat and open universes, but it avoids some of the
difficulties with closed universes.\footnote{Related discussions also
appear in Dawid (1999) and Kalyana Rama (1999).  The continued debate
of the difficulties of the Fischler-Susskind proposal in closed
universes (Wang and Abdalla, 1999, 2000; Cruz and Lepe, 2001) is, in
our view, rendered nugatory by the covariant entropy bound.}

The arguments for bounds of this type return to the Susskind process,
the gedankenexperiment by which the spherical entropy bound was
derived (Sec.~\ref{sec-susskind}).  A portion of the universe is
converted to a black hole; the second law of thermodynamics is
applied.  One then tries to understand what might prevent this
gedankenexperiment from being carried out.

For example, regions larger than the horizon are expanding too rapidly
to be converted to a black hole---they cannot be ``held together''
(Veneziano, 1999b).  Also, if a system is already inside a black hole,
it can no longer be converted to one.  Hence, one would not expect the
bound to hold in collapsing regions, such as the interior of black
holes or a collapsing universe (Easther and Lowe, 1999; Kaloper and
Linde, 1999).

This reasoning does expose some of the limitations of the spacelike
entropy bound (namely those that are illustrated by the explicit
counterexamples given in Sec.~\ref{sec-fail1} and \ref{sec-fail3}).
However, it fails to identify sufficient conditions under which the
bound is actually reliable.  Kaloper and Linde (1999) give
counterexamples to any statement of the type ``The area of the
particle (apparent, Hubble) horizon always exceeds the entropy
enclosed in it'' (Sec.~\ref{sec-na}).

In the following section we will introduce the covariant entropy
bound, which is formulated in terms of light-sheets.  In
Sec.~\ref{sec-tests} we will present evidence that this bound has
universal validity.  Starting from this general bound, one can find
sufficient conditions under which a spacelike formulation is valid
(Sec.~\ref{sec-spt}, \ref{sec-cosmocor}).  However, the conditions
themselves will involve the light-sheet concept in an essential way.
Not only is the spacelike formulation less general than the
light-sheet formulation; the range of validity of the former cannot be
reliably identified without the latter.

We conclude that the spacelike entropy bound is violated by realistic
matter systems.  In cosmology, its range of validity cannot be
intrinsically defined.

\section{The covariant entropy bound}
\label{sec-bb}

In this section we present a more successful generalization of
Eq.~(\ref{eq-spheb}): the covariant entropy bound.

There are two significant formal differences between the covariant
bound and the spacelike bound, Eq.~(\ref{eq-seb}).  The spacelike
formulation starts with a choice of spatial volume $V$.  The volume,
in turn, defines a boundary $B =\partial V$, whose area $A$ is then
claimed to be an upper bound on $S(V)$, the entropy in $V$.  The
covariant bound proceeds in the opposite direction.  A codimension 2
surface $B$ serves as the starting point for the construction of a
codimension 1 region $L$.  This is the first formal difference.  The
second is that $L$ is a null hypersurface, unlike $V$ which is
spacelike.

\begin{figure}[h] \centering
\includegraphics[width=7cm]{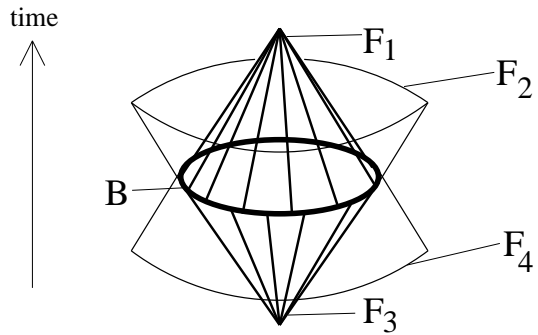}
\caption{The four null hypersurfaces orthogonal to a spherical surface
$B$.  The two cones $F_1$, $F_3$ have negative expansion and hence
correspond to light-sheets.  The covariant entropy bound states that
the entropy on each light-sheet will not exceed the area of $B$.  The
other two families of light rays, $F_2$ and $F_4$, generate the skirts
drawn in thin outline.  Their cross-sectional area is increasing, so
they are not light-sheets.  The entropy of the skirts is not related
to the area of $B$.---Compare this figure to Fig.~\ref{fig-spheb}.}
\label{fig-spheresheets}
\end{figure}

More precisely, $L$ is a {\em light-sheet}.  It is constructed by
following light rays that emanate from the surface $B$, as long as
they are not expanding.  There are always at least two suitable
directions away from $B$ (Fig.~\ref{fig-spheresheets}).  When 
light rays self-intersect, they start to expand.  Hence, light-sheets
terminate at focal points.

The {\sl covariant entropy bound} states that {\em the entropy on any
light-sheet of a surface $B$ will not exceed the area of $B$}:
\begin{equation}
S[L(B)]\leq {A(B)\over 4}.
\label{eq-bb}
\end{equation}
We will give a more formal definition at the end of this section.

We begin with some remarks on the conjectural nature of the bound, and
we mention related earlier proposals.  We will explain the geometric
construction of light-sheets in detail, giving special attention to
the considerations that motivate the condition of non-expansion
($\theta\leq 0$).  We give a definition of entropy on light-sheets,
and we discuss the extent to which the limitations of classical
general relativity are inherited by the covariant entropy bound.  We
then summarize how the bound is formulated, applied, and tested.
Parts of this section follow Bousso (1999a).

\subsection{Motivation and background}
\label{sec-mot}

There is no fundamental derivation of the covariant entropy bound.  We
present the bound because there is strong evidence that it holds
universally in nature.  The geometric construction is well-defined and
covariant.  The resulting entropy bound can be saturated, but no
example is known where it is exceeded.

In Sec.~\ref{sec-fmw} plausible relations between entropy and energy
are shown to be sufficient for the validity of the bound.  But these
relations do not at present appear to be universal or fundamental.  In
special situations, the covariant entropy bound reduces to the
spherical entropy bound, which is arguably a consequence of black hole
thermodynamics.  But in general, the covariant entropy bound cannot be
inferred from black hole physics; quite conversely, the generalized
second law of thermodynamics may be more appropriately regarded as a
consequence of the covariant bound (Sec.~\ref{sec-gia}).

The origin of the bound remains mysterious.  As discussed in the
introduction, this puzzle forms the basis of the holographic
principle, which asserts that the covariant entropy bound betrays the
number of degrees of freedom of quantum gravity (Sec.~\ref{sec-hp}).

Aside from its success, little motivation for a light-like formulation
can be offered.  Under the presupposition that {\em some\/} general
entropy bound waits to be discovered, one is guided to light rays by
circumstantial evidence.  This includes the failure of the spacelike
entropy bound (Sec.~\ref{sec-seb}), the properties of the Raychaudhuri
equation (Sec.~\ref{sec-ray}), and the loss of a dynamical dimension
in the light cone formulation of string theory
(Sec.~\ref{sec-further}).

Whatever the reasons, the idea that light rays might be involved in
relating a region to its surface area---or, rather, relating a surface
area to a light-like ``region''---arose in discussions of the
holographic principle from the beginning.  

Susskind (1995b) suggested that the horizon of a black hole can be
mapped, via light rays, to a distant, flat holographic screen, citing
the focussing theorem (Sec.~\ref{sec-ray}) to argue that the
information thus projected would satisfy the holographic bound.
Corley and Jacobson (1996) pointed out that the occurrence of focal
points, or {\em caustics}, could invalidate this argument, but showed
that one caustic-free family of light rays existed in Susskind's
example.  They further noted that both past and future directed
families of light rays can be considered.

Fischler and Susskind (1998) recognized that a light-like formulation
is crucial in cosmological spacetimes, because the spacelike entropy
bound fails.  They proposed that any spherical surface $B$ in FRW
cosmologies (see Sec.~\ref{sec-cosmo}) be related to (a portion of) a
light cone that comes from the past and ends on $B$.  This solved the
problem discussed in Sec.~\ref{sec-fail2} for flat and open universes but
not the problem of small areas in closed or recollapsing universes
(see Secs.~\ref{sec-fail1}, \ref{sec-fail3}).

The covariant entropy bound (Bousso, 1999a) can be regarded as a
refinement and generalization of the Fischler-Susskind proposal.  It
can be applied in arbitrary spacetimes, to any surface $B$ regardless
of shape, topology, and location.  It considers all four null
directions orthogonal to $B$ without prejudice.  It introduces a new
criterion, the contraction of light rays, both to select among the
possible light-like directions and to determine how far the light rays
may be followed.  For any $B$, there will be at least two ``allowed''
directions and hence two light-sheets, to each of which the bound applies
individually.

\subsection{Light-sheet kinematics}
\label{sec-kin}

Compared to the previously discussed bounds, Eqs.~(\ref{eq-spheb}) and
(\ref{eq-seb}), the non-trivial ingredient of the covariant entropy
bound lies in the concept of light-sheets.  Given a surface, a
light-sheet defines an adjacent spacetime region whose entropy should
be considered.  What has changed is not the formula, $S\leq A/4$, but
the prescription that determines where to look for the entropy $S$
that enters that formula.  Let us discuss in detail how light-sheets
are constructed.

\subsubsection{Orthogonal null hypersurfaces}
\label{sec-kin1}

A given surface $B$ possesses precisely four orthogonal null
directions (Fig.~\ref{fig-spheresheets}).  They are sometimes referred
to as {\em future directed ingoing}, {\em future directed outgoing},
{\em past directed ingoing}, and {\em past directed outgoing}, though
``in'' and ``out'' are not always useful labels.  Locally, these
directions can be represented by null hypersurfaces $F_1,\ldots,F_4$
that border on $B$.  The $F_i$ are generated by the past and the
future directed light rays orthogonal to $B$, on either side of $B$.

For example, suppose that $B$ is the wall of a (spherical) room in
approximately flat space, as shown in Fig.~\ref{fig-spheresheets}, at
$t=0$.  (We must keep in mind that $B$ denotes a surface at some
instant of time.)  Then the future directed light rays towards the
center of the room generate a null hypersurface $F_1$, which looks
like a light cone.  A physical way of describing $F_1$ is to imagine
that the wall is lined with light bulbs that all flash up at $t=0$.
As the light rays travel towards the center of the room they generate
$F_1$.

Similarly, one can line the outside of the wall with light bulbs.
Future directed light rays going to the outside will generate a second
null hypersurface $F_2$.  Finally, one can also send light rays
towards the past.  (We might prefer to think of these as arriving from
the past, i.e., a light bulb in the center of the room flashed at an
appropriate time for its rays to reach the wall at $t=0$.)  In any
case, the past directed light rays orthogonal to $B$ will generate two
more null hypersurfaces $F_3$ and $F_4$.

In Fig.~\ref{fig-spheresheets}, the two ingoing cones $F_1$ and $F_3$,
and the two outgoing ``skirts'', $F_2$ and $F_4$, are easily seen to
be null and orthogonal to $B$.  However, the existence of four null
hypersurfaces bordering on $B$ is guaranteed in Lorentzian geometry
independently of the shape and location of $B$.  They are always
uniquely generated by the four sets of surface-orthogonal light rays.

At least two of the four null hypersurfaces $F_1,\ldots,F_4$ will be
selected as light-sheets, according to the condition of non-positive
expansion discussed next.

\subsubsection{Light-sheet selection}
\label{sec-kin2}

Let us return to the example where $B$ is the wall of a spherical
room.  If gravity is weak, one would expect that the area $A$ of $B$
will be a bound on the entropy in the room (Sec.~\ref{sec-susskind}).
Clearly, $A$ cannot be related in any way to the entropy in the
infinite region outside the room; that entropy could be arbitrarily
large.  It appears that we should select $F_1$ or $F_3$ as
light-sheets in this example, because they correspond to our intuitive
notion of ``inside''.

The question is how to generalize this notion.  It is obvious that one
should compare an area only to entropy that is in some sense
``inside'' the area.  However, consider a closed universe, in which
space is a three-sphere.  As Sec.~\ref{sec-fail1} has illustrated, we
need a criterion that prevents us from considering the large part of
the three-sphere to be ``inside'' a small two-sphere $B$.

What we seek is a local condition, which will select whether some
direction away from $B$ is an inside direction.  This condition should
reduce to the intuitive, global notion---inside is where infinity is
not---where applicable.  An analogy in Euclidean space leads to a
useful criterion, the {\em contraction condition}.

\begin{figure}[h] \centering
\includegraphics[width=8.5cm]{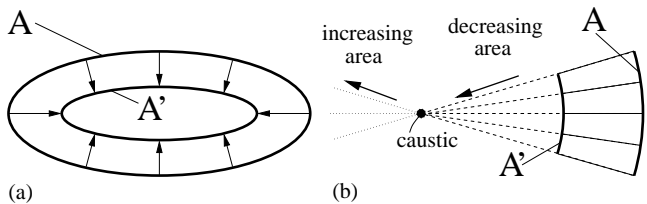}
\caption{Local definition of ``inside''.  (a) Ingoing rays
perpendicular to a convex surface in a Euclidean geometry span
decreasing area.  This motivates the following local definition.  (b)
Inside is the direction in which the cross-sectional area decreases
($A'\leq A$).  This criterion can be applied to light rays orthogonal
to any surface.  After light rays locally intersect, they begin to
expand.  Hence, light-sheets must be terminated at caustics.}
\label{fig-inside}
\end{figure}
Consider a convex closed surface $B$ of codimension one and area $A$
in flat Euclidean space, as shown in Fig.~\ref{fig-inside}a.  Now
construct all the geodesics intersecting $B$ orthogonally.  Follow
each geodesic an infinitesimal proper distance $dl$ to one of the two
sides of $B$.  The set of points thus obtained will span a similarly
shaped surface of area $A'$.  If $A'<A$, let us call the chosen
direction the ``inside''.  If $A'>A$, we have gone ``outside''.

Unlike the standard notion of ``inside'', the contraction criterion
does not depend on any knowledge of the global properties of $B$ and
of the space it is embedded in.  It can be applied independently to
arbitrarily small pieces of the surface.  One can always construct
orthogonal geodesics and ask in which direction they contract.  It is
local also in the orthogonal direction; the procedure can be repeated
after each infinitesimal step.

Let us return to Lorentzian signature, and consider a codimension 2
spatial surface $B$.  The contraction criterion cannot be used to find
a {\em spatial\/} region ``inside'' $B$.  There are infinitely many
different spacelike hypersurfaces $\Sigma$ containing $B$.  Which side
has contracting area could be influenced by the arbitrary choice of
$\Sigma$.

However, the four {\em null\/} directions $F_1,\ldots,F_4$ away from
$B$ are uniquely defined.  It is straightforward to adapt the
contraction criterion to this case.  Displacement by an infinitesimal
spatial distance is meaningless for light rays, because two points on
the same light ray always have distance zero.  Rather, an appropriate
analogue to length is the affine parameter $\lambda$ along the
light ray (see the Appendix).  Pick a particular direction $F_i$.
Follow the orthogonal null geodesics away from $B$ for an
infinitesimal affine distance $d\lambda$.  The points thus constructed
span a new surface of area $A'$.  If $A' \leq A$, then the direction
$F_i$ will be considered an ``inside'' direction, or {\em light-sheet
direction}.

By repeating this procedure for $i=1,\ldots,4$, one finds all null
directions that point to the ``inside'' of $B$ in this technical
sense.  Because the light rays generating opposite pairs of null
directions (e.g., $F_1$ and $F_4$) are continuations of each other, it
is clear that at least one member of each pair will be considered
inside.  If the light rays are locally neither expanding nor
contracting, both members of a pair will be called ``inside''.  Hence,
there will always be at least two light-sheet directions.  In
degenerate cases, there may be three or even four.

Mathematically, the contraction condition can be formulated thus:
\begin{equation}
\theta(\lambda)\leq 0~~~~\mbox{for}~\lambda=\lambda_0,
\label{eq-theta0}
\end{equation}
where $\lambda$ is an affine parameter for the light rays generating
$F_i$ and we assume that $\lambda$ increases in the direction away
from $B$.  $\lambda_0$ is the value of $\lambda$ on $B$.  The {\em
expansion}, $\theta$, of a family of light rays is discussed in detail
in Sec.~\ref{sec-ray}.  It can be understood as follows.  Consider a
bunch of infinitesimally neighboring light rays spanning a surface
area ${\cal A}$.  Then
\begin{equation}
\theta(\lambda) \equiv \frac{d{\cal A}/d\lambda}{\cal A}.
\label{eq-tda}
\end{equation}

As in the Euclidean analogy, this condition can be applied to each
infinitesimal surface element separately and so is local.  Crucially,
it applies to open surfaces as well as to closed ones.  This
represents a significant advance in the generality of the formulation.

For oddly shaped surfaces or very dynamical spacetimes, it is possible
for the expansion to change sign along some $F_i$.  For example, this
will happen for smooth concave surfaces in flat space.  Because of the
locality of the contraction criterion, one may split such surfaces
into pieces with constant sign, and continue the analysis for each
piece separately.  This permits us to assume henceforth without loss
of generality that the surfaces we consider have continuous
light-sheet directions.

For the simple case of the spherical surface in Minkowski space, the
condition (\ref{eq-tda}) reproduces the intuitive answer.  The area is
decreasing in the $F_1$ and $F_3$ directions---the past and future
directed light rays going to the center of the sphere.  We will call
any such surface, with two light-sheet directions on the same spatial
side, {\em normal}.

In highly dynamical geometries, the expansion or contraction of space
can be the more important effect on the expansion of light rays.  Then
it will not matter which spatial side they are directed at.  For
example, in an expanding universe, areas get small towards the past,
because the big bang is approached.  A sufficiently large sphere will
have two past directed light-sheets, but no future directed ones.  A
surface of this type is called {\em anti-trapped}.  Similarly, in a
collapsing universe or inside a black hole, space can shrink so
rapidly that both light-sheets are future directed.  Surfaces with
this property are {\em trapped}.

\begin{figure}[h] \centering
\includegraphics[width=5cm]{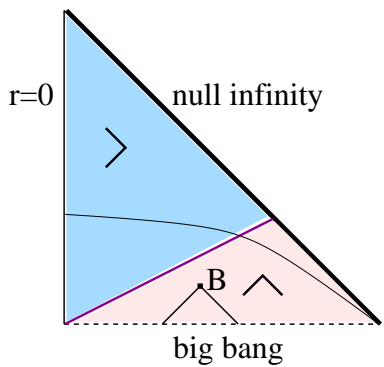}
\caption{Penrose diagram for an expanding universe (a flat or open FRW
universe, see Sec.~\ref{sec-cosmo}).  The thin curve is a slice of
constant time.  Each point in the interior of the diagram represents a
sphere.  The wedges indicate light-sheet directions.  The apparent
horizon (shown here for equation of state $p=\rho$) divides the normal
spheres near the origin from the anti-trapped spheres near the big
bang.  The light-sheets of any sphere $B$ can be represented by
inspecting the wedge that characterizes the local domain and drawing
lines away from the point representing $B$, in the direction of the
wedge's legs.}
\label{fig-flatfrw}
\end{figure}

In a Penrose diagram (Appendix), a sphere is represented by a point.
The four orthogonal null directions correspond to the four legs of an
``X'' centered on this point.  Light-sheet directions can be indicated
by drawing only the corresponding legs (Bousso 1999a).  Normal,
trapped, and anti-trapped surfaces are thus denoted by wedges of
different orientation (see Figs.~\ref{fig-flatfrw}, \ref{fig-clos}a,
and \ref{fig-star}).

\subsubsection{Light-sheet termination}
\label{sec-kin3}

From now on we will consider only inside directions, $F_j$, where $j$
runs over two or more elements of $\{1,2,3,4\}$.  For each $F_j$, a
light-sheet is generated by the corresponding family of light rays.  In
the example of the spherical surface in flat space, the light-sheets
are cones bounded by $B$, as shown in Fig.~\ref{fig-spheresheets}.

Strictly speaking, however, there was no particular reason to stop at
the tip of the cone, where all light rays intersect.  On the other
hand, it would clearly be desastrous to follow the light rays
arbitrarily far.  They would generate another cone which would grow
indefinitely, containing unbounded entropy.  One must enforce, by some
condition, that the light-sheet is terminated before this happens.  In
all but the most special cases, the light rays generating a
light-sheet will not intersect in a single point, so the condition
must be more general.

A suitable condition is to demand that the expansion be non-positive
{\em everywhere\/} on the light-sheet, and not only near $B$:
\begin{equation}
\theta(\lambda)\leq 0,
\label{eq-theta}
\end{equation}
for all values of the affine parameter on the light-sheet.

By construction (Sec.~\ref{sec-kin2}) the expansion is initially
negative or zero on any light-sheet.  Raychaudhuri's equation
guarantees that the expansion can only decrease..  (This will be shown
explicitly in Sec.~\ref{sec-ray}.)  The only way $\theta$ can become
positive is if light rays intersect, for example at the tip of the
light cone.  However, it is not necessary for all light rays to
intersect in the same point.  By Eq.~(\ref{eq-tda}), the expansion
becomes positive at any {\em caustic}, that is, any place where a
light ray crosses an infinitesimally neighboring light ray in the
light-sheet (Fig.~\ref{fig-inside}b).

Thus, Eq.~(\ref{eq-theta}) operates independently of any symmetries in
the setup.  It implies that light-sheets end at {\em
caustics}.\footnote{If the null energy condition (Appendix) is
violated, the condition (\ref{eq-theta}) can also terminate
light-shees at non-caustic points.}  In general, each light ray in a
light-sheet will have a different caustic point, and the resulting
caustic surfaces can be very complicated.  The case of a light cone is
special in that all light rays share the same caustic point at the
tip.  An ellipsoid in flat space will have a self-intersecting
light-sheet that may contain the same object more than once (at two
different times).  Gravitational back-reaction of matter will make the
caustic surfaces even more involved.

Non-local self-intersections of light rays do not lead to violations
of the contraction condition, Eq.~(\ref{eq-theta}).  That is, the
light-sheet must be terminated only where a light ray intersects its
neighbor, but not necessarily when it intersects another light ray
coming from a different portion of the surface $B$.  One can consider
modifications of the light-sheet definition where any
self-intersection terminates the light-sheet (Tavakol and Ellis, 1999;
Flanagan, Marolf, and Wald, 2000).  Since this modification can only
make light-sheets shorter, it can weaken the resulting bound.
However, in most applications, the resulting light-sheets are easier
to calculate (as Tavakol and Ellis, in particular, have stressed) and
still give useful bounds.%
\footnote{Low (2002) has argued that the future directed light-sheets
in cosmological spacetimes can be made arbitrarily extensive by
choosing a closed surface containing sufficiently flat pieces.  Low
concludes that the covariant entropy bound is violated in standard
cosmological solutions, unless it is modified to terminate
light-sheets also at non-local self-intersections.---This reasoning
overlooks that any surface element with local curvature radius larger
than the apparent horizon possesses only past directed light-sheets
(Bousso, 1999a; see Sec.~\ref{sec-ah}).  Independently of the
particular flaw in Low's argument, the conclusion is also directly
invalidated by the proof of Flanagan, Marolf, and Wald (2000).  (This
is just as well, as the modification advocated by Low would not have
solved the problem; non-local intersections can be suppressed by
considering open surfaces.)}

The condition, Eq.~(\ref{eq-theta}), subsumes Eq.~(\ref{eq-theta0}),
which applied only to the initial value of $\lambda$.  It is
satisfying that both the direction and the extent of light-sheets are
determined by the same simple condition, Eq.~(\ref{eq-theta}).

\subsection{Defining entropy}
\label{sec-edef}

\subsubsection{Entropy on a fixed light-sheet}
\label{sec-els}

The geometric construction of light-sheets is well-defined.
But how is ``the entropy on a light-sheet'', $S_{\rm matter}$,
determined?  Let us begin with an example where the definition of
$S_{\rm matter}$ is obvious.  Suppose that $B$ is a sphere around an
isolated, weakly gravitating thermodynamic system.  Given certain
macroscopic constraints, for example an energy or energy range,
pressure, volume, etc., the entropy of the system can be computed
either thermodynamically, or statistically as the logarithm of the
number of accessible quantum states.

To good approximation, the two light-sheets of $B$ are a past and a
future light cone. Let us consider the future directed light-sheet.
The cone contains the matter system completely (Fig.~\ref{fig-ball}c),
in the same sense in which a $t=\mbox{const}$ surface contains the
system completely (Fig.~\ref{fig-ball}a).  A light-sheet is just a
different way of taking a snapshot of a matter system---in light cone
time.  (In fact, this comes much closer to how the system is actually
observed in practice.)  Hence, the entropy on the light-sheet is
simply given by the entropy of the matter system.

A more problematic case arises when the light-sheet intersects only a
portion of an isolated matter system, or if there simply are no
isolated systems in the spacetime.  A reasonable (statistical) working
definition was given by Flanagan, Marolf, and Wald (2000), who
demanded that long wavelength modes which are not fully contained on
the light-sheet should not be included in the entropy.

In cosmological spacetimes, entropy is well approximated as a
continuous fluid.  In this case, $S_{\rm matter}$ is the integral of
the entropy density over the light-sheet (Secs.~\ref{sec-fmw},
\ref{sec-cosmo}).

One would expect that the gravitational field itself can encode
information perturbatively, in the form of gravitational waves.
Because it is difficult to separate such structure from a ``background
metric'', we will not discuss this case here.%
\footnote{Flanagan, Marolf, and Wald (2000) pointed out that
perturbative gravitational entropy affects the light-sheet by
producing shear, which in turn accelerates the focussing of light rays
(Sec.~\ref{sec-lsd}).  This suggests that the inclusion of such
entropy will not lead to violations of the bound.  Related research is
currently pursued by Bhattacharya, Chamblin, and Erlich
(2002).}

We have formulated the covariant entropy bound for matter systems in
classical geometry and have not made provisions for the inclusion of
the semi-classical Bekenstein entropy of black holes.  There is
evidence, however, that the area of event horizons can be included in
$S_{\rm matter}$.  However, in this case the light-sheet must not be
continued to the interior of the black hole.  The Bekenstein entropy
of the black hole already contains the information about objects that
fell inside; it must not be counted twice (Sec.~\ref{sec-bhc}).

\subsubsection{Entropy on an arbitrary light-sheet}
\label{sec-es}

So far we have treated the light-sheet of $B$ as a fixed null
hypersurface, e.g., in the example of an isolated thermodynamic
system.  Different microstates of the system, however, correspond to
different distributions of energy.  This is a small effect on average,
but it does imply that the geometry of light-sheets will vary with the
state of the system in principle.

In many examples, such as cosmological spacetimes, one can calculate
light-sheets in a large-scale, averaged geometry.  In this
approximation, one can estimate $S_{\rm matter}$ while holding the
light-sheet geometry fixed.

In general, however, one can at best hold the surface $B$ fixed,%
\footnote{We shall take this to mean that the internal metric of the
surface $B$ is held fixed.  It may be possible to relax this further,
for example by specifying only the area $A$ along with suitable
additional restrictions.}
but not the light-sheet of $B$.  We must consider $S_{\rm matter}$ to
be the entropy on {\em any\/} light-sheet of $B$.  Sec.~\ref{sec-csh},
for example, discusses the collapse of a shell onto an apparent black
hole horizon.  In this example, a part of the spacetime metric is
known, including $B$ and the initial expansions $\theta_i$ of its
orthogonal light rays.  However, the geometry to the future of $B$ is
not presumed, and different configurations contributing to the entropy
lead to macroscopically different future light-sheets.

In a static, asymptotically flat space the specification of a
spherical surface reduces to the specification of an energy range.
The enclosed energy must lie between zero and the mass of a black hole
that fills in the sphere.  Unlike most other thermodynamic quantities
such as energy, however, the area of surfaces is well-defined in
arbitrary geometries.

In the most general case, one may specify only a surface $B$ but no
information about the embedding of $B$ in any spacetime.  One is
interested in the entropy of the ``fundamental system''
(Sec.~\ref{sec-funds}), i.e., the number of quantum states associated
with the light-sheets of $B$ in {\em any\/} geometry containing $B$.
This leaves too much freedom for Eq.~(\ref{eq-bb}) to be checked
explicitly.  The covariant entropy bound essentially becomes the full
statement of the holographic principle (Sec.~\ref{sec-hp}) in this
limit.

\subsection{Limitations}
\label{sec-schlim}

Here we discuss how the covariant entropy bound is tied to a regime of
approximately classical spacetimes with reasonable matter content.
The discussion of the ``species problem'' (Sec.~\ref{sec-species})
carries over without significant changes and will not be repeated.

\subsubsection{Energy conditions}
\label{sec-econds}

In Sec.~\ref{sec-gb} we showed that the entropy of a ball of radiation
is bounded by $A^{3/4}$, and hence is less than its surface area.  For
larger values of the entropy, the mass of the ball would exceed its
radius, so it would collapse to form a black hole.  But what if matter
of negative energy was added to the system?  This would offset the
gravitational backreaction of the gas without decreasing its entropy.
The entropy in any region could be increased at will while keeping the
geometry flat.

This does not automatically mean that the holographic principle (and
indeed, the generalized second law of thermodynamics) is wrong.  A way
around the problem might be to show that instabilities develop that
will invalidate the setup we have just suggested.  But more to the
point, the holographic principle is expected to be a property of the
real world.  And to a good approximation, matter with negative mass
does not exist in the real world.\footnote{We discuss quantum effects
and a negative cosmological constant below.}

Einstein's general relativity does not restrict matter content, but
tells us only how matter affects the shape of spacetime.  Yet, of all
the types of matter that could be added to a Lagrangian, few actually
occur in nature.  Many would have pathological properties or
catastrophic implications, such as the instability of flat space.  

In a unified theory underlying gravity and all other forces, one would
expect that the matter content is dictated by the theory.  String
theory, for example, comes packaged with a particular field content in
its perturbative limits.  However, there are many physically
interesting spacetimes that have yet to be described in string theory
(Sec.~\ref{sec-strings}), so it would be premature to consider only
fields arising in this framework.

One would like to test the covariant entropy bound in a broad class of
systems, but we are not interested whether the bound holds for matter
that is entirely unphysical.  It is reasonable to exclude matter whose
energy density appears negative to a light ray, or which permits the
superluminal transport of energy.%
\footnote{This demand applies to every matter component separately
(Bousso, 1999a).  This differs from the role of energy conditions in
the singularity theorems (Hawking and Ellis, 1973), whose proofs are
sensitive only to the total stress tensor.  The above example shows
that the total stress tensor can be innocuous when components of
negative and positive mass are superimposed.  An interesting question
is whether instabilities lead to a separation of components, and thus
to an eventual violation of energy conditions on the total stress
tensor.  We would like to thank J.~Bekenstein and A.~Mayo for raising
this question.}
In other words, let us demand the null energy condition as well as the
causal energy condition.  Both conditions are spelled out in the
Appendix, Eqs.~(\ref{eq-nec}) and (\ref{eq-cec}).  They are believed
to be satisfied classically by all physically reasonable forms of
matter.%
\footnote{The dominant energy condition has sometimes been demanded
instead of Eq.~(\ref{eq-nec}) and Eq.~(\ref{eq-cec}).  It is a
stronger condition that has the disadvantage of excluding a negative
cosmological constant (Bousso, 1999a).---One can also ask whether, in
a reversal of the logical direction, entropy bounds can be used to
infer energy conditions that characterize physically acceptable matter
(Brustein, Foffa, and Mayo, 2002).}

Negative energy density is generally disallowed by these conditions,
with the exception of a negative cosmological constant.  This is
desirable, because a negative cosmological constant does not lead to
instabilities or other pathologies.  It may well occur in the
universe, though it is not currently favored by observation.  Unlike
other forms of negative energy, a negative cosmological constant
cannot be used to cancel the gravitational field of ordinary
thermodynamic systems, so it should not lead to difficulties with the
holographic principle.

Quantum effects can violate the above energy conditions.  Casimir
energy, for example, can be negative.  However, the relation between
the magnitude, size, and duration of such violations is severely
constrained (see, e.g., Ford and Roman, 1995, 1997, 1999; Flanagan,
1997; Fewster and Eveson, 1998; Fewster, 1999; further references are
found in Borde, Ford, and Roman, 2001).  Even where they occur, their
gravitational effects may be overcompensated by those of ordinary
matter.  It has not been possible so far to construct a counterexample
to the covariant entropy bound using quantum effects in ordinary
matter systems.

\subsubsection{Quantum fluctuations}
\label{sec-qfluc}

What about quantum effects in the geometry itself?  The holographic
principle refers to geometric concepts such as area, and orthogonal
light rays.  As such, it can be applied only where spacetime is
approximately classical.  This contradicts in no way its deep relation
to quantum gravity, as inferred from the quantum aspects of black
holes (Sec.~\ref{sec-bh}) and demonstrated by the AdS/CFT
correspondence (Sec.~\ref{sec-ads}).

In the real world, $\hbar$ is fixed, so the regime of classical
geometry is generically found in the limit of low curvature and large
distances compared to the Planck scale, Eq.~(\ref{eq-lpl}).  Setting
$\hbar$ to $0$ would not only be unphysical; as Lowe (1999) points
out, it would render the holographic bound, $Akc^3/4G\hbar$, trivial.

Lowe (1999) has argued that a naive application of the bound
encounters difficulties when effects of quantum gravity become
important.  With sufficient fine tuning, one can arrange for an
evaporating black hole to remain in equilibrium with ingoing radiation
for an arbitarily long time.  Consider the future directed outgoing
light-sheet of an area on the black hole horizon.  Lowe claims that
this light-sheet will have exactly vanishing expansion and will
continue to generate the horizon in the future, as it would in a
classical spacetime.  This would allow an arbitrarily large amount of
ingoing radiation entropy to pass through the light-sheet, in
violation of the covariant entropy bound.

If a light-sheet lingers in a region that cannot be described by
classical general relativity without violating energy conditions for
portions of the matter, then it is outside the scope of the present
formulation of the covariant entropy bound.  The study of light-sheets
of this type may guide the exploration of semi-classical
generalizations of the covariant entropy bound.  For example, it may
be appropriate to associate the outgoing Hawking radiation with a
negative entropy flux on this light-sheet (Flanagan, Marolf, and Wald,
2000).\footnote{More radical extensions have been proposed by
Markopoulou and Smolin (1999) and by Smolin (2001).}

However, Bousso (2000a) argued that a violation of the covariant
entropy bound has not been demonstrated in Lowe's example.  In any
realistic situation small fluctuations in the energy density of
radiation will occur.  They are indeed inevitable if information is to
be transported through the light-sheet.  Thus the expansion along the
light-sheet will fluctuate.  If it becomes positive, the light-sheet
must be terminated.  If it fluctuates but never becomes positive, then
it will be negative on average.  In that case an averaged version of
the focussing theorem implies that the light rays will focus within a
finite affine parameter.

The focussing is enhanced by the $-\theta^2/2$ term in Raychaudhuri's
equation, (\ref{eq-ray}), which contributes to focussing whenever
$\theta$ fluctuates about zero.  Because of these effects, the
light-sheets considered by Lowe (1999) will not remain on the horizon,
but will collapse into the black hole.  New families of light rays
continually move inside to generate the event horizon.  It is possible
to transport unlimited entropy through the black hole horizon in this
case, but not through any particular light-sheet.

\subsection{Summary}
\label{sec-presc}

In any $D$-dimensional Lorentzian spacetime $M$, the covariant
entropy bound can be stated as follows.

{\em Let $A(B)$ be the area of an arbitrary $D-2$ dimensional spatial
surface $B$ (which need not be closed).  A $D-1$ dimensional
hypersurface $L$ is called a light-sheet of $B$ if $L$ is generated by
light rays which begin at $B$, extend orthogonally away from $B$, and
have non-positive expansion,
\begin{equation}
\theta\leq 0,
\end{equation}
everywhere on $L$.  Let $S$ be the entropy on any light-sheet of $B$.
Then}
\begin{equation}
S \leq {A(B)\over 4}.
\label{eq-bb2}
\end{equation}

Let us restate the covariant entropy bound one more time, in a
constructive form most suitable for applying and testing the bound, as
we will in Sec.~\ref{sec-tests}.

\begin{enumerate}

\item{Pick any $D-2$ dimensional spatial surface $B$, and determine
its area $A(B)$.  There will be four families of light rays projecting
orthogonally away from $B$: $F_1 \ldots F_4$.}

\item{Usually additional information is available, such as the
macroscopic spacetime metric everywhere or in a neighborhood of
$B$.\footnote{The case where no such information is presumed seems too
general to be practally testable; see the end of Sec.~\ref{sec-es}.}
Then the expansion $\theta$ of the orthogonal light rays can be
calculated for each family.  Of the four families, at least two will
not expand ($\theta\leq 0$).  Determine which.}

\item{Pick one of the non-expanding families, $F_j$.  Follow each
light ray no further than to a caustic, a place where it intersects
with neighboring light rays.  The light rays form a $D-1$ dimensional
null hypersurface, a light-sheet $L(B)$.}

\item{Determine the entropy $S[L(B)]$ of matter on the light-sheet
$L$, as described in Sec.~\ref{sec-els}.\footnote{In particular, one
may wish to include in $S$ quantum states which do not all give rise
to the same macroscopic spacetime geometry, keeping fixed only the
intrinsic geometry of $B$.  In this case, step 3 has to be repeated
for each state or class of states with different geometry.  Then
$L(B)$ denotes the collection of all the different light-sheets
emanating in the $j$-th direction.}}

\item{The quantities $S[L(B)]$ and $A(B)$ can then be compared.  The
covariant entropy bound states that the entropy on the light-sheet
will not exceed a quarter of the area: $S[L(B)] \leq \frac{A(B)}{4}$.
This must hold for any surface $B$, and it applies to each
non-expanding null direction, $F_j$, separately.}

\end{enumerate}

The first three steps can be carried out most systematically by using
geometric tools which will be introduced at the beginning of
Sec.~\ref{sec-ray}.  In simple geometries, however, they often require
little more than inspection of the metric.

The light-sheet construction is well-defined in the limit where
geometry can be described classically.  It is conjectured to be valid
for all physically realistic matter systems.  In the absence of a
fundamental theory with definite matter content, the energy conditions
given in Sec.~\ref{sec-econds} approximately delineate the boundaries
of an enormous arena of spacetimes and matter systems, in which the
covariant entropy bound implies falsifiable, highly non-trivial
limitations on information content.

In particular, the bound is predictive and can be tested by
observation, in the sense that the entropy and geometry of real matter
systems can be determined (or, as in the case of large cosmological
regions, at least estimated) from experimental measurements.

\section{The dynamics of light-sheets}
\label{sec-lsd}

Entropy requires energy.  In Sec.~\ref{sec-qftwrong}, this notion gave
us some insight into a mechanism underlying the spherical entropy
bound.  Let us briefly repeat the idea.  When one tries to excite too
many degrees of freedom in a spherical region of fixed boundary area
$A$, the region becomes very massive and eventually forms a black hole
of area no larger than $A$.  Because of the second law of
thermodynamics, this collapse must set in before the entropy exceeds
$A/4$.  Of course, it can be difficult to verify this quantitatively
for a specific system; one would have to know its detailed properties
and gravitational back-reaction.

In this section, we identify a related mechanism underlying the
covariant entropy bound.  Entropy costs energy, energy focusses light,
focussing leads to the formation of caustics, and caustics prevent
light-sheets from going on forever.  As before, the critical link in
this argument is the relation between entropy and energy.
Quantitatively, it depends on the details of specific matter systems
and cannot be calculated in general.  Indeed, this is one of the
puzzles that make the generality of the covariant entropy bound so
striking.  

In many situations, however, entropy can be approximated by a local
flow of entropy density.  With plausible assumptions on the relation
between the entropy and energy density, which we review, Flanagan,
Marolf, and Wald (2000) proved the covariant entropy bound.

We also present the spacelike projection theorem, which identifies
conditions under which the covariant bound implies a spacelike bound
(Bousso, 1999a).

\subsection{Raychaudhuri's equation and the focussing theorem}
\label{sec-ray}

A family of light rays, such as the ones generating a light-sheet, is
locally characterized by its expansion, shear, and twist, which are
defined as follows.

Let $B$ be a surface of $D-2$ spatial dimensions, parametrized by
coordinates $x^\alpha$, $\alpha=1,\ldots,D-2$.  Pick one of the four
families of light rays $F_1,\ldots,F_4$ that emanate from $B$ into the
past and future directions to either side of $B$
(Fig.~\ref{fig-spheresheets}).  Each light ray satisfies the equation
for geodesics (Appendix):
\begin{equation}
\frac{dk^a}{d\lambda} + \Gamma^a_{~bc} k^b k^c =0,
\label{eq-geodesic}
\end{equation}
where $\lambda$ is an affine parameter.  The tangent vector $k^a$ is
defined by
\begin{equation}
k^a = \frac{dx^a}{d\lambda}
\end{equation}
and satisfies the null condition $k^a k_a=0$.  The light rays generate
a null hypersurface $L$ parametrized by coordinates
$(x^\alpha,\lambda)$.  This can be rephrased as follows.  In a
neighborhood of $B$, each point on $L$ is unambiguously defined by the
light ray on which it lies ($x^\alpha$) and the affine distance from
$B$ ($\lambda$).

Let $l^a$ be the null vector field on $B$ that is orthogonal to $B$
and satisfies $k^a l_a=-2$.  (This means that $l^a$ has the same time
direction as $k^a$ and is tangent to the orthogonal light rays
constructed on the other side of $B$.)  The induced $D-2$ dimensional
metric on the surface $B$ is given by
\begin{equation}
h_{ab} = g_{ab} + \frac{1}{2} \left(k_a l_b + k_b l_a\right).
\label{eq-induce}
\end{equation}
In a similar manner, an induced metric can be found for all other
spatial cross-sections of $L$.

The {\em null extrinsic curvature},
\begin{equation}
B_{ab} = h^c_{~a} h^d_{~b} \nabla_c k_d,
\label{eq-bab}
\end{equation}
contains information about the {\em expansion}, $\theta$, {\em shear},
$\sigma_{ab}$, and {\em twist}, $\omega_{ab}$, of the family of
light rays, $L$:
\begin{eqnarray}
\label{eq-thetadef}
\theta &=& h^{ab} B_{ab}, \\
\sigma_{ab} &=& \frac{1}{2} \left( B_{ab}+B_{ba} \right) -
\frac{1}{D-2}\theta h_{ab}, \\
\omega_{ab} &=& \frac{1}{2} \left( B_{ab}-B_{ba} \right).
\end{eqnarray}
Note that all of these quantities are functions of
$(x^\alpha,\lambda)$.

At this point, one can inspect the initial values of $\theta$ on $B$.
Where they are positive, one must discard $L$ and choose a different
null direction for the construction of a light-sheet.

The Raychaudhuri equation describes the change of the expansion along
the light rays:
\begin{equation}
\frac{d\theta}{d\lambda} = -\frac{1}{D-2} \theta^2 -
\sigma_{ab}\sigma^{ab} + \omega_{ab}\omega^{ab} - 8\pi T_{ab} k^a k^b.
\label{eq-ray}
\end{equation}
For a surface-orthogonal family of light rays, such as $L$, the twist
vanishes (Wald, 1984).  The final term, $-T_{ab} k^a k^b$, will be
non-positive if the null energy condition is satisfied by matter,
which we assume (Sec.~\ref{sec-econds}).  Then the right hand side of
the Raychaudhuri equation is manifestly non-positive.  It follows that
the expansion never increases.

By solving the differential inequality
\begin{equation}
\frac{d\theta}{d\lambda} \leq -\frac{1}{D-2} \theta^2,
\label{eq-rayn}
\end{equation}
one arrives at the {\em focussing theorem}:\footnote{In the context of
the AdS/CFT correspondence (Sec.~\ref{sec-ads}), the role of focussing
theorem in the construction of light-sheets has been related to the
c-theorem (Balasubramanian, Gimon, and Minic, 2000; Sahakian,
2000a,b).}  If the expansion of a family of light rays takes the
negative value $\theta_1$ at any point $\lambda_1$, then $\theta$ will
diverge to $-\infty$ at some affine parameter $\lambda_2\leq\lambda_1
+{D-2\over|\theta_1|}$.

The divergence of $\theta$ indicates that the cross-sectional area is
locally vanishing, as can be seen from Eq.~(\ref{eq-tda}).  As
discussed in Sec.~\ref{sec-kin3}, this is a caustic point, at which
infinitesimally neighboring light rays intersect.

By construction, the expansion on light-sheets is zero or negative.
If it is zero, the focussing theorem does not apply.  For example,
suppose that $B$ is a portion of the $xy$ plane in Minkowski space:
$z=t=0,~~x^2+y^2\leq 1$.  Then each light-sheet is infinitely large,
with everywhere vanishing expansion: $z=\pm t,~~x^2+y^2\leq 1$.
However, this is correct only if the spacetime is exactly Minkowski,
with no matter or gravitational waves.  In this case the light-sheets
contain no entropy in any case, so their infinite size leads to no
difficulties with the covariant entropy bound.

If a light-sheet encounters any matter (or more precisely, if $T_{ab}
k^a k^b >0$ anywhere on the light-sheet), then the light rays will be
focussed according to Eq.~(\ref{eq-ray}).  Then the focussing theorem
applies, and it follows that the light rays will eventually form
caustics, forcing the light-sheet to end.  This will happen even if no
further energy is encountered by the light rays, though it will occur
sooner if there is additional matter.

If we accept that entropy requires energy, we thus see at a
qualitative level that entropy causes light rays to focus.  Thus, the
presence of entropy hastens the termination of light-sheets.
Quantitatively, it appears to do so at a sufficient rate to protect
the covariant entropy bound, but slowly enough to allow saturation of
the bound.  This is seen in many examples, including those studied in
Sec.~\ref{sec-tests}.  The reason for this quantitative behavior is
not yet fundamentally understood.  (This just reformulates, in terms
of light-sheet dynamics, the central puzzle laid out in the
introduction and reiterated in Sec.~\ref{sec-hp}.)

\subsection{Sufficient conditions for the covariant entropy bound}
\label{sec-fmw}

Flanagan, Marolf, and Wald (2000; henceforth in this section, FMW)
showed that the covariant entropy bound is always satisfied if certain
assumptions about the relation between entropy density and energy
density are made.  In fact, they proved the bound under either one of
two sets of assumptions.  We will state these assumptions and discuss
their plausibility and physical significance.  We will not reproduce
the two proofs here.

The first set of conditions are no easier to verify, in any given
spacetime, than the covariant entropy bound itself.  Light-sheets
have to be constructed, their endpoints found, and entropy can be
defined only by an analysis of modes.  The first set of conditions
should therefore be regarded as an interesting reformulation of the
covariant entropy bound, which may shed some light on its relation to
the Bekenstein bound, Eq.~(\ref{eq-bekbound}).  

The second set of conditions involves relations between locally
defined energy and entropy densities only.  As long as the entropy
content of a spacetime admits a fluid approximation, one can easily
check whether these conditions hold.  In such spacetimes, the second
FMW theorem obviates the need to construct all light-sheets and verify
the bound for each one.

Neither set of conditions is implied by any fundamental law of
physics.  The conditions do not apply to some physically realistic
systems (which nevertheless obey the covariant entropy bound).
Furthermore, they do not permit macroscopic variations of spacetime,
precluding a verification of the bound in its strongest sense
(Sec.~\ref{sec-es}).

Thus, as FMW point out, the two theorems do not constitute a
fundamental explanation of the covariant entropy bound.  By
eliminating a large class of potential counter-examples, they do
provide important evidence for the validity of the covariant entropy
bound.  The second set can significantly shortcut the verification of
the bound in cosmological spacetimes.  Moreover, the broad validity of
the FMW hypotheses may itself betray an aspect of an underlying
theory.

\subsubsection{The first FMW theorem}
\label{sec-fmw1}

The first set of assumptions is

\begin{itemize}

\item{Associated with each light-sheet $L$ in spacetime there is an
entropy flux 4-vector $s^a_L$ whose integral over $L$ is the
entropy flux through $L$.}

\item{The inequality
\begin{equation}
\left|s_L^a k^a\right|\leq
{\pi}(\lambda_\infty-\lambda) T_{ab} k^a k^b
\label{eq-fmw1}
\end{equation}
holds everywhere on $L$.  Here $\lambda_\infty$ is the value of the
affine parameter at the endpoint of the light-sheet.}

\end{itemize}

The entropy flux vector $s_L^a$ is defined non-locally by demanding
that only modes that are fully captured on $L$ contribute to the
entropy on $L$.  Modes that are partially contained on $L$ do not
contribute.  This convention recognizes that entropy is a non-local
phenomenon.  It is particularly useful when light-sheets penetrate a
thermodynamic system only partially, as discussed in
Sec.~\ref{sec-els}.

This set of assumptions can be viewed as a kind of ``light ray
equivalent'' of Bekenstein's bound, Eq.~(\ref{eq-bekbound}), with the
affine parameter playing the role of the circumferential radius.
However, it is not clear whether one should expect this condition to
be satisfied in regions of dominant gravity.  Indeed, it does not
apply to some weakly gravitating systems (Sec.~\ref{sec-gia}).

FMW were actually able to prove a stronger form of the covariant
entropy bound from the above hypotheses.  Namely, suppose that the
light-sheet of a surface of area $A$ is constructed, but the
light rays are not followed all the way to the caustics.  The
resulting light-sheet is, in a sense, shorter than necessary, and one
would expect that the entropy on it, $S$, will not saturate the bound.
The final area spanned by the light rays, $A'$, will be less than $A$
but non-zero (Fig.~\ref{fig-inside}b).

FMW showed, with the above assumptions, that a tightened bound results
in this case:
\begin{equation}
S\leq\frac{A-A'}{4}.
\label{eq-strongbound}
\end{equation}
Note that this expression behaves correctly in the limit where the
light-sheet is maximized [$A'\to 0$; one recovers Eq.~(\ref{eq-bb2})]
and minimized ($A'\to A$; there is no light-sheet and hence no
entropy).

The strengthened form, Eq.~(\ref{eq-strongbound}), of the covariant
entropy bound, Eq.~(\ref{eq-bb2}), appears to have broad, but not
completely general validity (Sec.~\ref{sec-gia}).

\subsubsection{The second FMW theorem}
\label{sec-fmw2}

Through a rather non-trivial proof, FMW showed that the covariant
entropy bound can also be derived from a second set of assumptions,
namely:

\begin{itemize}

\item{The entropy content of spacetime is well approximated by an
absolute entropy flux vector field $s^a$.}

\item{For any null vector $k^a$, the inequalities
\begin{eqnarray}
(s_a k^a)^2 & \leq & {1\over 16\pi} T_{ab} k^a k^b, \\ 
\left| k^a k^b \nabla_a s_b \right| & \leq &
{\pi\over 4} T_{ab} k^a k^b
\end{eqnarray}
hold at everywhere in the spacetime.}

\end{itemize}

These assumptions are satisfied by a wide range of matter systems,
including Bose and Fermi gases below the Planck temperature.  It is
straightforward to check that all of the adiabatically evolving
cosmologies investigated in Sec.~\ref{sec-cosmo} conform to the above
conditions.  Thus, the second FMW theorem rules out an enormous class
of potential counterexamples, obviating the hard work of calculating
light-sheets.  (We will find light-sheets in simple cosmologies
anyway, both in order to gain intuition about how the light-sheet
formulation works in cosmology, and also because this analysis is
needed for the discussion of holographic screens in
Sec.~\ref{sec-screens}.)

Generally speaking, the notion of an entropy flux assumes that entropy
can be treated as a kind of local fluid.  This is often a good
approximation, but it ignores the non-local character of entropy and
does not hold at a fundamental level.

\subsection{Relation to other bounds and to the GSL}
\label{sec-rob}

\subsubsection{Spacelike projection theorem}
\label{sec-spt}

We have seen in Sec.~\ref{sec-fail} that the spacelike entropy bound
does not hold in general.  Taking the covariant entropy bound as a
general starting point, one may derive other, more limited
formulations, whose regimes of validity are defined by the assumptions
entering the derivation.  Here we use the light-sheet formulation to
recover the spacelike entropy bound, Eq.~(\ref{eq-seb}), along with
precise conditions under which it holds.  By imposing further
conditions, even more specialized bounds can be obtained; an example
valid for certain regions in cosmological spacetimes is discussed in
Sec.~\ref{sec-cosmocor} below.

\begin{figure}[h] \centering
\includegraphics[width=7cm]{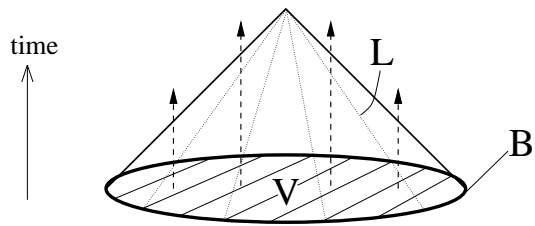}
\caption{Spacelike projection theorem.  If the surface $B$ has a
complete future directed light-sheet $L$, then the spacelike entropy
bound applies to any spatial region $V$ enclosed by $B$.}
\label{fig-spt}
\end{figure}
{\sl Spacelike projection theorem} (Bousso, 1999a).  {\em Let $B$ be a
closed surface.  Assume that $B$ permits at least one future directed
light-sheet $L$.  Moreover, assume that $L$ is {\em complete}, i.e.,
$B$ is its only boundary (Fig.~\ref{fig-spt}).  Let $S(V)$ be the
entropy in a spatial region $V$ enclosed by $B$ on the same side as
$L$.  Then}
\begin{equation}
S(V) \leq S(L) \leq \frac{A}{4}.
\label{eq-spt}
\end{equation}
{\sl Proof.} Independently of the choice of $V$ (i.e., the choice of a
time coordinate), all matter present on $V$ will pass through $L$.
The second law of thermodynamics implies the first inequality, the
covariant entropy bound implies the second.

What is the physical significance of the assumptions made in the
theorem?  Suppose that the region enclosed by $B$ is weakly
gravitating.  Then we may expect that all assumptions of the theorem
are satisfied.  Namely, if $B$ did not have a future directed
light-sheet, it would be anti-trapped---a sign of strong gravity.  If
$L$ had other boundaries, this would indicate the presence of a future
singularity less than one light-crossing time from $B$---again, a sign
of strong gravity.

Thus, for a closed, weakly gravitating, smooth surface $B$ we may
expect the spacelike entropy bound to be valid.  In particular, the
spherical entropy bound, deemed necessary for the validity of the GSL
in the Susskind process, follows from the covariant bound.  This can
be see by inspecting the assumptions in (Sec.~\ref{sec-spheb}), which
guarantee that the conditions of the spacelike projection theorem are
satisfied.

\subsubsection{Generalized second law and Bekenstein bound}
\label{sec-gia}

In fact, FMW showed that the covariant bound implies the GSL directly
for any process of black hole formation, such as the Susskind process
(Sec.~\ref{sec-susskind}).

Consider a surface $B$ of area $A$ on the event horizon of a black
hole.  The past directed ingoing light rays will have non-positive
expansion; they generate a light-sheet.  The light-sheet contains all
the matter that formed the black hole.  The covariant bound implies
that $S_{\rm matter} \leq A(B)/4 = S_{\rm BH}$.  Hence, the
generalized second law is satisfied for the process in which a black
hole is newly formed from matter.

Next, let us consider a more general process, the absorption of a
matter system by an existing black hole.  This includes the Geroch
process (Sec.~\ref{sec-geroch}).  Does the covariant bound
also imply the GSL in this case?

Consider a surface $B$ on the event horizon after the matter system,
of entropy $S_{\rm matter}$, has fallen in, and follow the
past-ingoing light-rays again.  The light-rays are focussed by the
energy momentum of the matter.  ``After'' proceeding through the
matter system, let us terminate the light-sheet.  Thus the light-sheet
contains precisely the entropy $S_{\rm matter}$.  The rays will span a
final area $A'$ (which is really the initial area of the event horizon
before the matter fell in).

According to an outside observer, the Bekenstein entropy of the black
hole has increased by $(A-A')/4$, while the matter entropy $S_{\rm
matter}$ has been lost.  According to the ``strengthened form'' of the
covariant entropy bound considered by FMW, Eq.~(\ref{eq-strongbound}),
the total entropy has not decreased.  The original covariant bound,
Eq.~(\ref{eq-bb2}), does not by itself imply the generalized second
law of thermodynamics, Eq.~(\ref{eq-gsl}), in this process.

Eq.~(\ref{eq-strongbound}) can also be used to derive a version of
Bekenstein's bound, Eq.~(\ref{eq-bekbound})---though, unfortunately, a
version that is too strong.  Consider the light-sheet of an
approximately flat surface of area $A$, bounding one side of a
rectangular thermodynamic system.  With suitable time-slicing, the
surface can be chosen to have vanishing null expansion, $\theta$.

With assumptions on the average energy density and the equation of
state, Raychaudhuri's equation can be used to estimate the final area
$A'$ of the light-sheet where it exits the opposite side of the matter
system.  The strengthened form of the covariant entropy bound,
Eq.~(\ref{eq-strongbound}) then implies the bound given in
Eq.~(\ref{eq-bekbound-wrong}).  However, for very flat systems this
bound can be violated (Sec.~\ref{sec-unruh})!

Hence, (\ref{eq-strongbound}) cannot hold in the same generality that
is claimed for the original covariant entropy bound,
Eq.~(\ref{eq-bb2}).\footnote{It follows that the first FMW hypotheses
do not hold in general.  An earlier counterexample to
Eq.~(\ref{eq-strongbound}), and hence to these hypotheses, was given
by Guedens (2000).}  However, the range of validity of
Eq.~(\ref{eq-strongbound}) does appear to be extremely broad.  In view
of the significance of its implications, it will be important to
better understand its scope.

We conclude that the covariant entropy bound implies the spherical
bound in its regime of validity, defines a range of validity for the
spacelike bound, and implies the GSL for black hole formation
processes.  The strengthened form of the covariant bound given by FMW,
Eq.~(\ref{eq-strongbound}), implies the GSL for absorption processes
and, under suitable assumptions, yields Bekenstein's bound [though in
a form that demonstates that Eq.~(\ref{eq-strongbound}) cannot be
universally valid].

The result of this section suggest that the holographic principle
(Sec.~\ref{sec-hp}) will take a primary role in the complex of ideas
we have surveyed.  It may come to be viewed as the logical origin not
only of the covariant entropy bound, but also of more particular laws
that hold under suitable conditions, such as the spherical entropy
bound, Bekenstein's bound, and the generalized second law of
thermodynamics.

\section{Applications and examples}
\label{sec-tests}

In this section, the covariant entropy bound is applied to a variety
of matter systems and spacetimes.  We demonstrate how the light-sheet
formulation evades the various difficulties encountered by the
spacelike entropy bound (Sec.~\ref{sec-fail}).

We apply the bound to cosmology and verify explicitly that it is
satisfied in a wide class of universes.  No violations are found
during the gravitational collapse of a star, a shell, or the whole
universe, though the bound can be saturated.

\subsection{Cosmology}
\label{sec-cosmo}

\subsubsection{FRW metric and entropy density}

Friedmann-Robertson-Walker (FRW) metrics describe homogeneous,
isotropic universes, including, to a good degree of approximation, the
portion we have seen of our own universe.  Often the metric is
expressed in the form
\begin{equation}
ds^2 = -dt^2 + a^2(t) \left( \frac{dr^2}{1-kr^2} + r^2 d\Omega^2
\right).
\label{eq-FRW1}
\end{equation}

\begin{figure}[h] \centering
\includegraphics[width=8.5cm]{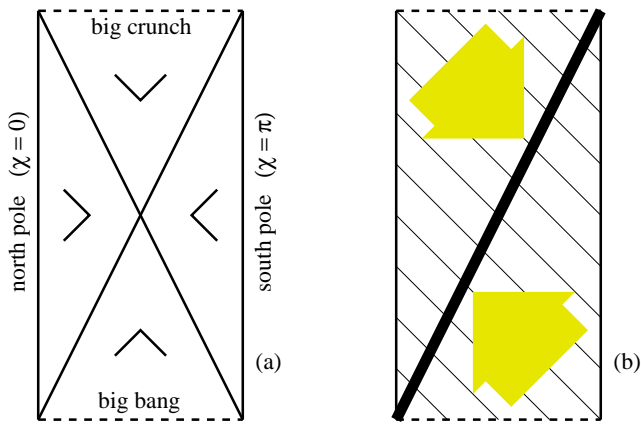}
\caption{(a) Penrose diagram for a closed FRW universe filled with
pressureless dust.  The three-sphere time slices are represented by
horizontal lines (not shown).  Two apparent horizons divide the
diagram into four wedge domains: normal spheres are found near the
poles, trapped (anti-trapped) spheres near the big bang (big crunch).
(b) The construction of a global holographic screen
(Sec.~\ref{sec-screens}) proceeds by foliating the spacetime into a
stack of light cones.  The information on each slice can be stored on
the maximal sphere, which lies on the apparent horizon.}
\label{fig-clos}
\end{figure}

We will find it more useful to use the conformal time $\eta$ and the
comoving coordinate $\chi$:
\begin{equation}
d\eta = {dt\over a(t)},~~~d\chi = {dr\over\sqrt{1-kr^2}}.
\end{equation} 
In these coordinates the FRW metric takes the form
\begin{equation}
ds^2 = a^2(\eta) \left[ -d\eta^2 + d\chi^2 + f^2(\chi) d\Omega^2
\right].
\label{eq-FRW2}
\end{equation}
Here $k = -1$, $0$, $1$ and $f(\chi) = \sinh \chi$, $\chi$, $\sin
\chi$ correspond to open, flat, and closed universes respectively.
Relevant Penrose diagrams are shown in Figs.~\ref{fig-flatfrw} and
\ref{fig-clos}a.

In cosmology, the entropy is usually described by an entropy density
$\sigma$, the entropy per physical volume:
\begin{equation}
S(V) = \int_V d^3x \sqrt{h}\, \sigma.
\end{equation}
For FRW universes, $\sigma$ depends only on time.  We will assume, for
now, that the universe evolves adiabatically.  Thus, the physical
entropy density is diluted by cosmological expansion:
\begin{equation}
\sigma(\eta) = {s\over a(\eta)^3}.
\end{equation}
The comoving entropy density $s$ is constant in space and time.

\subsubsection{Expansion and apparent horizons}
\label{sec-ah}

Let us verify that the covariant entropy bound is satisfied for each
light-sheet of any spherical surface $A$.  The first step is to
identify the light-sheet directions.  We must classify each sphere as
trapped, normal, or anti-trapped (Sec.~\ref{sec-kin2}).  Let us
therefore compute the initial expansion of the four families of
light rays orthogonal to an arbitrary sphere characterized by some
value of $(\eta,\chi)$.

We take the affine parameter to agree locally with $\pm 2\eta$ and use
Eq.~(\ref{eq-tda}).  Differentiation with respect to $\eta$ ($\chi$)
is denoted by a dot (prime).  Instead of labelling the families
$F_1,\ldots,F_4$, it will be more convenient to use the notation
$(\pm\pm)$, where the first sign refers to the time ($\eta$) direction
of the light rays and the second sign denotes whether they are
directed at larger or smaller values of $\chi$.

For the future directed families one finds
\begin{equation}
\theta_{+\pm} = {\dot{a}\over a} \pm {f'\over f}.
\label{eq-thp}
\end{equation}
The expansion of the past directed families is given by
\begin{equation}
\theta_{-\pm} = - {\dot{a}\over a} \pm {f'\over f}.
\label{eq-thm}
\end{equation}

Note that the first term in Eq.~(\ref{eq-thp}) is positive when the
universe expands and negative if it contracts.  The term diverges when
$a\to 0$, i.e., near singularities.  The second term is given by
$\cot\chi$ ($1/\chi$; $\coth\chi$) for a closed (flat; open) universe.
It diverges at the origin ($\chi\to 0$), and for a closed universe it
also diverges at the opposite pole ($\chi\to\pi$).  

The signs of the four quantities $\theta_{\pm\pm}$ depend on the
relative strength of the two terms.  The quickest way to classify
surfaces is to identify marginal spheres, where the two terms are of
equal magnitude.

The {\em apparent horizon\/} is defined geometrically as a sphere at
which at least one pair of orthogonal null congruences have zero
expansion.  It satisfies the condition
\begin{equation}
{\dot{a}\over a} = \pm {f'\over f},
\label{eq-ahcond}
\end{equation}
which can be used to identify its location $\chi_{\rm AH}(\eta)$ as a
function of time.  There is one solution for open and flat universes.
For a closed universe, there are generally two solutions, which are
symmetric about the equator [$\chi_{{\rm AH}'}(\eta) =
\pi-\chi_{\rm AH} (\eta)$].

The proper area of the apparent horizon is given by
\begin{equation}
A_{\rm AH}(\eta) = 4 \pi a(\eta)^2 f[\chi_{\rm AH}(\eta)]^2 = 
{4\pi a^2\over \left({\dot a\over a}\right)^2+k}.
\end{equation}
Using Friedmann's equation,
\begin{equation}
{\dot a^2\over a^2} = \frac{8\pi \rho a^2}{3} - k,
\end{equation}
one finds
\begin{equation}
A_{\rm AH}(\eta) = {3\over 2\rho(\eta)},
\label{eq-ah-rho}
\end{equation}
where $\rho$ is the energy density of matter.

At any time $\eta$, the spheres that are smaller than the apparent
horizon,
\begin{equation}
A< A_{\rm AH},
\end{equation}
are normal.  (See the end of Sec.~\ref{sec-kin2} for the definitions
of normal, trapped, and anti-trapped surfaces.)  Because the second
term $f'/f$ dominates in the expressions for the expansion, the
cosmological evolution has no effect on the light-sheet directions.
The two light-sheets will be a past and a future directed family
going to the same spatial side.  In a flat or open universe, they will
be directed towards $\chi=0$ (Fig.~\ref{fig-flatfrw}).  In a closed
universe, the light-sheets of a normal sphere will be directed towards
the nearest pole, $\chi=0$ or $\chi=\pi$ (Fig.~\ref{fig-clos}a).

For spheres greater than the apparent horizon
\begin{equation}
A> A_{\rm AH},
\end{equation}
the cosmological term $\dot a/a$ dominates in the expressions for the
expansion.  Then there are two cases.  Suppose that $\dot a>0$, i.e.,
the universe is expanding.  Then the spheres are anti-trapped.  Both
light-sheets are past directed, as indicated by a wedge opening to the
bottom in the Penrose diagram.  If $A>A_{\rm AH}$ and $\dot a<0$, then
both future directed families will have negative expansion.  This case
describes trapped spheres in a collapsing universe.  They are denoted
by a wedge opening to the top (Fig.~\ref{fig-clos}a).

\subsubsection{Light-sheets vs.~spatial volumes}
\label{sec-lvs}

We have now classified all spherical surfaces in all FRW universes
according to their light-sheet directions.  Before proceeding to a
detailed calculation of the entropy contained on the light-sheets, we
note that the violations of the spacelike entropy bound identified in
Secs.~\ref{sec-fail1} and \ref{sec-fail2} do not apply to the
covariant bound.

The area of a sphere at $\eta_0,\chi_0$ is given by
\begin{equation}
A(\eta_0,\chi_0) = 4\pi a(\eta_0)^2 f(\chi_0)^2.
\label{eq-afrw}
\end{equation}
To remind ourselves that the spacelike entropy bound fails in
cosmology, let us begin by comparing this area to the entropy enclosed
in the spatial volume $V(\chi_0)$ defined by $\chi\leq\chi_0$ at equal
time $\eta=\eta_0$.  With our assumption of adiabaticity, this depends
only on $\chi_0$:
\begin{equation}
S[V(\chi_0)] = 4\pi s \int_0^{\chi_0} d\chi f(\chi)^2.
\end{equation}

For a flat universe [$f(\chi) = \chi$], the area grows like $\chi_0^2$
but the entropy grows like $\chi_0^3$.  Thus, $S[V(\chi_0)]>A$ for
sufficiently large $\chi_0$. (This was pointed out earlier in
Sec.~\ref{sec-fail2}.)  For a closed universe ($f(\chi) =\sin\chi$),
$\chi$ ranges only from $0$ to $\pi$.  $S[V(\chi_0)]$ is monotonically
increasing in this range, but $A\to 0$ for $\chi_0\to\pi$.  Again, one
has $S[V(\chi_0)]>A$.  This is a special case of the problem discussed
in Sec.~\ref{sec-fail1}.

Why don't light-sheets run into the same difficulties?  Consider first
a large sphere in a flat universe (Fig.~\ref{fig-flatfrw}).  The
future-ingoing light rays cover the same amount of entropy as the
enclosed spatial volume.  However, for spheres greater than the
apparent horizon, the future-ingoing light rays are expanding and
hence do not form a light-sheet.  Only past directed light-sheets are
permitted.  The past-ingoing light rays, for example, will proceed
towards the origin.  However, if the sphere is greater than the
particle horizon ($\chi>\eta$), they will terminate at the big bang
($\eta=0$) and will not get all the way to $\chi=0$.  Instead of a
comoving ball $0\leq\chi'\leq\chi$, they will sweep out only a shell
of width $\eta$: $\chi-\eta\leq\chi'\leq\chi$.  Thus the entropy to
area ratio does not diverge for large $\chi$, but approaches a
constant value.

Small spheres ($A<A_{\rm AH}$) in a closed universe
(Fig.~\ref{fig-clos}a) permit only light-sheets that are directed to
the smaller enclosed region.  The light rays directed towards the
larger portion of the universe will be initially expanding and hence
do not form light-sheets.  Both in the flat and the closed case, we
see that the $\theta\leq 0$ contraction condition is of crucial
importance.

\subsubsection{Solutions with fixed equation of state}

The matter content of FRW universes is most generally described by a
perfect fluid, with stress tensor
\begin{equation}
T^a_{~b} = \mbox{diag}(-\rho,p,p,p).
\end{equation}
Let us assume that the pressure $p$ and energy density $\rho$ are
related by a fixed equation of state
\begin{equation}
p = w \rho.
\end{equation}
Our universe and many other more general solutions can be pieced
together from solutions obtained via this ansatz, because the
transitions between different effective equations of state are very
rapid.

For most of its lifetime, our universe was dominated by pressureless
dust and hence was characterized by $w=0$.  The early universe was
dominated by radiation, which is described by $w={1\over 3}$.  A
cosmological constant, which may have been present at very early
times and perhaps again today, corresponds to $w=-1$.  

With this ansatz for the matter content and the FRW ansatz for the
metric, Einstein's equation can be solved.  This determines the scale
factor in Eq.~(\ref{eq-FRW2}):
\begin{equation}
a(\eta) = a_0 \left[f({\eta\over q})\right]^q,
\label{eq-a}
\end{equation}
where
\begin{equation}
q = {2\over 1+3w},
\label{eq-q}
\end{equation}
and $f$ is the $\sin$ (the identity, $\sinh$) for a closed (flat,
open) universe, as in Eq.~(\ref{eq-FRW2}).  From Eq.~(\ref{eq-ahcond})
it follows that an apparent horizon is located at
\begin{equation}
\chi_{\rm AH}(\eta) = {\eta\over q}
\label{eq-ahq}
\end{equation}
in all cases.  An additional mirror horizon lies at $\pi-{\eta\over
q}$ in the closed case.

Having established the light-sheet directions as a function of $t$ and
$r$, we will now check whether the covariant entropy bound is
satisfied on all light-sheets.  The present treatment concentrates on
flat and closed ($k=0,1$) universes with $w\geq 0$.  However, we will
quote results for $w<0$, i.e., negative pressure (Kaloper and Linde,
1999), which involves additional subtleties.  We will also comment on
the inflationary case ($w=-1$).  We omit the open universes ($k=-1$)
because they do not give rise to qualitatively new features (Fischler
and Susskind, 1998).  Bousso (1999a) discusses closed universes in
detail.  The main additional features beyond the flat case are covered
in Secs.~\ref{sec-lvs} and \ref{sec-collapses}.  We will comment on
the inflationary case ($w=-1$) separately.

\subsubsection{Flat universe}
\label{sec-ff}

Let us consider all possible light-sheets of all spherical areas
($0<\chi<\infty$) at the time $\eta$ in a flat FRW universe,
\begin{equation}
A(\eta,\chi) = 4\pi a(\eta)^2 \chi^2.
\end{equation} 
If $\chi\leq\chi_{\rm AH}(\eta)$, the sphere is normal, and the
light-sheet directions are ($+-$) and ($--$).  If $\chi\geq\chi_{\rm
AH}$, the sphere is anti-trapped, with light-sheets ($-+$) and ($--$)
(Fig.~\ref{fig-flatfrw}).

We begin with the future-ingoing ($+-$) light rays.  They contract
towards the origin and generate a conical light-sheet whose
coordinates $(\chi',\eta')$ obey
\begin{equation}
\chi'+\eta' = \chi+\eta,
\end{equation}
This light-sheet contains the comoving entropy in the region
$0\leq\chi'\leq\chi$, which is given by
\begin{equation}
S_{+-} = {4\pi\over 3} s \chi^3.
\end{equation}
The ratio of entropy to area,
\begin{equation}
{S_{+-}\over A} = {s\chi\over 3a(\eta)^2},
\end{equation}
is maximized by the outermost normal
surface at any given time $\eta$, the sphere on the apparent horizon.
Thus we obtain the bound
\begin{equation}
{S_{+-}\over A} \leq {s \chi_{\rm AH}(\eta)\over 3 a(\eta)^2}.
\end{equation}

The past-ingoing ($--$) light-sheet of any surface with $\chi<\eta$
also reaches a caustic at $\chi=0$.  If $\chi>\eta$, then the
light-sheet is truncated instead by the big bang singularity at
$\eta=0$.  Then it will contain the comoving entropy in the region
$\chi-\eta\leq\chi'\leq\chi$.  The entropy to area ratio is given by
\begin{equation}
{S_{--}\over A} = {s\, \eta\over a(\eta)^2} \left(1-{\eta\over\chi}
  +{\eta^2\over 3\chi^2} \right).
\end{equation}
This ratio is maximized for large spheres ($\chi\to\infty$), yielding
the bound
\begin{equation}
{S_{--}\over A} \leq {s\, \eta\over a(\eta)^2}
\end{equation}
for the ($--$) light-sheets at time $\eta$.

Finally, we must consider the past-outgoing ($-+$) light-sheet of any
surface with $\chi>\chi_{\rm AH}$.  It is truncated by the big bang
singularity and contains the entropy within
$\chi\leq\chi'\leq\chi+\eta$.  The ratio of the entropy to the area,
\begin{equation}
{S_{-+}\over A} = {s\, \eta\over a(\eta)^2} \left(1+{\eta\over\chi}
  +{\eta^2\over 3\chi^2} \right),
\end{equation}
is maximized for the smallest possible value of $\chi$, the apparent
horizon.  We find the bound
\begin{equation}
{S_{-+}\over A} \leq {s\, \eta\over a(\eta)^2} \left(1+{\eta\over
\chi_{\rm AH}}  +{\eta^2\over 3 \chi_{\rm AH}^2} \right).
\end{equation}

We now use the solution for fixed equation of state, setting $a_0=1$
for convenience:
\begin{equation}
a(\eta) = \left({\eta\over q}\right)^q,~~~
\chi_{\rm AH}(\eta)={\eta\over q}.
\end{equation}
Up to factors of order unity, the bounds for all three types of
light-sheets at time $\eta$ agree:
\begin{equation}
{S\over A} \leq s\, \eta^{1-2q}.
\label{eq-soaflat}
\end{equation}
Note that one Planck distance corresponds to the comoving coordinate
distance $\Delta\chi=a(\eta)^{-1}$.  At the Planck time, $\eta \sim
a(\eta) \sim O(1)$.  Hence, $s$ is roughly the amount of entropy
contained in a single Planck volume at one Planck time after the big
bang.  This is the earliest time and shortest distance scale one can
hope to discuss without a full quantum gravity description.  It is
reasonable to assume that a Planck volume contains no more than one
bit of information:
\begin{equation}
s\lesssim 1.
\end{equation}

Eq.~(\ref{eq-soaflat}) then implies that the covariant entropy bound,
Eq.~(\ref{eq-bb2}), is satisfied at the Planck time.  Moreover, the
bound will continue to be satisfied by all light-sheets of all spheres
at later times ($\eta>1$), if $q \geq {1\over 2}$.  In terms of the
parameter $w$, this corresponds to the condition
\begin{equation}
w\leq 1.
\end{equation}
This result was obtained by Fischler and Susskind (1998) who also
assumed $w\geq 0$.

Kaloper and Linde (1999) showed more generally that the entropy bound
will be satisfied at all times if $-1<w\leq 1$, provided that the
bound is satisfied at the Planck time.\footnote{Like Fischler and
Susskind (1998), this work precedes the covariant entropy bound
(Bousso, 1999a).  Hence it considers only the ($--$) case, which
corresponds to the Fischler-Susskind proposal.  We have seen that the
entropy range on other light-sheets does not differ significantly in
the flat case.  Of course, the {\em absence\/} of a ($--$) light-sheet
for some surfaces in other universes is crucial for the validity of
the covariant entropy bound (see, e.g., Secs.~\ref{sec-lvs},
\ref{sec-collapses}).---Davies (1987) obtained $w \geq -1$ as a
condition for the growth of the apparent horizon in an inflating
universe.}  The case $w=-1$ corresponds to de~Sitter space, in which
there is no initial singularity.  Since a cosmological constant does
not carry entropy, the bound is trivially satisfied in this case.  In
summary, all light-sheets of all surfaces in any flat FRW universe
with equation of state satisfying
\begin{equation}
-1\leq w\leq 1
\label{eq-frwcond}
\end{equation}
satisfy the covariant entropy bound, Eq.~(\ref{eq-bb2}).

This condition is physically very reasonable.  It follows from the
causal energy condition, which prohibits the superluminal flow of
energy.  We assumed in Sec.~\ref{sec-econds} that this condition holds
along with the null energy condition.  The definitions of all relevant
energy conditions are reviewed in the Appendix.

\subsubsection{Non-adiabatic evolution and mixed equations of state}
\label{sec-na}

So far, we have assumed that the universe evolves adiabatically.  In
order to relax this assumption, one generally has to abandon the FRW
solution given above and find the exact geometry describing a
cosmology with increasing entropy.  However, the global solution will
not change significantly if we rearrange matter on scales smaller than
the apparent horizon.

Consider the future-ingoing light-sheet of the present apparent
horizon, $L_{+-}[\eta_0, \chi_{\rm AH}(\eta_0)]$.  All
entropy we generate using the matter available to us inside the
apparent horizon, will have to pass through this light-sheet.  An
efficient way to generate entropy is to form black holes.  Building on
a related discussion by Bak and Rey (2000a), Bousso (1999a) showed
that the highest entropy is obtained in the limit where all matter is
converted into a few big black holes.  In this limit,
$S_{+-}/A[\eta_0,\chi_{\rm AH}(\eta_0)]$ approaches $1/4$ from below.
Hence the covariant bound is satisfied and can be saturated.

According to the inflationary model of the early universe (see, e.g.,
Linde, 1990), a different non-adiabatic process occurred at the end of
inflation.  At the time of reheating, matter is produced and a large
amount of entropy is generated.  One might be concerned that the
holographic principle is violated by inflation (Easther and Lowe,
1999), or that it places severe constraints on acceptable models
(Kalyana Rama and Sarkar, 1999).

Before inflation ended, however, there was almost no entropy.  Hence,
all past directed light-sheets can be truncated at the reheating
surface, $\eta=\eta_{\rm reheat}$.  The energy density at reheating is
expected to be significantly below the Planck density.  The
light-sheets will be cut shorter than in our above discussion, which
assumed that standard cosmology extended all the way back to the
Planck era.  Hence, inflation leads to no difficulties with the
holographic principle.\footnote{Fabinger (2001) has suggested a bound
on entanglement entropy, assuming certain inflationary models apply.}

Kaloper and Linde (1999) studied a particularly interesting cosmology,
a flat FRW universe with ordinary matter, $w_1\geq 0, \rho_1>0$, as
well as a small negative cosmological constant, $w_2=-1, \rho_2<0$.
The universe starts matter dominated, but the cosmological constant
eventually takes over the evolution.  It slows down and eventually
reverses the expansion.  In a time symmetric fashion, matter
eventually dominates and the universe ends in a future singularity.

The Kaloper-Linde universe provides a tough testing ground for
proposals for a cosmological holographic principle.  As in any flat
FRW universe, spacelike holography breaks down for sufficiently large
surfaces.  Moreover, as in any collapsing universe, this occurs even
if one restricts to surfaces within the particle horizon, or the
Hubble horizon.  Most interestingly, the ``apparent horizon'' proposal
of Bak and Rey (2000a) fails in this cosmology.  This can be understood
by applying the spacelike projection theorem to cosmology, as we
discuss next.

The holographic principle in anisotropic models was discussed by
Fischler and Susskind (1998) and by Cataldo {\em et al.} (2001).
Inhomogeneous universes have been considered by Tavakol and Ellis
(1999); see also Wang, Abdallah, and Osada (2000).

\subsubsection{A cosmological corollary}
\label{sec-cosmocor}

Let us return to a question first raised in Sec.~\ref{sec-range}.
What is the largest volume in a cosmological spacetime to which the
spacelike holographic principle can be applied?  The spacelike
projection theorem (Sec.~\ref{sec-spt}) guarantees that the spacelike
entropy bound will hold for surfaces that admit a future directed,
complete light-sheet.  Let us apply this to cosmology.  Surfaces on or
within the apparent horizon are normal and hence admit a future
directed light-sheet.  However, the completeness condition is not
trivial and must be demanded separately.  In the Kaloper-Linde
universe, for example, the future light-sheets of sufficiently late
surfaces on the apparent horizon are truncated by the future
singularity.

We thus arrive at the following corollary to the spacelike projection
theorem (Bousso, 1999a): {\em The area of any sphere within the
apparent horizon exceeds the entropy enclosed in it, if the future
light-sheet of the sphere is complete.}

\subsection{Gravitational collapse}
\label{sec-collapses}

Any argument for an entropy bound based on the generalized second law
of thermodynamics must surely become invalid in a collapse regime.
When a system is already inside its own Schwarzschild radius, it can
no longer be converted into a black hole of equal surface area.

Indeed, the example of the collapsing star (Sec.~\ref{sec-fail3}), and
the conclusions reached by various analyses of collapsing universes
(Sec.~\ref{sec-range}) would seem to discourage hopes of finding a
non-trivial holographic entropy bound that continues to hold in
regions undergoing gravitational collapse.  Surprisingly, to the
extent that it has been tested, the covariant entropy bound does
remain valid in such regions.

Whenever possible, the validity of the bound is most easily verified
by showing that a given solution satisfies the local hypotheses of
Flanagan, Marolf, and Wald (2000).  Otherwise, light-sheets must be
found explicitly.

Ideally, one would like to investigate systems with high entropy, in
dynamical, collapsing spacetime regions.  Generically, such regions
will be extremely inhomogeneous, which makes the practical calculation
of light-sheets difficult.  However, one should keep in mind that
other proposals for general entropy bounds, such as the spacelike
entropy bound, are quickly invalidated by simple, easily tractable
counterexamples that make use of gravitational collapse.

It is remarkable, from this point of view, that the covariant bound
has not met its demise by any of the standard collapse solutions that
are readily available in the literature.  To illustrate how the
covariant bound evades violation, we will review its application to
two simple examples, a collapsing star and a closed universe.

We will also consider a particular setup that allows the calculation
of light-sheets deep inside a black hole formed by the collapse of a
spherical shell.  In this example one has good quantitative control
over the collapse of a system of arbitrarily high entropy.

\subsubsection{Collapsing universe}
\label{sec-cun}

Let us begin with a very simple example, the adiabatic recollapse of a
closed FRW universe.  In this case the recollapsing phase is just the
time-reversal of the expanding phase.  The light-sheet directions are
similarly reversed (Fig.~\ref{fig-clos}a).  Small spheres near the
poles are normal, but larger spheres, which are anti-trapped during
expansion will be trapped during collapse.  Their light-sheets are
future directed and hence are typically truncated by the future (big
crunch) singularity.

Because the solution is symmetric under time reversal, the validity of
the covariant entropy bound in the collapse phase follows from its
validity in the expanding phase.  The latter can be verified
straightforwardly.  For anti-trapped spheres, the calculation
(Fischler and Susskind, 1998) is similar to the analysis of the flat
case (Sec.~\ref{sec-ff}).  For small spheres one needs to pay special
attention to choosing the correct inside directions (see
Sec.~\ref{sec-lvs}).

\subsubsection{Collapsing star}
\label{sec-cst}

Next, we return to the collapsing star of Sec.~\ref{sec-fail3}.  Why
don't the arguments demonstrating the break-down of other entropy
bounds extend to the covariant entropy bound?

The metric in and around a collapsing star is well described by the
Oppenheimer-Snyder solution (Misner, Thorne, and Wheeler, 1973).  In
this solution, the star is modelled by a suitable portion of a
collapsing closed FRW universe.  That is, one considers the coordinate
range
\begin{equation}
0\leq\chi\leq\chi_0,~~\eta>q{\pi\over 2},
\end{equation}
in the metric of Eq.~(\ref{eq-FRW2}).  Here, $q$ depends on the
equation of state in the star according to Eq.~(\ref{eq-q}).  Also,
$\chi_0<\pi/2$, so that the star does not overclose the universe.
Outside the star, space is empty.  Birkhoff's theorem dictates that
the metric will be given by a portion of the Schwarzschild solution,
Eq.~(\ref{eq-schsch}).

\begin{figure}[h] \centering
\includegraphics[width=8.5cm]{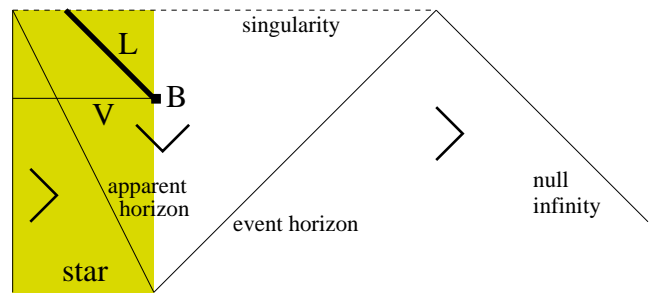}
\caption{Penrose diagram of a collapsing star (shaded).  At late
times, the area of the star's surface becomes very small ($B$).  The
enclosed entropy (in the spatial region $V$) stays finite, so that the
spacelike entropy bound is violated.  The covariant entropy bound
avoids this difficulty because only future directed light-sheets are
allowed. $L$ is truncated by the future singularity; it does not
contain the entire star.}
\label{fig-star}
\end{figure}
The corresponding Penrose diagram is shown in Fig.~\ref{fig-star}.
The light-sheet directions are obtained from the corresponding
portions of the Penrose diagrams for the closed universe
(Fig.~\ref{fig-clos}a) and for the Schwarzschild solution.  At
sufficiently late times, the apparent horizon reaches the surface of
the star.  At this moment, the star forms a black hole.  The surface
of the star is trapped at all later times.  Hence, it admits only
future directed light-sheets near the future singularity.

According to Eq.~(\ref{eq-afrw}), the surface area of the star is
given by
\begin{equation}
A(\chi_0,\eta) = A_{\rm max} \left( \sin{\eta\over q} \right)^q.
\end{equation}
Recall that $q$ is positive and of order unity for realistic equations
of state.  At the time of maximum expansion, $A=A_{\rm max} \equiv
4\pi a_0^2 \sin^2 \chi_0$.  The future singularity corresponds to the
time $\eta=q\pi$.

Let $B$ be the star's surface at a time $\eta_0>q\pi-\chi_0$.  The
future directed ingoing light-sheet will be truncated by the future
singularity at $\chi=\chi_0-(q\pi-\eta_0)$, i.e., it will not traverse
the star completely (Fig.~\ref{fig-star}).  Hence, it will not contain
the full entropy of the star.  For very late times, $\eta_0\to\pi$,
the surface area approaches zero, $A(\chi_0,\eta_0) \to 0$.  The
spacelike entropy bound is violated, $S(V)>A(B)$, because the entropy
of the star does not decrease (Sec.~\ref{sec-fail3}).  But the entropy
$S(L)$ on the ingoing light-sheet, $L$, vanishes in this limit,
because $L$ probes only a shallow outer shell, rather than the
complete star.

Light-sheet truncation by future singularities is but one of several
mechanisms that conspire to protect the covariant entropy bound during
gravitational collapse (see Bousso, 1999a).

\subsubsection{Collapsing shell}
\label{sec-csh}

Consider a small black hole of radius $r_0= 2m$.  In the future of the
collapse event that formed this black hole, the apparent black hole
horizon is a null hypersurface with spacelike, spherical
cross-sections of area $A=4\pi r_0^2$.

Let us pick a particular sphere $B$ of area $A$ on the apparent
horizon.  By definition, the expansion of the past directed ingoing
and the future directed outgoing light rays vanishes near $B$, so both
are allowed light-sheet directions.  

The former light-sheet contains all of the infalling matter that
formed the black hole, with entropy $S_{\rm orig}$.  The covariant
entropy bound, in this case, is the statement of the generalized
second law: the horizon entropy, $A/4$, is greater than the lost
matter entropy, $S_{\rm orig}$.  The future directed ingoing
light rays will be contracting.  They will contain entropy $S_{\rm
orig}$ or less, so the covariant bound is satisfied once more.

We will be interested in the future directed outgoing light-sheet,
$L$.  It will continue to generate the apparent horizon of the black
hole.  Indeed, if no more matter ever enters the black hole, this
apparent horizon coincides with the event horizon, and the light-sheet
will continue forever at zero expansion.

Suppose, however, that more matter eventually falls into the black
hole.  When this happens, the apparent horizon moves out to a larger
value $r>r_0$.  (It will be generated by a new set of light rays that
were formerly expanding.) The light-sheet $L$, however, will begin to
collapse, according to Eq.~(\ref{eq-ray}).  The covariant entropy
bound predicts that the light rays will reach a singularity, or a
caustic, before encountering more entropy than $A/4$.

This is a remarkable prediction.  It claims that one cannot collapse
more entropy through a (temporary) black hole horizon than it already
has.  This claim has been tested (Bousso, 1999a).  Here we summarize
only the method and results.

Far outside the black hole, one can assemble a shell of matter
concentric with the black hole.  By choosing the initial radius of
this shell to be sufficiently large, one can suppress local
gravitational effects and give the shell arbitary total mass, $M$, and
width, $w$.

Let us assume that the shell is exactly spherically symmetric, even at
the microscopic level.  This suppresses the deflection of radial
light rays into angular directions, rendering the eventual calculation
of $L$ tractable.  Moreover, it permits an estimate of the entropy of
the shell.

In weakly gravitating systems, Bekenstein's bound,
Eq.~(\ref{eq-bekbound}), has much empirical support (Bekenstein, 1981,
1984; Schiffer and Bekenstein, 1989).  There is independent evidence
that the bound is always obeyed and can be nearly saturated by
realistic, weakly gravitating matter systems.

Because all excitations are carried by radial modes, the shell can be
divided along radial walls.  This yields several weakly gravitating
systems of largest length scale $w$.  To each, Bekenstein's bound
applies.  After reassembling the shell, one finds that its total
entropy is bounded by
\begin{equation}
S\leq 2\pi Mw.
\label{eq-shell}
\end{equation}
In principle, there are no restrictions on either $M$ or $w$, so the
amount of entropy that can be collapsed onto the black hole is
unlimited.  

Now consider the adiabatic collapse of the shell.  When the inner
surface of the shell has shrunk to area $A$, the shell will first be
reached by the light rays generating $L$.  As the light rays penetrate
the collapsing shell, they are focussed by the shell's stress tensor.
Their expansion becomes negative.  Eventually they reach a caustic.

In order to violate the bound with a shell of large entropy, one would
like to ensure that all of the shell's entropy, $S$, will actually be
contained on $L$.  Thus, one should demand that the light rays must
not reach a caustic before they have fully crossed the shell and
reemerged on the outer surface of the shell.  

Inspection of the collapse solution, however, reveals that this
requirement restricts the shell's mass and width,
\begin{equation}
M w \leq r_0^2/2.
\end{equation}
By Eq.~(\ref{eq-shell}), this also limits the entropy of the shell:
\begin{equation}
S \leq \pi r_0^2= {A\over 4}.
\end{equation}
The entropy on the light-sheet $L$ may saturate the covariant bound,
but it will not violate it.

\subsection{Nearly null boundaries}
\label{sec-nearnull}

In Sec.~\ref{sec-fail4} it was shown that any isolated, weakly
gravitating matter system can be surrounded with a closed surface of
arbitarily small area, in violation of the spacelike entropy bound,
Eq.~(\ref{eq-seb}).

\begin{figure}[h] \centering
\includegraphics[width=8cm]{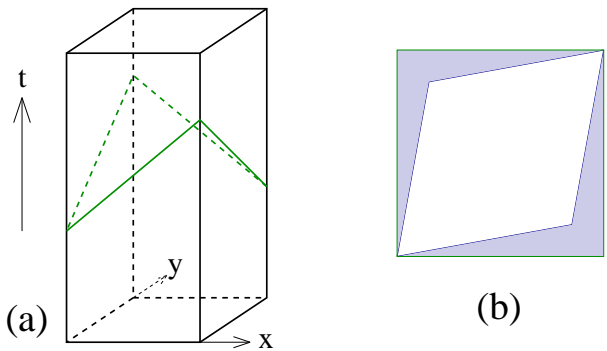}
\caption{(a) A square system in 2+1 dimensions, surrounded by a
surface $B$ of almost vanishing length $A$. (b) [Here the time
dimension is projected out.] The light-sheet of $B$ intersects only
with a negligible (shaded) fraction of the system.}
\label{fig-wiggly}
\end{figure}

In order to capture the key advantage of the light-sheet formulation,
Eq.~(\ref{eq-bb2}), we find it simplest to consider a square-shaped
system occupying the region $0\leq x,y \leq a$ in 2+1 dimensional
Minkowski space; $t$ is the time coordinate in the system's rest frame
(Fig.~\ref{fig-wiggly}a).  The boundary length of the system at $t=0$
is
\begin{equation}
A_0 = 4a.
\end{equation}

Let us define a new boundary $B$ by a zig-zag curve consisting of the
following four segments: $y=0$, $t=\beta x$ for $0\leq x\leq a$;
$x=a$, $t=\beta (a-y)$ for $0\leq y\leq a$; $y=a$, $t=\beta (a-x)$ for
$0\leq x \leq a$; and $x=0$, $t=\beta y$ for $0\leq y \leq a$.  This
can be regarded as the boundary of the system in some non-standard
time-slicing.  Its length is Lorentz-contracted relative to the
boundary in the rest frame:
\begin{equation}
A(B) = A_0 \sqrt{1-\beta^2}.
\end{equation}
The length of $B$ vanishes in the limit as $\beta\to 1$.

The future-ingoing light-sheet $L(B)$ can be computed by piecing
together the light-sheets of all four segments.  The light-sheet of
the first segment is obtained by translating the segment in the
direction $(1, \beta, \sqrt{1-\beta^2})$.  (It is instructive to
verify that this generates an orthogonal null hypersurface of
vanishing expansion.  The curvature of spacetime is neglected in order
to isolate the effect of ``wiggling'' the boundary.)  For
$\beta^2>{1\over 2}$, this light-sheet covers a fraction
$\frac{\sqrt{1-\beta^2}}{2\beta}$ of the total system
(Fig.~\ref{fig-wiggly}b).  The light-sheets of the other segments are
similarly computed.

To leading order in $(1-\beta)$, the total fraction of the system
covered by $L(B)$,
\begin{equation}
{V(\beta)\over V_0} = \frac{2\sqrt{1-\beta^2}}{\beta},
\end{equation}
vanishes at
the same rate as the boundary length.

The future-ingoing light-sheet is not complete in this case; it has
boundaries running through the interior of the system.  Hence, the
assumptions of the spacelike projection theorem are not satisfied.
(This is not just an artifact of the sharp edges of $B$.  If $B$ was
smoothed at the edges, it would contain a segment on which only past
directed light rays would be contracting.  Thus, $B$ would not admit a
future directed light-sheet everywhere, and the spacelike projection
theorem would still not apply.)

\section{The holographic principle}
\label{sec-hp}

\subsection{Assessment}

The previous sections have built a strong case for a holographic
principle.\footnote{All of the following points are
independent of the considerations of economy and unitarity that
motivated 't~Hooft's and Susskind's holographic principle
(Sec.~\ref{sec-hsi}).  However, those arguments emerge strengthened,
since a key difficulty, the absence of a general entropy bound, has
been overcome (Sec.~\ref{sec-limitations}).  One can no longer object
that more than $A/4$ degrees of freedom might be needed to describe
the physics, say, in strongly gravitating regions.}

\begin{itemize}

\item{The covariant entropy bound is well-defined
(Sec.~\ref{sec-bb}).  The light-sheet construction establishes a
precise relation between surfaces and adjacent hypersurfaces.  The
area of the former must be compared to the entropy contained on the
latter.  Thus the bound is testable, in an arena limited only by the
range of semi-classical gravity, the approximate framework we are
compelled to use until a general quantum theory of gravity becomes
available.  Like any law of physics, it can of course be tested only
to the extent that the relevant quantities and constructs (here, area,
light-sheets, and entropy) are practically computable.  But
importantly, the bound will not become ill-defined in a regime which
is otherwise physically well-understood.}

\item{The bound has been examined and found to hold in a wide range
of examples, some of which we reviewed in Sec.~\ref{sec-tests}.  No
physically realistic counterexample has been found.  This is
remarkable especially in view of the ease with which the general
validity of some alternative proposals can be excluded
(Sec.~\ref{sec-seb}).}

\item{The bound is non-trivial.  Naively one would expect the maximal
entropy to grow with the volume of spatial regions.  Instead, it is
set by the area of surfaces.}

\item{The bound refers to statistical entropy.\footnote{A conventional
thermodynamic interpretation is clearly not tenable.  Most
thermodynamic quantities are not defined in general spacetimes.
Moreover, the time direction imprinted on thermodynamic entropy
conflicts with the invariance of the covariant entropy bound under
reversal of time (Bousso, 1999a).} Since it involves no assumptions
about the microscopic properties of matter, it places a fundamental
limit on the number of degrees of freedom in nature.}

\item{The bound is not explained by other laws of physics that are
presently known.  Unlike its less general predecessors (e.g., the
spherical entropy bound, Sec.~\ref{sec-spheb}), the covariant bound
cannot be regarded merely as a consequence of black hole
thermodynamics.  Arguments involving the formation of black holes
cannot explain an entropy bound whose scope extends to the deep
interior of black holes and to cosmology.---We conclude that {\em the
bound is an imprint of a more fundamental theory.}}

\item{Yet, the covariant bound is closely related to the black hole
entropy and the generalized second law, long considered important
clues to quantum gravity.  Though the bound does not itself follow
from thermodynamics, it implies other bounds which have been argued to
be necessary for upholding the second law (Sec.~\ref{sec-rob}).  We
also note that the bound essentially involves the quantum states of
matter.---We conclude that {\em the fundamental theory responsible for
the bound unifies matter, gravity, and quantum mechanics.}}

\item{The bound relates information to a single geometric quantity
(area).  The bound's simplicity, in addition to its generality, makes
the case for its fundamental significance compelling.---We conclude
that {\em the area of any surface $B$ measures the information content
of an underlying theory describing all possible physics on the
light-sheets of $B$}.\footnote{An entropy bound in terms of a more
complex combination of physical quantities (e.g., Brustein and
Veneziano, 2000), even if it holds generally, would not seem to betray
a concrete relation of this kind.}}

\end{itemize}

\subsection{Formulation}

Let us combine the three conclusions drawn above (in italics) and
formulate the {\sl holographic principle} (Bousso, 1999a,b).

{\em The covariant entropy bound is a law of
physics which must be manifest in an underlying theory.  This theory
must be a unified quantum theory of matter and spacetime.  From it,
Lorentzian geometries and their matter content must emerge in such a
way that the number of independent quantum states describing the
light-sheets of any surface $B$ is manifestly bounded by the
exponential of the surface area:}
\begin{equation}
{\cal N}[L(B)] \leq e^{A(B)/4}.
\end{equation}
(See Secs.~\ref{sec-not}, \ref{sec-presc} for notation.)

Implicit in the phrase ``quantum states'' is the equivalence, in
quantum theory, of the logarithm of the dimension ${\cal N}$ of
Hilbert space and the amount of information stored in the quantum
system.  As it is not obvious that quantum mechanics will be primary
in a unified theory, a more neutral formulation of the holographic
principle may be preferable:

{\em $N$, the number of degrees of freedom (or the number of bits
times $\ln 2$) involved in the description of $L(B)$, must not exceed
$A(B)/4$.}

\subsection{Implications}
\label{sec-imp}

The holographic principle implies a radical reduction in the number of
degrees of freedom we use to describe nature.  It exposes quantum
field theory, which has degrees of freedom at every point in space, as
a highly redundant effective description, in which the true number of
degrees of freedom is obscured (Sec.~\ref{sec-qftwrong}).

The holographic principle challenges us to formulate a theory in which
the covariant entropy bound is manifest.  How can a {\em holographic
theory\/} be constructed?  Physics appears to be local to a good
approximation.  The number of degrees of freedom in any local theory
is extensive in the volume.  Yet, the holographic principle dictates
that the information content is in correspondence with the area of
surfaces.  How can this tension be resolved?  There appear to be two
main lines of approach, each casting the challenge in a different
form.

One type of approach aims to retain locality.  A local theory could be
rendered holographic if an explicit gauge invariance was identified,
leaving only as many physical degrees of freedom as dictated by the
covariant entropy bound.  The challenge, in this case, is to implement
such an enormous and rather peculiar gauge invariance.

For example, 't~Hooft (1999, 2000a, 2001a,b,c) is pursuing a local
approach in which quantum states arise as limit cycles of a classical
dissipative system (see also van de Bruck, 2000).  The emergence of an
area's worth of physical degrees of freedom has yet to be demonstrated
in such models.

A second type of approach regards locality as an emergent phenomenon
without fundamental significance.  In this case, the holographic
data are primary.  The challenge is not only to understand their
generation and evolution.  One must also explain how to translate
underlying data, in a suitable regime, into a classical spacetime
inhabited by local quantum fields.  In a successful construction, the
geometry must be shaped and the matter distributed so as to satisfy
the covariant entropy bound.  Because holographic data are most
naturally associated with the area of surfaces, a serious difficulty
arises in understanding how locality can emerge in this type of
approach.

The AdS/CFT correspondence (Sec.~\ref{sec-ads}) lends credence to the
second type of approach.  However, because it benefits from several
peculiarities of the asympotically AdS universes to which it applies
(Sec.~\ref{sec-screens}), it has offered little help to researchers
pursuing such approaches more broadly.

Some of the proposals and investigations discussed in Secs.~\ref{sec-toe}
and \ref{sec-ds} can be associated to the second type.

Which type of approach one prefers will depend, to a great extent, on
which difficulty one abhors more: the elimination of most degrees of
freedom, or the recovery of locality.  The dichotomy is hardly strict;
the two alternatives are not mutually exclusive.  A successful theory
may admit several equivalent formulations, thus reconciling both
points of view.

Since light-sheets are central to the formulation of the holographic
principle, one would expect null hypersurfaces to play a primary role
in the classical limit of an underlying holographic theory (though
this may not be apparent in descriptions of weakly time-dependent
geometries; see Sec.~\ref{sec-ads}).

\section{Holographic screens and holographic theories}
\label{sec-hsht}

We will begin this section by discussing which aspects of the
holographic principle have already been realized in string theory.  We
assess how general the class of universes is in which the holographic
principle is thus implemented.  In this context, we will present the
most explicit example of a holographic theory presently known.  The
AdS/CFT correspondence defines quantum gravity---albeit in a limited
set of spacetimes.  Anti-de Sitter space contains a kind of
holographic screen, a distant hypersurface on which holographic data
can be stored and evolved forward using a conformal field theory.

We will then review the construction of holographic screens in general
spacetimes, including those without boundary.  Using light-sheets, it
is always possible to find such screens.  However, a theory that
generally describes the generation and evolution of holographic data
remains elusive.  The structure of screens offers some clues about the
difficulties that must be addressed.  We will list a number of
approaches.

We will also discuss the application of the covariant entropy bound to
universes with positive vacuum energy.  In this class of spacetimes
the holographic principle appears to place a particularly strong
constraint on an underlying description.

\subsection{String theory and the holographic principle}
\label{sec-strings}

\subsubsection{A work in progress}

String theory naturally produces a unified quantum description of
gravity and matter fields.  Its framework has proven self-consistent
in remarkably non-trivial ways, given rise to powerful mathematical
structures, and solved numerous physical problems.  One might wonder
what the holographic principle is still needed for.  If a good theory
is available, why search further?  What is left to do?

String theory has developed in an unconventional way.  It began as a
formula whose physical interpretation in terms of strings was
understood only later.  The theory was first misunderstood as a
description of hadrons, and only later recognized as a quantum theory
of gravity.  It forms part of a rigid mathematical structure whose
content and physical implications continue to be explored.

String theory\footnote{We shall take related eleven-dimensional
theories to be included in this term.} has yet to address many of the
most pressing questions one would like to ask of a fundamental theory.
These include phenomenological issues: Why does the world have four
large dimensions?  What is the origin of the stardard model?  How is
supersymmetry broken?  More importantly, there are conceptual
difficulties.  It is unclear how the theory can be applied to
realistic cosmological spacetimes, and how it might describe most
black holes and singularities of general relativity.

String theory's most notable recent successes hinged on the discovery
of a new set of objects in the theory, D-branes (Polchinksi, 1995).
Before D-branes, string theory's list of open questions was longer
than it is today.  This serves as a reminder that unsolved problems
need not signal the failure of string theory.  Neither should they be
dismissed as mere technical difficulties.  Instead, they may indicate
that there are still crucial parts of the theory that have not been
discovered.

There is little evidence that string theory, in its current form,
represents more than a small portion, or a limiting case, of a bigger
theoretical structure.  Nor is it clear that the exploration of this
structure will continue to proceed most efficiently from
within.\footnote{In particular, Banks (2000b) has argued that there
may be no sense in which all isolated ``vacua'' of the theory can be
smoothly connected.}

An intriguing success of the covariant entropy bound is its validity
in highly dynamical geometries, whose description has proven
especially difficult in string theory.  This suggests that the
holographic principle may offer useful guidance to the further
development of the theory.

Its present limitations prevent string theory from explaining the
general validity of the covariant entropy bound.  The theory is not
under control in many situations of interest, for example when
supersymmetry is broken.  Moreover, many solutions of physical
relevance, including most of the examples in this text, do not appear
to be admitted by string theory in its current form.

\subsubsection{Is string theory holographic?}

These restrictions aside, one may ask whether the holographic
principle is manifest in string theory.  Let us consider, for a
moment, only spacetimes that string theory can describe, and in which
the holographic principle is also well-defined (i.e., geometry is
approximately classical).  Is the number of degrees of freedom
involved in the string theory description set by the area of surfaces?

In perturbative string theory, the holographic principle is only
partly realized.  Effects associated with holography include the
independence of the wave function on the longitudinal coordinate in
the light cone frame, and the growth of the size of states with their
momentum (see the reviews cited in Sec.~\ref{sec-further}; Thorn, {\em
op.\ cit.}; Klebanov and Susskind, 1988; Susskind, 1995b; see also
Susskind, 1995a).  

A number of authors have studied the extent to which string theory
exhibits the non-locality implied by the holographic principle (Lowe,
Susskind, and Uglum, 1994; Lowe {\em et al.}, 1995).  These
investigations are closely related to the problem of understanding of
the unitarity of black hole evaporation from the point of view of
string theory, in particular through the principle of black hole
complementarity (Sec.~\ref{sec-bhc}).

The entropy bound of one bit per Planck area, however, is not explicit
in perturbative string theory.  Susskind (1995b) showed that the
perturbative expansion breaks down before the bound is violated (see
also Banks and Susskind, 1996).  One would expect the holographic
principle to be fully manifest only in a non-perturbative formulation
of the theory.

Since the holographic principle was conceived, non-perturbative
definitions of string theory have indeed become available for two
special classes of spacetimes.  Remarkably, in the AdS/CFT
correspondence, the number of degrees of freedom agrees manifestly
with the holographic principle, as we discuss below.  In Matrix theory
(Banks {\em et al.}, 1997) the corresponding arguments are somewhat
less precise.  This is discussed, e.g., by Bigatti and Susskind
(1997), and by Banks (1998, 1999), where further references can be
found.\footnote{A significant non-perturbative result closely related
to the holographic principle is the microscopic derivation of the
entropy of certain black holes in string theory (Strominger and Vafa,
1996).}

The holographic principle may not only aid the search for other
non-perturbative definitions of string theory.  It could also
contribute to a background-independent formulation that would
illuminate the conceptual foundation of string theory.

\subsection{AdS/CFT correspondence}
\label{sec-ads}

An example of the AdS/CFT correspondence concerns type IIB string
theory in an asymptotically AdS$_5\times \mathbf{S}^5$ spacetime (the
{\em bulk}), with $n$ units of five-form flux on the
five-sphere\footnote{There is a notational conflict with most of the
literature, where $N$ denotes the size of the gauge group.  In this
review, $N$ is reserved for the number of degrees of freedom
(Sec.~\ref{sec-ndof}).} (Maldacena, 1998; Gubser, Klebanov, and
Polyakov, 1998; Witten, 1998).  This theory, which includes gravity,
is claimed to be non-perturbatively defined by a particular conformal
field theory without gravity, namely 3+1 dimensional supersymmetric
Yang-Mills theory with gauge group $U(n)$ and 16 real supercharges.
We will refer to this theory as the {\em dual CFT}.

The metric of AdS$_5\times \mathbf{S}^5$ is
\begin{equation}
ds^2 = R^2 \left[ - \frac{1+r^2}{1-r^2} dt^2 + \frac{4}{(1-r^2)^2}
\left( dr^2 + r^2 d\Omega_3^2 \right) + d\Omega_5^2 \right],
\label{eq-ads-metric}
\end{equation}
where $d\Omega_d$ denotes the metric of a $d$-dimensional unit sphere.
The radius of curvature is related to the flux by the formula
\begin{equation}
R = n^{1/4},
\label{eq-rflux}
\end{equation}
in units of the ten-dimensional Planck length.

The proper area of the three-spheres diverges as $r \rightarrow 1$.
After conformal rescaling (Hawking and Ellis, 1973), the spacelike
hypersurface, $t=\mbox{const}, 0 \leq r < 1$ is an open ball, times a
five-sphere.  (The conformal picture for AdS space thus resembles the
worldvolume occupied by a spherical system, as in Fig.~\ref{fig-ball}.)
Because the five-sphere factor has constant physical radius, and the
scale factor vanishes as $r\to 1$, the five-sphere is scaled to a
point in this limit.  Thus, the conformal boundary of space is a
three-sphere residing at $r=1$.

It follows that the conformal boundary of the spacetime is ${\mathbb
R} \times \mathbf{S}^3$.  This agrees with the dimension of the CFT.
Hence, it is often said that the dual CFT ``lives'' on the boundary of
AdS space.

The idea that data given on the boundary of space completely describe
all physics in the interior is suggestive of the holographic
principle.  It would appear that the dual CFT achieves what local
field theory in the interior could not do.  It contains an area's
worth of degrees of freedom, avoiding the redundancy of a local
description.  However, to check quantitatively whether the holographic
bound really manifests itself in the dual CFT, one must compute the
CFT's number of degrees of freedom, $N$.  This must not exceed the
boundary area, $A$, in ten-dimensional Planck units.

The proper area of the boundary is divergent.  The number of degrees
of freedom of a conformal field theory on a sphere is also divergent,
since there are modes at arbitrarily small scales.  In order to make a
sensible comparison, Susskind and Witten (1998) regularized the bulk
spacetime by removing the region $1-\delta<r<1$, where $\delta\ll 1$.
This corresponds to an infrared cutoff.  The idea is that a modified
version of the AdS/CFT correspondence still holds for this truncated
spacetime.  

The area of the $\mathbf{S}^3 \times \mathbf{S}^5$ boundary surface%
\footnote{Unlike Susskind and Witten (1998), we do not compactify the
bulk to five dimensions in this discussion; all quantities refer to a
ten-dimensional bulk.  Hence the area is eight-dimensional.}
is approximately given by
\begin{equation}
A\approx {R^8\over\delta^3}
\end{equation}
In order to find the number of degrees of freedom of the dual CFT, one
has to understand how the truncation of the bulk modifies the CFT.
For this purpose, Susskind and Witten (1998) identified and exploited
a peculiar property of the AdS/CFT correspondence: infrared effects in
the bulk correspond to ultraviolet effects on the boundary.  

There are many detailed arguments supporting this so-called {\em UV/IR
relation\/} (see also, e.g., Balasubramanian and Kraus, 1999; Peet and
Polchinski, 1999).  Here we give just one example.  A string stretched
across the bulk is represented by a point charge in the dual CFT.  The
energy of the string is linearly divergent near the boundary.  In the
dual CFT this is reflected in the divergent self-energy of a point
charge.  The bulk divergence is regularized by an infrared cut-off,
which renders the string length finite, with energy proportional to
$\delta^{-1}$.  In the dual CFT, the same finite result for the
self-energy is achieved by an ultraviolet cutoff at the short distance
$\delta$.

We have scaled the radius of the three-dimensional conformal sphere to
unity.  A short distance cut-off $\delta$ thus partitions the sphere
into $\delta^{-3}$ cells.  For each quantum field, one may expect to
store a single bit of information per cell.  A $U(n)$ gauge theory
comprises roughly $n^2$ independent quantum fields, so the total
number of degrees of freedom is given by
\begin{equation} 
N \approx \frac{n^2}{\delta^3}.
\end{equation}
Using Eq.~(\ref{eq-rflux}) we find that the CFT number of degrees of
freedom saturates the holographic bound,
\begin{equation}
N \approx A,
\end{equation}
where we must keep in mind that this estimate is only valid to within
factors of order unity.

Thus, the number of CFT degrees of freedom agrees with the number of
physical degrees of freedom contained on any light-sheet of the
boundary surface $\mathbf{S}^3 \times \mathbf{S}^5$.  One must also
verify that there is a light-sheet that contains all of the entropy in
the spacetime.  If all light-sheets terminated before reaching $r=0$,
this would leave the possibility that there is additional information
in the center of the universe which is not encoded by the CFT.  In
that case, the CFT would not provide a complete description of the
full bulk geometry---which is, after all, the claim of the AdS/CFT
correspondence.

The boundary surface is normal (Bousso, 1999b), so that both past and
future ingoing light-sheets exist.  In an asymptotically AdS$_5\times
\mathbf{S}^5$ spacetime without past or future singularities, either
of these light-sheets will be complete.  Thus one may expect the CFT
to describe the entire spacetime.%
\footnote{If there are black holes in the spacetime, then the
future directed light-sheet may cross the black hole horizon and end
on the future singularity.  Then the light-sheet may miss part of the
interior of the black hole.  One can still argue that the CFT
completely describes all physics accessible to an observer at
infinity.  A light-sheet can be terminated at the black hole horizon,
with the horizon area added to its entropy content.  The data on a
horizon, in turn, are complementary to the information in the black
hole interior (Susskind, Thorlacius, and Uglum, 1993).}

Thus, the CFT state on the boundary (at one instant of time) contains
holographic data for a complete slice of the spacetime.  The full
boundary of the spacetime includes a time dimension and is given by
$\mathbb R\times \mathbf{S}^3\times \mathbf{S}^5$.  Each moment of
time defines an $\mathbf{S}^3\times \mathbf{S}^5$ boundary area, and
each such area admits a complete future directed light-sheet.  The
resulting sequence of light-sheets foliate the spacetime into a stack
of light cones (each of which looks like the cone in
Fig.~\ref{fig-ball}c).  There is a slice-by-slice holographic
correspondence between bulk physics and dual CFT data.  By the
spacelike projection theorem (Sec.~\ref{sec-spt}), the same
correspondence holds for the spacelike slicing shown in
Fig.~\ref{fig-ball}a.

Thus, a spacelike formulation of the holographic principle is mostly
adequate in AdS.  In recent years there has been great interest in
models of our universe in which four-dimensional gauge fields are
holographic duals to the physics of an extra spatial dimension (see,
e.g., Randall and Sundrum, 1999a,b)---a kind of ``inverse
holography''.  Such models can be realized by introducing codimension
one objects, {\em branes}, into a five-dimensional bulk spacetime.
If the bulk is Anti-de Sitter space, the holographic correspondence is
expected to be a version of the AdS/CFT correspondence.  In a very
general class of models (Karch and Randall, 2001), the brane fields
are dual only to a portion of the bulk.  Attempts to apply the
spacelike holographic principle lead to contradictions in this case,
and the use of the light-sheet formulation is essential (Bousso and
Randall, 2001).

To summarize, the AdS/CFT correspondence exhibits the following
features:
\begin{itemize} 
\item{There exists a slicing of the spacetime such that the state of
the bulk on each slice is fully described by data not exceeding $A$
bits, where $A$ is the area of the boundary of the slice.}
\item{There exists a theory without redundant degrees of freedom, the
CFT, which generates the unitary evolution of boundary data from slice
to slice.}
\end{itemize}

Perhaps due to the intense focus on the AdS/CFT correspondence in
recent years, the holographic principle has come to be widely regarded
as synonymous with these two properties.  Their partial failure to
generalize to other spacetimes has sometimes been confused with a
failure of the holographic principle.  We emphasize, therefore, that
neither property is sufficient or necessary for the holographic
principle, as defined in Sec.~\ref{sec-hp}.

Assuming the validity of the covariant entropy bound in arbitrary
spacetimes, Bousso (1999b) showed that a close analogue of the first
property always holds.  The second, however, is not straightforwardly
generalized.  It should not be regarded as a universal consequence of
the holographic principle, but as a peculiarity of Anti-de Sitter
space.

\subsection{Holographic screens for general spacetimes}
\label{sec-screens}

\subsubsection{Construction}

Any spacetime, including closed universes, contains a kind of
holographic boundary, or screen.  It is most easily obtained by
slicing the spacetime into light cones.  The total entropy on each
light cone can be holographically stored on the largest surface
embedded in the cone.  Our construction follows Bousso (1999b); see
also Bigatti and Susskind (2000), Bousso (2000a).

Consider the past light cone, ${\cal L}^-$ (technically, the boundary
of the past), of a point $p$ in any spacetime satisfying the null
energy condition.  The following considerations will show that ${\cal
L}^-$ consists of one or two light-sheets.

The area spanned by the light rays will initially increase with affine
parameter distance $\lambda$ from $p$.  In some cases, for example
AdS, the area keeps increasing indefinitely.  For any surface
$B(\lambda_1)$ the holographic principle implies that the total number
of degrees of freedom on the portion $0\leq\lambda\leq\lambda_1$ is
bounded by $A(B(\lambda_1))/4$.  One can express this by saying that
$B(\lambda_1)$ is a {\em holographic screen}, a surface on which the
information describing all physics on the enclosed light cone portion
can be encoded at less than one bit per Planck area.  If the light
cone is extended indefinitely, it will reach the conformal boundary of
spacetime, where its area diverges.  In this limit one obtains a
holographic screen for the entire light cone.

A second possibility is that the area does not increase forever with
the affine parameter.  Instead, it may reach a maximum, after which it
starts to contract.  The focussing theorem (Sec.~\ref{sec-ray})
implies that contracting light rays will eventually reach caustics or
a singularity of the spacetime.  Let us continue the light cone until
such points are reached.

Let $B$ be the apparent horizon, i.e., the spatial surface with
maximum area on the light cone.  $B$ divides the light cone into two
portions.  By construction, the expansion of light rays in both
directions away from $B$ vanishes locally and is non-positive
everywhere.  (We will not be concerned with the second pair of null
directions, which does not coincide with the light cone.)  Hence, both
portions are light-sheets of $B$.  It follows that the total number of
degrees of freedom on the light cone is bounded by the area of its
largest spatial surface:
\begin{equation}
N \leq {A(B)\over 2}.
\end{equation}
The denominator is 2 because the holographic bound ($A/4$) applies
separately to each light-sheet, and the light cone consists of two
light-sheets.

Consider, for example, a universe that starts with an initial
singularity, a big bang.  Following light rays backwards in time, our
past light cone grows at first.  Eventually, however, it must shrink,
because all areas vanish as the big bang is approached.

One can summarize both cases by the statement that a holographic
screen for all the data on a light cone is the surface where its
spatial area is largest.  A global holographic screen for the entire
spacetime can now be constructed as follows.

One picks a worldline $P(t)$ and finds the past light cone ${\cal
L}^-(t)$ of each point.  The resulting stack of light cones foliates
the spacetime.%
\footnote{A few remarks are in order.  1.~A foliation can also be
obtained from future light cones, or from more general null
hypersurfaces.  2.~ Depending on global structure, the past light
cones may foliate only the portion of the spacetime visible to the
observer.  Suitable extensions permit a global foliation by other null
hypersurfaces.  3.~If light rays generating the past light cone of $p$
intersect, they leave the boundary of the past of $p$ and become
timelike separated from $p$.  To obtain a good foliation, one should
terminate such light rays even if they intersect with non-neighboring
light rays, as suggested by Tavakol and Ellis (1999).  This can only
shorten the light-sheet and will not affect our conclusions.}
Each cone has a surface of maximal area, $B(t)$.  These surfaces form
a hypersurface in the spacetime or on its boundary.  Cone by cone, the
information in the spacetime bulk can be represented by no more than
$A(t)$ bits on the screen, where $A(t)$ is the area of $B(t)$.

In suitably symmetric spacetimes, the construction of holographic
screens is simplified by a Penrose diagram.  The spacetime must first
be divided into ``wedge domains'', as shown in Fig.~\ref{fig-clos}a
for a closed universe.  (A light cone foliation corresponds to a set
of parallel lines at 45 degrees to the vertical.  The remaining
ambiguity corresponds to the choice of past or future light cones.)
In order to get to a holographic screen, one has to follow each line
in the direction of the tip of the wedge.  Either one ends up at a
boundary, or at an apparent horizon, where the wedge flips.

The example shown in Fig.~\ref{fig-clos}b is remarkable because it
demonstrates that holographic screens can be constructed for closed
universes.  Thus, an explanation of the origin of the holographic
principle should not ultimately hinge upon the presence of a boundary
of spacetime, as it does in the AdS/CFT correspondence.

Using the general method given above, global holographic screens have
been constructed explicitly for various other spacetimes (Bousso,
1999b), including Minkowski space, de~Sitter space, and various FRW
universes.  In many cases, they do form a part of the boundary of
spacetime, for example in asymptotically AdS, Minkowski, and
de~Sitter spacetimes.%
\footnote{Some subtleties arise in the de~Sitter case which allow,
alternatively, the use of a finite area apparent horizon as a screen
(Bousso 1999b).  See also Sec.~\ref{sec-ds}}
For several examples, Penrose diagrams with wedges and screens are
found in Bousso (1999b) and Bigatti and Susskind (2000).

\subsubsection{Properties and implications}

Some of the properties of the boundary of AdS, such as its area and
its behavior under conformal transformations, can be used to infer
features of the dual CFT.  Properties of global holographic screens
can similarly provide clues about holographic theories underlying
other classes of spacetimes (Bousso, 1999b).

In AdS, the global holographic screen is unique.  It is the direct
product of a spatial sphere at infinity with the real time axis.  If
the sphere is regulated, as in Sec.~\ref{sec-ads} above, its area can
be taken to be constant in time.  None of these properties are
necessarily shared by the global screens of other spacetimes.  Let us
identify some key differences and discuss possible implications.

\begin{itemize} 

\item{In general, global holographic screens are highly non-unique.  For
example, observers following different worldlines correspond to
different stacks of light cones; their screens do not usually agree.}

\item{One finds that spacetimes with horizons can have disconnected
screen-hypersurfaces.  This occurs, for example, in the collapse of a
star to form a black hole (Bousso, 1999b).  Consider light cones
centered at $r=0$.  The past light cones are all maximal on ${\cal
I}^-$.  The future light cones are maximal on ${\cal I}^+$ only if
they start outside the event horizon.  Future light cones from points
inside the black hole are maximal on an apparent horizon in the black
hole interior.  Thus, there is one screen in the past, but two
disconnected screens in the future.

These two features may be related to black hole complementarity
(Sec.~\ref{sec-limitations}), which suggests that the choice of an
observer (i.e., a causally connected region) is a kind of gauge choice
in quantum gravity.  Related questions have recently been raised in
the context of de~Sitter space, where black hole complementarity
suggests a restriction to one causal region (Sec.~\ref{sec-ds}).  They
also play a central role in the framework for a holographic theory of
cosmology pursued by Banks (2000c) and Banks and Fischler (2001a,b).}

\item{The area of the maximal surface generically varies from cone to
cone: $A(t) \neq \mbox{const}$.  For example, the area of the apparent
horizon in a flat FRW universe vanishes at the big bang and increases
monotonically, diverging for late-time cones (Fig.~\ref{fig-flatfrw}).
In a closed FRW universe, the area of the apparent horizon increases
while the universe expands and decreases during the collapsing phase
(Fig.~\ref{fig-clos}b).

This behavior poses a challenge, because it would seem that the number
of degrees of freedom of a holographic theory can vary with
time.\footnote{Strominger (2001b) has recently suggested that the
growth of a screen might be understood as inverse RG flow in a dual
field theory.}  The shrinking of a screen raises concerns about a
conflict with the second law (Kaloper and Linde, 1999).  However, the
following observation suggests that the parameter $t$ should not be
uncritically given a temporal interpretation on a screen
hypersurface.}

\item{The maximal surfaces do not necessarily form timelike
(i.e. Lorentzian signature) hypersurfaces.  In de Sitter space, for
example, the global screens are the two conformal spheres at past and
future infinite time.  Both of these screens have Euclidean
signature.\footnote{This does not mean that holography reduces to
ordinary Cauchy evolution.  Holographic encoding does not make use of
equations of motion.  There is always a projection, slice by slice, of
holographic data onto the screen.  Moreover, the limit of 1 bit per
Planck area, central to holography, plays no role in Cauchy
evolution.}  The same is true for the apparent horizons in spacetimes
with a $w>1/3$ equation of state (Fig.~\ref{fig-flatfrw}).}

\item{Screens can be located in the spacetime interior.  Screens near
the boundary have the advantage that metric perturbations and quantum
fluctuations fall off in a controlled way.  The common large distance
structure of different asymptotically Anti-de Sitter spacetimes, for
example, makes it possible to describe a whole class of universes as
different states in the same theory.  

The shape of interior screens, on the other hand, is affected by small
variations of the spacetime.  The apparent horizon in cosmological
solutions, for example, will depend on the details of the matter
distribution.  Thus it is not clear how to group cosmological
spacetimes into related classes (see, however, Sec.~\ref{sec-ds}).}

\end{itemize} 

The AdS/CFT correspondence realizes the holographic principle
explicitly in a quantum gravity theory.  The points just mentioned
show that, intricate though it may be, this success benefits from
serendipidous simplifications.  In more general spacetimes, it remains
unclear how the holographic principle can be made manifest through a
theory with explicitly holographic degrees of freedom.  In particular,
one can argue that the screen should not be presumed; all information
about the geometry should come out of the theory itself.

Nevertheless, the existence of global holographic screens in general
spacetimes is an encouraging result.  It demonstrates that there is
always a way of projecting holographic data, and it provides novel
structures.  The understanding of their significance remains an
important challenge.

\subsection{Towards a holographic theory}
\label{sec-toe}

We have convinced ourselves of a universal relation between areas,
light-sheets, and information.  The holographic principle instructs us
to embed this relation in a suitable quantum theory of gravity.  It
suggests that null hypersurfaces, and possibly global screens, will be
given a special role in the regime where classical geometry emerges.
How far have we come in this endeavor?

The extent to which holography is explicit in string theory and
related frameworks has been discussed in Secs.~\ref{sec-strings} and
\ref{sec-ads}.  We have also mentioned the local approach being
developed by 't~Hooft (Sec.~\ref{sec-imp}).

An effectively lower-dimensional description is evident in the quantum
gravity of $2+1$ dimensional spacetimes (Witten, 1988; see also van
Nieuwenhuizen, 1985; Achucarro and Townsend, 1986; Brown and Henneaux,
1986.  See Carlip, 1995, for a review of $2+1$ gravity).  As in the
light cone formulation of string theory, however, the entropy bound is
not manifest.  Ho\v{r}ava (1999) has proposed a Chern-Simons
formulation of (eleven-dimensional) M-theory, arguing that the
holographic entropy bound is thus implemented.  Light-like directions
do not appear to play a special role in present Chern-Simons
approaches.

The importance of null hypersurfaces in holography resonates with the
twistor approach to quantum gravity (see the review by Penrose and
MacCallum, 1972), but this connection has not yet been substantiated.
Jacobson (1995) has investigated how Einstein's equation can be
recovered from the geometric entropy of local Rindler horizons.
Markopoulou and Smolin (1998, 1999) have proposed to construct a
manifestly holographic quantum theory of gravity based on the
formalism of spin networks.  Smolin (2001) discusses related
approaches to an implementation of the holographic principle and
provides further references.

Banks (2000c) and Banks and Fischler (2001a,b) have sketched a
preliminary framework for holographic theories of cosmological
spacetimes.  After discretizing time, one considers a network of
screens obtained from a discrete family of observers. In other words,
one constructs the past light cones of a discrete set of points spread
throughout the spacetime.  The maximal area on each light cone
determines the dimension of a Hilbert space describing the enclosed
portion of the spacetime.  light cone intersections and inclusion
relations give rise to a complicated network of Hilbert spaces, whose
dimensions encode geometric information.  A theory is sought which
will give rise to spacetime geometry by inverting these steps.  The
rules for the generation of Hilbert space networks, and the
construction of a suitable time evolution operator, are not yet
understood.

Banks and Fischler (2001a) have also argued that considerations of
entropy determine the inital state of a big bang universe.  By
Eqs.~(\ref{eq-soaflat}) and (\ref{eq-frwcond}), maximally stiff
matter, with equation of state $p=\rho$, has marginal properties under
the holographic principle.  This motivates a model based on the
initial domination of a $p=\rho$ fluid, from which Banks and Fischler
are aiming to obtain new perspectives on a number of standard
cosmological problems.

It has recently been noticed (Banks, 2000a; Fischler, 2000a,b) that
the holographic principle has particularly strong implications in
certain universes with a positive cosmological constant.  As we
discuss next, this could be of help in characterizing a holographic
theory for a class of spacetimes that may include our universe.

\subsection{Holography in de~Sitter space}
\label{sec-ds}

Generally the holographic principle restricts the number of degrees of
freedom, $N$, only relative to some specified surface.  There are
spacetimes, however, where the holographic principle implies an
absolute upper limit on $N$.  This follows in particular if it is
possible to find a global holographic screen whose area never exceeds
$N$.  Physically, there is not ``enough room'' in such universes to
generate entropy greater than $N$.  In particular, they cannot
accommodate black holes with area greater than of order $N$.

An absolute entropy bound could be viewed as a hint about
characteristics of the quantum description of a whole class of
spacetimes.  The most radical conclusion would be to look for theories
that come with only $e^N$ of states (Banks, 2000a; Bousso, 2000b;
Fischler, 2000a,b; Dyson, Lindesay, and Susskind, 2002).\footnote{For
a speculation on the origin of the number $N$, see Mena Marugan and
Carneiro (2001).} This is quite unusual; even the Hilbert space of a
single harmonic oscillator contains infinitely many states.

If a continuous deformation of Cauchy data can take a universe with
maximal entropy $N$ to one with $N'\neq N$, it is hard to argue that
they should be described by two entirely different theories.  Hence,
this approach will be compelling only if physical criteria can be
found which characterize a class of spacetimes with finite $N$,
independently of initial data.

As we discuss below, a suitable class may be the universes that become
similar to de~Sitter space asymptotically in the future.  However, we
will not find this criterion entirely satisfactory.  We will comment
on its problems and possible generalizations.

\subsubsection{de~Sitter space}

The maximally symmetric spacetime with positive curvature is de~Sitter
space.  It is a solution to Einstein's equation with a positive
cosmological constant, $\Lambda$, and no other matter.  Using $w=-1$
in Eqs.~(\ref{eq-FRW2}), (\ref{eq-a}), and (\ref{eq-q}), the metric
can be written as a closed FRW universe,
\begin{equation}
ds^2 = \frac{a_0^2}{\sin^2 \eta}\left( -d\eta^2+ d\chi^2 +
\sin^2\chi\, d\Omega_{D-2}^2\right).
\label{eq-dsmetric}
\end{equation}
The curvature radius is related to the cosmological constant by
\begin{equation}
a_0^2 = {(D-1)(D-2)\over 2\Lambda}.
\label{eq-alambda}
\end{equation}
For simplicity, we will take $D=4$ unless stated otherwise.

\begin{figure}[h] \centering
\includegraphics[width=7cm]{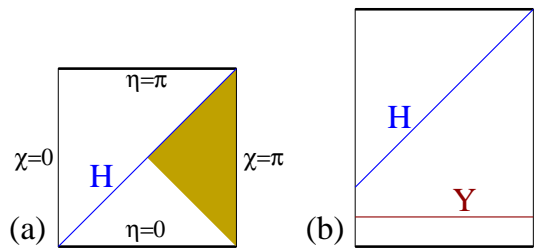}
\caption{(a) Penrose diagram for empty de~Sitter space.  $H$ is the
future event horizon of an observer on the south pole ($\chi=\pi$).
The shaded region is the ``southern diamond''.  (b) Penrose diagram
for a generic solution that asymptotes to de~Sitter in the past and
future (dS$^\pm$).  The future event horizon has complete time slices
in its past, such as $Y$.}
\label{fig-ds}
\end{figure}

The spatial three-spheres contract from infinite size to size $a_0$
($0<\eta\leq \pi/2$), then re-expand ($\pi/2\leq\eta<\pi$).  The
Penrose diagram is a square, with spacelike conformal boundaries at
$\eta=0$, $\pi$.  A light ray emitted on the north pole ($\chi=0$) at
early times ($\eta\ll 1$) barely fails to reach the south pole
($\chi=\pi$) in the infinite future (Fig.~\ref{fig-ds}a).  

The light rays at $\eta=\chi$ reach neither the north nor the south
pole in finite affine time.  They generate a null hypersurface $H$, of
constant cross-sectional area.  (All spatial sections of $H$ are
spheres of radius $a_0$.)  $H$ is the future event horizon of an
observer at the south pole.  It bounds the region from which signals
can reach the observer.  There is a past event horizon
($\eta=\pi-\chi$) which bounds the region to which the southern
observer can send a signal.

The intersection of both regions forms the ``southern diamond'', the
region that can be probed by the observer.  It is covered by a static
coordinate system:\footnote{The coordinates $r$ and $t$ defined here
differ from those defined at the beginning of Sec.~\ref{sec-cosmo}.}
\begin{equation}
ds^2 = a_0^2 \left[-(1-r^2) dt^2 + {dr^2\over 1-r^2}
+ r^2 d\Omega^2 \right].
\end{equation}

Note that the location of event horizons in de~Sitter space depends on
a choice of observer ($r=0$).  Despite this difference, Gibbons and
Hawking (1977) showed that the future event horizon of de~Sitter space
shares many properties with the event horizons of black holes.
Classically, objects that fall across the event horizon cannot be
recovered.  This would seem to endanger the second law of
thermodynamics, in the sense discussed in Sec.~\ref{sec-gsl}.

Mirroring the reasoning of Sec.~\ref{sec-gsl}, one concludes that the
horizon must be assigned a semi-classical Bekenstein-Hawking entropy
equal to a quarter of its area,
\begin{equation}
S_{\rm dS} = \pi a_0^2 = \frac{3\pi}{\Lambda}.
\label{eq-dsvac}
\end{equation}
Gibbons and Hawking (1977) showed that an observer in de~Sitter space
will detect thermal radiation coming from the horizon, at a
temperature $T=1/2\pi a_0$.\footnote{See the end of
Sec.~\ref{sec-unruh} for references to Bekenstein and Unruh-Wald
bounds arising in de~Sitter space.}

In pure de~Sitter space, there is no matter entropy, so the total
entropy is given by Eq.~(\ref{eq-dsvac}).

\subsubsection{dS$^\pm$ spacetimes}

So far we have discussed empty de Sitter space.  Generally one is
interested in describing a larger class of spacetimes, which might be
characterized by asymptotic conditions.  Let us consider spacetimes
that approach de~Sitter space asymptotically both in the past and in
the future.  We denote this class by dS$^\pm$.  Its quantum
description has recently attracted much attention (e.g.,
Balasubramanian, Ho\v{r}ava, and Minic, 2001; Strominger, 2001a;
Witten, 2001; for extensive lists of references, see, e.g.,
Balasubramanian, de Boer, and Minic, 2001; Spradlin and Volovich,
2001).  Implications of the holographic principle in other
accelerating universes have been considered by Hellerman, Kaloper, and
Susskind (2001) and Fischler {\em et al.} (2001); see also Banks and
Dine (2001), Carneiro da Cunha (2002).

If de~Sitter space is not completely empty, the Penrose diagram will
be deformed.  In the asymptotic regions matter is diluted, but in the
interior of the spacetime it can have significant density.  Gao and
Wald (2000) showed under generic assumptions%
\footnote{Among other technical requirements, 
the spacetime must be geodesically complete and non-empty.  Strictly,
the presence of both asymptotic regions is not sufficient to guarantee
geodesic completeness, because black holes can form.  One would not
expect the geodesic incompleteness due to black hole singularities to
invalidate the above conclusions, however.}
that the back-reaction of matter makes the height of the diagram
greater than its width.  Then the future event horizon will cross the
entire space and converge in the north (Fig.~\ref{fig-ds}b).  Because
the spacetime approaches empty de~Sitter space in the future, the
horizon will asymptote to a surface $B$, a sphere of radius $a_0$
surrounding the south pole.  There will be no matter inside this
sphere at late times.  All matter will have passed through the future
event horizon.

The future event horizon can be regarded as a light-sheet of the
surface $B$.  This implies that the entropy of all matter on any
earlier Cauchy slices cannot exceed a quarter of the area $A(B) = 4\pi
a_0^2$.  With Eq.~(\ref{eq-alambda}) we find that
\begin{equation}
S_{\rm global} \leq \frac{3\pi}{\Lambda}.
\end{equation}
In particular, this holds for the total entropy in the asymptotic
past.  

We will not be concerned with the unobservable future region behind
the event horizon.  We conclude that {\em in a dS$^\pm$ spacetime, the
global entropy cannot exceed ($3\pi$ times) the inverse cosmological
constant.}  

This may seem a surprising result, since the initial equal-time slices
can be taken arbitrarily large, and an arbitrary amount of entropy can
be placed on them.  However, if the matter density becomes larger than
the energy density of the cosmological constant during the collapsing
phase, the universe will collapse to a big crunch.  Then there will be
no future infinity, in contradiction to our assumption.

\subsubsection{dS$^+$ spacetimes}

An even larger class of spacetimes is characterized by the condition
that they approach de~Sitter space in the asymptotic future.  No
restrictions are made on the behavior in the past.  This class will be
labelled dS$^+$.  In addition to all of the dS$^\pm$ universes, it
includes, for example, flat FRW universes that start with a big bang
singularity and are dominated by matter or radiation for some time.
At late times, all matter is diluted, only a cosmological constant
remains, and the metric approaches that of empty de~Sitter space.

Recent astronomical data (Riess {\em et al.}, 1998; Perlmutter {\em et
al.}, 1999) favor a non-zero value of $\Lambda\sim 10^{-120}$.  If
this really corresponds to a fixed cosmological constant, our own
universe is in the dS$^+$ class.  This makes the study of
de~Sitter-like spacetimes, in particular the dS$^+$ class of
universes, especially significant.

The global entropy at early times is unbounded in this class.  In the
dS$^\pm$ class, constraints arise, roughly speaking, because all
matter has to ``fit'' through a throat three-sphere at $\eta=\pi/2$.
In the dS$^+$ class, there is no need for a contracting phase.  The
universe can be everywhere expanding, with non-compact spacelike
hypersurfaces of infinite total entropy.

However, an observer's vision is cloaked by the de~Sitter event
horizon that forms at late times.  Let us ask only how much entropy
can be detected by any single observer (Banks, 2000a; Fischler,
2000a,b).  This is easy to answer because the final entropy is known.
At late times, there is no matter and only a de~Sitter event horizon,
so the total entropy will be given by Eq.~(\ref{eq-dsvac}).  By the
generalized second law of thermodynamics, the entropy at all other
times will be less or equal.

It follows that {\em in a dS$^+$ spacetime, the entropy available to
any observer cannot exceed ($3\pi$ times) the inverse cosmological
constant.}  The restriction to a single observer is natural in view of
black hole complementarity (Sec.~\ref{sec-bhc}).

\subsubsection{Other universes with positive $\Lambda$}

Although they comprise a broad class, it still somewhat unnatural to
restrict one's attention to dS$^+$ universes.  Because of exposure
to thermal radiation, an observer in de~Sitter space cannot last
forever.  It is as unphysical to talk about arbitrarily long times as
it is to compare the observations of causally disconnected observers.

Moreover, fluctuations in the Gibbons-Hawking radiation cause black
holes to form.  If they are too big, they can cause a big crunch---a
collapse of the entire spacetime.  But even the persistent production
of ordinary black holes means that any observer who is not otherwise
thermalized will fall into a black hole.  In short, quantum effects
will prevent any observer from reaching ${\cal I}^+$.

So how can spacetimes with an absolute entropy bound be usefully
characterized?  With assumptions involving spherical symmetry, the
covariant entropy bound implies that that the observable entropy in
any universe with $\Lambda>0$ is bounded by $3\pi/\Lambda$ (Bousso,
2000b).  In addition to all dS$^+$ spacetimes, this class includes,
for example, closed recollapsing FRW universes in which the
cosmological constant is subdominant at all times.  This result relies
on the ``causal diamond'' definition of an observable region.  It
would seem to suggest that $\Lambda>0$ may be a sufficient condition
for the absolute entropy bound, $S\leq 3\pi/\Lambda$.

At least in $D>4$, however, one can construct product manifolds with
fluxes, which admit entropy greater than that of $D$-dimensional
de~Sitter space with the same cosmological constant (Bousso, DeWolfe,
and Myers, 2002).  A fully satisfactory classification of spacetimes
with finite entropy remains an outstanding problem.

\section*{Acknowledgments}

I would like to thank S.~Adler, M.~Aganagic, T.~Banks, J.~Bekenstein,
W.~Fischler, E.~Flanagan, S.~Hughes, T.~Jacobson, D.~Marolf, R.~Myers,
A.~Peet, J.~Polchinski, M.~Srednicki, and L.~Susskind for helpful
comments on drafts of this text.  This work was supported in part by
the National Science Foundation under Grant No.\ PHY99-07949.

\appendix \section{General relativity}

In this Appendix, we summarize most of the geometric terminology that
pervades this paper.  No attempts at completeness and precision are
made; in particular, we will ignore issues of smoothness.  The
textbooks of Hawking and Ellis (1973), Misner, Thorne, and Wheeler
(1973), and Wald (1984) may be consulted for a more thorough
discussion of this material.

\paragraph{Metric, examples, and Einstein's equation}

General relativity describes the world as a classical spacetime ${\cal
M}$ with $D-1$ spatial dimensions and one time dimension.
Mathematically, ${\cal M}$ is a manifold whose shape is described by a
metric $g_{ab}$ of Lorentzian signature $(-,+,\ldots,+)$.  In a
coordinate system $(x^0,\ldots,x^{D-1})$, the invariant distance $ds$
between infinitesimally neighboring points is given by
\begin{equation}
ds^2 = g_{ab}(x^0,\ldots,x^{D-1}) dx^a dx^b.
\end{equation}
Summation over like indices is always implied.

For example, the flat spacetime of special relativity (Minkowski
space) in $D=4$ has the metric
\begin{eqnarray} 
ds^2 & = & -dt^2 + dx^2 + dy^2 + dz^2\\
     & = & -dt^2 + dr^2 + r^2 d\Omega^2
\end{eqnarray} 
in Cartesian or spherical coordinates, respectively.  A Schwarzschild
black hole of mass $M$ is described by the metric
\begin{equation}
ds^2 = -\left(1-{2M\over r}\right) dt^2 + \left(1-{2M\over
r}\right)^{-1} dr^2 + r^2 d\Omega^2.
\end{equation}
The black hole horizon, $r=2M$, is a regular hypersurface, though this
is not explicit in these coordinates.  There is a singularity at
$r=0$.

Einstein's equation,
\begin{equation} 
G_{ab} = 8\pi T_{ab},
\end{equation}
relates the shape of space to its matter content.  The Einstein
tensor, $G_{ab}$, is a nonlinear construct involving the metric and
its first and second partial derivatives.  The stress tensor,
$T_{ab}$, is discussed further below.

\paragraph{Timelike, spacelike, and null curves}

A curve is a map from (a portion of) $\mathbb R$ into ${\cal M}$.  In
a coordinate system it is defined by a set of functions
$x^a(\lambda)$, $\lambda \in \mathbb R$.  At each point the curve has
a tangent vector, ${dx^a\over d\lambda}$.

A vector $v^a$ pointing up or down in time is called {\em timelike}.
It has negative norm, $g_{ab} v^a v^b <0$.  Massive particles (such as
observers) cannot attain or exceed the speed of light.  They follow
{\em timelike curves}, or {\em worldlines}, i.e., their tangent vector
is everywhere timelike.  A vector $k^a$ is called {\em null\/} or {\em
light-like\/} if its norm vanishes.  light rays follow {\em null
curves\/} through spacetime; their tangent vector is everywhere null.
{\em Spacelike vectors\/} have positive norm.  Spacelike curves
connect points that can be regarded as simultaneous (in some
coordinate system).  No physical object or information follows
spacelike curves; this would require superluminal speed.

\paragraph{Geodesic curves}

Curves that are ``as straight as is possible'' in a given curved
geometry are called {\em geodesics}.  They satisfy the {\em geodesic
equation},
\begin{equation}
{d^2x^a\over d\lambda^2} + \Gamma^a_{~bc} {dx^b\over d\lambda}
{dx^c\over d\lambda} = \alpha {dx^a\over d\lambda}.
\end{equation}
(The {\em Christoffel symbols}, $\Gamma^a_{~bc}$, are obtained from
the metric and its first derivatives.)  Any geodesic can be
reparametrized ($\lambda \to \lambda'$) so that $\alpha$ vanishes.  A
parameter with which $\alpha=0$ is called {\em affine}.

Unless non-gravitational forces act, a massive particle follows a
timelike geodesic.  Similarly, light rays don't just follow any null
curve; they generate a {\em null geodesic}.  We use the terms
``light ray'' and ``null geodesic'' interchangeably.

Two points are {\em timelike separated\/} if there exists a timelike
curve connecting them.  Then they can be regarded as subsequent events
on an observer's worldline.  Two points are {\em null separated\/} if
they are connected only by a light ray.  Two points are {\em spacelike
separated\/} if it is impossible for any object or signal to travel
from one point from the other, i.e., if they are connected only by
spacelike curves.

\paragraph{Visualization and light cones}

In all depictions of spacetime geometry in this paper, the time
direction goes up, and light rays travel at 45 degrees.  The light
rays emanating from a given event $P$ (e.g., when a bulb flashes) thus
form a cone, the {\em future light cone}.  light rays arriving at $P$
from the past form the {\em past light cone\/} of $P$.  They limit the
spacetime regions that an observer at $P$ can send a signal to, or
receive a signal from.

Events that are timelike separated from $P$ are in the interior of the
light cones.  Null separated events are on one of the light cones, and
spacelike separated events are outside the light cones.  The worldline
of a massive particle is always at an angle of less than 45 degrees
with the vertical axis.  A moment of time can be visualized as a
horizontal plane.

\paragraph{Surfaces and hypersurfaces}

In this text, the term {\em surface\/} always denotes a $D-2$
dimensional set of points, all of which are spacelike separated from
each other.  For example, a soap bubble at an instant of time is a
surface.  Its whole history in time, however, is not a surface.

A {\em hypersurface\/} $H$ is a $D-1$ dimensional subset of the
spacetime (with suitable smoothness conditions).  $H$ has $D-1$
linearly independent tangent vectors, and one normal vector, at every
point.  If the normal vector is everywhere timelike (null, spacelike),
then $H$ is called a {\em spacelike (null, timelike) hypersurface}.

Physically, a spacelike hypersurface can be interpreted as ``the world
at some instant of time''; hence, it is also called a {\em
hypersurface of equal time}, or simply, a {\em time slice}
(Fig.~\ref{fig-spheb}).  A timelike hypersurface can be interpreted as
the history of a surface.  A soap bubble, for example, inevitably
moves forward in time.  Each point on the bubble follows a timelike
curve.  Together, these curves form a timelike hypersurface.

Null hypersurfaces play a central role in this review, because the
holographic principle relates the area of a surface to the number of
degrees of freedom on a light-sheet, and light-sheets are null
hypersurfaces.  If a soap bubble could travel at the speed of light,
each point would follow a light ray.  Together, the light rays would
form a null hypersurface.  A particularly simple example of a null
hypersurface is a light cone.

More generally, a null hypersurface is generated by the light rays
orthogonal to a surface.  This is discussed in detail in
Sec.~\ref{sec-kin}.  As before, ``null'' is borderline between
``spacelike'' and ``timelike''.  This gives null hypersurfaces great
rigidity; under small deformations, they lose their causal character.
This is why any surface has only four orthogonal null hypersurfaces,
but a continuous set of timelike or spacelike hypersurfaces.

\paragraph{Penrose diagrams}

Many spacetimes contain infinite distances in time, or in space, or
both.  They have four or more dimensions, and they are generally not
flat.  All of these features make it difficult to draw a spacetime on
a piece of paper.

However, often one is less interested in the details of a spacetime's
shape than in global questions.  Are there observers that can see the
whole spacetime if they wait long enough?  Are parts of the spacetime
hidden behind horizons, unable to send signals to an asymptotic region
(i.e., are there black holes)?  Does the spacetime contain
singularities, places where Einstein's equation predicts its own
breakdown?  If so, are they timelike, so that they can be probed, or
spacelike, so that they lie entirely in the past or in the future?

Penrose diagrams are two-dimensional figures that capture certain
global features of a geometry while discarding some metric
information.  The ground rules are those of all spacetime diagrams:
time goes up, and light rays travel at 45 degrees.  An important new
rule is that (almost) every point represents a sphere.  This arises as
follows.

We assume that the spacetime ${\cal M}$ is at least approximately spherically
symmetric.  Then the only non-trivial coordinates are radius and time,
which facilitates the representation in a planar diagram.  Usually
there is a vertical edge on one side of the diagram where the radius
of spheres goes to zero.  This edge is the worldline of the origin of
the spherical coordinate system.  All other points in the diagram
represent $(D-2)$ spheres.  (In a closed universe, the spheres shrink
to zero size on two opposite poles, and the diagram will have two such
edges.  There are also universes where the spheres do not shrink to
zero anywhere.)

A {\em conformal transformation\/} takes the physical metric, $g_{ab}$,
to an {\em unphysical metric}, $\tilde g_{ab}$:
\begin{equation}
g_{ab}\to \tilde g_{ab} = \Omega^2 g_{ab}.
\end{equation}
The conformal factor, $\Omega$, is a function on the spacetime
manifold ${\cal M}$.  The unphysical metric defines an unphysical
spacetime $\tilde M$.

A conformal transformation changes distances between points.  However,
it is easy to check that it preserves causal relations.  Two points
that are spacelike (null, timelike) separated in the spacetime ${\cal
M}$ will have the same relation in the unphysical spacetime $\tilde
{\cal M}$.

Penrose diagrams exploit these properties.  A Penrose diagram of
${\cal M}$ is really a picture of an unphysical spacetime $\tilde
{\cal M}$ obtained by a suitable conformal transformation.  The idea
is to pick a transformation that will remove inconvenient aspects of
the metric.  The causal structure is guaranteed to survive.  Here are
two examples.

A judicious choice of the function $\Omega$ will map asymptotic
regions in ${\cal M}$, where distances diverge, to finite regions in
$\tilde {\cal M}$.  An explicit example is given by
Eq.~(\ref{eq-dsmetric}).  By dropping the overall conformal factor and
suppressing the trivial directions along the $(D-2)$ sphere, one
obtains the unphysical metric depicted in the Penrose diagram
(Fig.~\ref{fig-ds}).  The asymptotic infinities of de~Sitter space are
thus shown to be spacelike.  Moreover, the spacetime can now be
represented by a finite diagram.

A neighborhood of a singularity in the spacetime ${\cal M}$ can be
``blown up'' by the conformal factor, thus exposing the causal
structure of the singularity.  An example is the closed FRW universe,
Eq.~(\ref{eq-FRW2}); let us take $w\geq 0$ in Eq.~(\ref{eq-q}) and
(\ref{eq-a}).  Again, the prefactor can be removed by a conformal
transformation, which shows that the big bang and big crunch
singularities are spacelike (Fig.~\ref{fig-clos}).

Conformal transformations yielding Penrose diagrams of other
spacetimes are found, e.g., in Hawking and Ellis (1973), and Wald
(1984).

\paragraph{Energy conditions}

The stress tensor, $T_{ab}$, is assumed to satisfy certain conditions
that are deemed physically reasonable.  The {\em null energy
conditon\/}\footnote{``null convergence condition'' in Hawking and
Ellis (1973)} demands that
\begin{equation}
T_{ab} k^a k^b \geq 0 \mbox{~for all null vectors~} k^a.
\label{eq-nec}
\end{equation} 
This means that light rays are focussed, not anti-focussed, by
matter (Sec.~\ref{sec-ray}).
The {\em causal energy condition\/} is
\begin{equation} 
T_{ab} v^b T^{ac} v_c \leq 0 \mbox{~for all timelike vectors~} v^a.
\label{eq-cec}
\end{equation}
This means that energy cannot flow faster than the speed of light.

In Sec.~\ref{sec-econds}, the null and causal conditions are both
demanded to hold for any component of matter, in order to outline a
classical, physically acceptable regime of spacetimes in which the
covariant entropy bound is expected to hold.  The {\em dominant energy
condition\/} is somewhat stronger; it combines the causal energy
condition, Eq.~(\ref{eq-cec}), with the {\em weak energy condition},
\begin{equation}
T_{ab} v^a v^b \geq 0 \mbox{~for all timelike vectors~} v^a.
\end{equation}
 
In cosmology and in many other situations, the stress tensor takes the
form of a perfect fluid with energy density $\rho$ and pressure $p$:
\begin{equation}
T_{ab} = \rho u_a u_b + p (g_{ab} +u_a u_b),
\end{equation}
where the unit timelike vector field $u^a$ indicates the direction of
flow.  In a perfect fluid, the above energy conditions are equivalent
to the following conditions on $p$ and $\rho$.
\begin{eqnarray}
\mbox{null e.~c.:~~}   &  \rho \geq -p \\
\mbox{causal e.~c.:~~} &  |\rho| \geq |p| \\
\mbox{null and causal:~~} &  |\rho| \geq |p| \mbox{~and} \\
& \rho <0 \mbox{~only if~} \rho=-p \\
\mbox{weak e.~c.:~~} &  \rho \geq -p \mbox{~and~} \rho\geq 0 \\
\mbox{dominant e.~c.:~~} &  \rho \geq |p|
\end{eqnarray}
With the further assumption of a fixed equation of state, $p=w\rho$,
conditions on $w$ can be derived.

\bibliographystyle{myrmp} \bibliography{all}

\end{document}